\documentclass[epj]{svjour}  

\usepackage{color}
\usepackage[pdftex]{graphicx}
\usepackage{amssymb, amsfonts, amsmath}
\usepackage[utf8]{inputenc}
\usepackage{pgf,siunitx} 
\usepackage{etoolbox}    
\apptocmd{\sloppy}{\hbadness 10000\relax}{}{}

\graphicspath{{plots/}}

\renewcommand{\l}{\left}
\renewcommand{\r}{\right}
\newcommand{\cf}{{\it{cf. }}}
\newcommand{\eg}{{\it{e.g. }}}
\newcommand{\ie}{{\it{i.e. }}}
\newcommand{\g}[1]{\gamma_{#1}} 
\newcommand{\DB}[1]{\stackrel{\leftarrow}{D}_{#1}} 
\newcommand{\DF}[1]{\stackrel{\rightarrow}{D}_{#1}} 
\newcommand{\DBF}[1]{\stackrel{\leftrightarrow}{D}_{#1}} 

\newcommand{\diag}{\mathrm{diag}}  
\newcommand{\bra}[1]{\left< #1 \right|} 
\newcommand{\ket}[1]{\left| #1 \right>} 

\newcommand{\gev}{\,\mathrm{GeV}}

\newcommand{\mev}{\,\mathrm{MeV}}
\newcommand{\fm}{\,\mathrm{fm}}

\newcommand{\stat}{\mathrm{stat}}

\newcommand{\tsep}{t_\mathrm{sep}}
\newcommand{\tins}{t_\mathrm{ins}}

\newcommand{\Ctwopt}[2]{C^\mathrm{2pt}(\vec{#2}, #1)}

\newcommand{\ffMagneticConv}[1][Q^2]{{G^{{\text{Conv.}}}_{{M}}(#1)}}
\newcommand{\ffMagneticPEVA}[1][Q^2]{{G^{{\text{PEVA}}}_{{M}}(#1)}}

\newcommand{\indSdbl}[0]{{u_{\indSpr}}}

\newcommand{\indSne}[0]{{n}}

\newcommand{\indSpr}[0]{{p}}

\newcommand{\indSsing}[0]{{d_{\indSpr}}}

\begin{document}
\title{Excited states in nucleon structure calculations}
\subtitle{}
\author{Konstantin Ottnad\inst{1}}
\institute{PRISMA$^+$~Cluster~of~Excellence and Institut~f\"ur~Kernphysik, Johannes~Gutenberg-Universität~Mainz, 55099~Mainz, Germany, \email{kottnad@kph.uni-mainz.de}}

\date{Received: date / Revised version: date}

\abstract{
Excited state contributions represent a formidable challenge for hadron structure calculations in lattice QCD. For physical systems that exhibit an exponential signal-to-noise problem they often hinder the extraction of ground state matrix elements, introducing a major source of systematic error in lattice calculations of such quantities. The development of methods to treat the contribution of excited states and the current status of related lattice studies are reviewed with focus on nucleon structure calculations that are notoriously affected by excited state contamination.
}

\maketitle

\section{Introduction}
\label{sec:introduction}
The\, lattice\, formulation\, of\, Quantum\, Chromodynamics (QCD) is the only known {\it ab initio} method to study the properties of hadrons from first principles. Through the development of more powerful computers and algorithms over the last decades it has become possible to perform precise lattice calculations for a broad range of applications. However, in order to make contact with experimental results from a controlled extrapolations of lattice results to the physical point, it is crucial to not only improve statistical precision but also to achieve control over systematics. A lot of progress has been made regarding chiral, continuum and finite size extrapolations which are now part of many current lattice studies, as simulations at physical quark mass and including multiple values of the lattice spacing and volumes have become feasible in the last few years. Still, in the presence of an exponential signal-to-noise problem, which is typical for observables involving baryons, a fourth systematic effect often remains a challenge, \ie residual excited state contamination. The reason for this is that the signal is lost in noise at Euclidean time separations that are still insufficient to gain ground state dominance. Therefore, isolation of the ground state cannot be achieved in a direct way without taking further measures. \par

Most notably affected by this kind of systematic effect are lattice QCD calculations of nucleon structure, that cover a rich variety of observables. Among the most basic such observables are nucleon charges that can be computed from forward matrix elements at zero momentum transfer. An important example is the nucleon axial charge $g_A^{u-d}=1.2724(23)$~\cite{Tanabashi:2018oca} which is experimentally measured in neutron $\beta$-decay and often serves as a benchmark observable for nucleon structure calculations in lattice QCD. On the other hand, there is less experimental information available on scalar and tensor charges which may give beyond the Standard Model (BSM) contribution to the nucleon $\beta$-decay~\cite{Ivanov:2018vmz}. Therefore, control over excited states is all the more important in lattice QCD calculations of these observables which may provide crucial input on searches for BSM physics as they are relevant to dark matter searches~\cite{DelNobile:2013sia}. Furthermore, the tensor charge plays a role in BSM searches for $CP$-violation as it controls the contribution of quark electric dipole moments to the neutron electric dipole moment~\cite{Bhattacharya:2015esa}. Another example for an observable of great phenomenological interest at zero momentum transfer is the average quark momentum fraction which contributes to the nucleon spin decomposition~\cite{Ji:1996ek}. At non-vanishing momentum transfer various form factors are studied, \eg electromagnetic and axial form factors and the corresponding radii that parametrize the slope of these form factors at vanishing momentum. In particular, the proton radius has received a lot of attention in the last decade due to the so-called proton radius puzzle, see \eg refs.~\cite{Mergell:1995bf,Bernauer:2010wm,Pohl:2010zza,Mohr:2012tt,Lorenz:2012tm,Lorenz:2014yda,Antognini:1900ns}, although this has likely been resolved by now~\cite{Hammer:2019uab} and the most recent measurement from electron-proton scattering~\cite{Xiong:2019umf} indeed agrees with the very precise results obtained from measurements of the Lamb shift measured for muonic hydrogen~\cite{Pohl:2010zza,Antognini:1900ns}. Axial form factors in turn are experimentally less well-known~\cite{Bernard:2001rs,AguilarArevalo:2010zc,Hill:2017wgb} but may provide critical input for future experiments related to neutrino physics~\cite{Alvarez-Ruso:2017oui,Kronfeld:2019nfb}, which makes them attractive observables to be studied in lattice QCD. \par

The aim of this article is to review the state-of-the-art of methods that are applied to tame residual excited state contamination in modern nucleon structure calculations and to perform a critical assessment of the efficacy of these methods. Therefore, further systematics related to chiral, continuum and finite size extrapolations that affect these calculations are not within the scope of this review. From a physics point of view, a subset of nucleon structure observables will be considered for which excited state contamination is known to be an important, or even the dominant systematic. The selection comprises observables that can be computed with sufficiently good statistical precision and for which dedicated studies of excited state systematics and related methods can be found in the literature. \par

The review is organized as follows: In the next section basic methods for nucleon structure calculations in lattice QCD are summarized and observables relevant to this review are defined. In sect.~\ref{sec:excited_states} theoretical aspects of excited states and their expected effects on such calculations are discussed. The following sections are dedicated to the review of various methods that have been employed in the treatment of excited state contamination and their respective applications in recent lattice studies. More specifically, methods that aim at additional suppression of excited states by summation over the operator insertion are reviewed in sect.~\ref{sec:summation_method}, while sects.~\ref{sec:multi_state_fits}~and~\ref{sec:variational_techniques} deal with the various application of multi-state fits and the variational approach, respectively. In the end, a summary on the advantages and shortcomings of the individual methods and a brief outlook on possible future developments is given. \par

\section{Nucleon structure calculations in lattice QCD} \label{sec:nucleon_structure}
The calculation of nucleon properties in lattice QCD is based on the numerical evaluation of $n$-point functions in discretized Euclidean spacetime. For the simple example of a two-point function
\begin{equation}
 \Ctwopt{t}{p}=\sum_{\vec{x}} e^{i \vec{p} \vec{x}} \langle \chi(\vec{x}, t) \chi^\dag(\vec{0}, 0) \rangle \,,
 \label{eq:2pt} 
\end{equation}
with source at the origin and sink at $(\vec{x}, t)$, the corresponding spectral decomposition
\begin{equation}
 \Ctwopt{t}{p}=\sum_{k} \frac{1}{2E_k(\vec{p})} \l| \bra{\Omega} \chi \ket{k, \vec{p}} \r| e^{-E_k(\vec{p}) t} \,,
 \label{eq:2pt_spectral_decomposition}
\end{equation}
contains all possible hadronic states $\ket{k, \vec{p}}$ with integer label $k$ and energies $E_k(\vec{p})$ that share the same continuum quantum numbers compatible with the choice of the interpolating operator $\chi(\vec{x}, t)$. This typically includes excitations of the ground state as well as matching multi-particle states. The reason for this is that the wave functions of the hadronic states $\ket{k}$ are unknown and thus it is impossible to directly construct interpolating operators that couple only to a desired state $k'$ while the remaining overlap factors $\bra{0} \chi \ket{k}$ for $k\neq k'$ vanish exactly. Moreover, rotational symmetry is broken at finite values of the lattice spacing which may lead to additional mixing between operators that would otherwise fall into different irreducible representation in the continuum limit. Therefore, the extraction of the ground state energies and overlap factors requires to compute $\Ctwopt{t}{\vec{p}}$ at large Euclidean time separation $t$, such that all higher terms are exponentially suppressed 
\begin{align}
 \Ctwopt{t}{p} \rightarrow& \l| \bra{\Omega} \chi \ket{0, \vec{p}} \r|^2 e^{-E_0(\vec{p}) t} \notag \\ 
  &\times \l(1 + \mathcal{O}\l(e^{-(E_{1}(\vec{p})-E_{0}(\vec{p})) t}\r) \r) \,.
 \label{eq:2pt_leading_term}
\end{align}
This motivates the definition of the effective energy
\begin{equation}
 E_\mathrm{eff}(\vec{p},t,\tau) = \frac{1}{\tau} \log\frac{\Ctwopt{t}{\vec{p}}}{\Ctwopt{t+\tau}{\vec{p}}} \,,
 \label{eq:eff_mass}
\end{equation}
which is called effective mass for $\vec{p}=0$. For large values of $t$ it asymptotically approaches a plateau from which the energy can be extracted. The parameter $\tau$ is commonly set to $\tau/a=1$. In practice, the plateau can be identified once the residual slope has become negligible compared to the statistical error of the individual data points. \par

\subsection{Nucleon matrix elements and ratio method} \label{subsec:NME}
While hadron masses can be readily obtained from two-point functions, the study of the structure of hadrons from lattice QCD relies on the computation of matrix elements. With respect to excited states a particularly relevant application are nucleon matrix elements (NMEs)
\begin{align}
 &\bra{N(p_f,s_f)} \mathcal{O}^X_{\mu_1...\mu_n} \ket{N(p_i,s_i)} \notag \\
 &= \bar{u}(p_f, s_f) W^X_{\mu_1...\mu_n}(Q^2)  u(p_i,s_i) \,,
 \label{eq:NME}
\end{align}
where $N(p_i,s_i)$, $N(p_f,s_f)$ denote nucleon states with initial (final) state momentum $p_i$ ($p_f$) and spin $s_i$ ($s_f$). On the right hand side of $u(p_i,s_i)$, $\bar{u}(p_f, s_f)$ are the corresponding Dirac spinors, while $W^X_{\mu_1...\mu_n}(Q^2)$ contains form factors and kinematic factors. The form factor decomposition depends on the choice of the operator insertion that we restrict to a bilinear operator of the form
\begin{equation}
 \mathcal{O}^X_{\mu_1...\mu_n} = \bar{q}(x) \Gamma^X_{\mu_1...\mu_n} q(x).
 \label{eq:generic_insertion}
\end{equation}
where $\Gamma^X_{\mu_1...\mu_n}$ collects Dirac matrices and possibly derivatives, and $q(x)$, $\bar{q}(x)$ denote quark fields. Actual examples of operator insertions relevant to nucleon structure calculations will be discussed in subsect.~\ref{subsec:observables}. In practice, most lattice calculations are performed assuming exact isospin symmetry in the light quark sector. In this case, the insertion of the isovector combination
\begin{equation}
 \mathcal{O}^{X,u-d}_{\mu_1...\mu_n} = \bar{u}(x) \Gamma^X_{\mu_1...\mu_n}  u(x) - \bar{d}(x) \Gamma^X_{\mu_1...\mu_n} d(x) \,, 
 \label{eq:isovector_op}
\end{equation}
leads to the cancellation of quark-disconnected diagrams that are inherently more noisy then quark-connected contributions. For the isoscalar counterpart $\mathcal{O}^{X,u+d}_{\mu_1...\mu_n}$ this cancellation does not occur, which further increases the signal-to-noise problem. Moreover, for studies of observables that probe the contributions of strange and charm quarks to nucleon structure such as \eg strange and charm electromagnetic form factors, quark-disconnected diagrams give the only contribution apart from possible mixing with quark-connected contributions under renormalization. \par

The lattice determination of NMEs as defined in eq.~(\ref{eq:NME}) generally requires the computation of spin-projected two-point functions
\begin{equation}
 \Ctwopt{t-t_i}{p}=\Gamma_\mathrm{2pt}^{\alpha\beta} \sum_{\vec{x}} e^{i \vec{p} \cdot (\vec{x}-\vec{x}_i)} \langle \chi_\alpha(\vec{x},t) \bar{\chi}_\beta(\vec{x}_i,t_i)\rangle\,,
  \label{eq:2pt_xspace}
\end{equation}
and three-point functions
\begin{align}
  &C^X_{\mu_1...\mu_n}(\vec{p}_f, \vec{p}_i, t_\mathcal{O}-t_i, t_f-t_i) \notag \\
    &\quad=\Gamma_\mathrm{3pt}^{\alpha\beta} \sum_{\vec{x}_f, \vec{x}_\mathcal{O}}  e^{i \vec{p}' \cdot (\vec{x}_{f}-\vec{x}_\mathcal{O})} e^{i \vec{p} \cdot (\vec{x}_\mathcal{O}-\vec{x}_{i})} \notag \\
    &\qquad\times \langle \chi_\alpha(\vec{x}_f, t_f) \mathcal{O}^X_{\mu_1...\mu_n}(\vec{x}_\mathcal{O}, t_\mathcal{O}) \bar{\chi}_\beta(\vec{x}_i, t_i)\rangle \,.
  \label{eq:3pt_xspace}
\end{align}
where $\chi_\alpha(\vec{x}_f, t_f)$ and $\bar{\chi}_\beta(\vec{x}_i, t_i)$) denote interpolating operators for the final and initial nucleon state, $\Gamma^{\alpha\beta}_\mathrm{2pt}$ and $\Gamma^{\alpha\beta}_\mathrm{3pt}$ are suitable spin projectors and the operator is inserted at $x_\mathcal{O} = (\vec{x}_\mathcal{O}, t_\mathcal{O})$. A common choice for the nucleon interpolating field is given by e.g.
\begin{equation}
 \chi_\alpha(\vec{x},t) = \epsilon_{abc} (u_a^T(\vec{x},t) C\gamma_5 d_b(\vec{x}, t)) u_{c,\alpha}(\vec{x},t)  \,,
 \label{eq:standard_interpolator}
\end{equation}
where $C$ is the charge conjugation matrix and $u$, $d$ denote up- and down quark fields that are usually smeared. \par

Assuming that the source time is zero, the two-point function in momentum space for $t>0$ reads
\begin{align}
 \Ctwopt{t}{p} &= \Gamma_\mathrm{2pt}^{\alpha\beta} \langle \chi_\alpha(\vec{p}, t) \bar{\chi}_\beta(\vec{p}, 0)\rangle \notag \\
               &= \sum_k A_k(\vec{p}) A_k^*(\vec{p}) e^{-E_k(\vec{p}) t} \,,
 \label{eq:2pt_pspace}
\end{align}
where the overlap of the interpolating operator $\chi$ with the $k$-th state $A_k(\vec{p})=\bra{\Omega} \chi \ket{k, \vec{p}}$ of momentum $\vec{p}$ has been introduced in the spectral decomposition in the second line. Similarly, the three-point function can be expressed as
\begin{align}
 &C^X_{\mu_1...\mu_n}(\vec{p}_f, \vec{p}_i, \tins, \tsep) \notag \\
  &\quad = \Gamma_\mathrm{3pt}^{\alpha\beta} \langle \chi_\alpha(\vec{p}_f, \tsep)  \mathcal{O}^X_{\mu_1...\mu_n}(\vec{p}_f-\vec{p}_i, \tins) \bar{\chi}_\beta(\vec{p}_i, 0) \rangle \notag \\
  &\quad = \sum_{k,l} A_k(\vec{p}_f) A_l(\vec{p}_i)^* \bra{k, \vec{p}_f} \mathcal{O}^X_{\mu_1...\mu_n} \ket{l, \vec{p}_i} \notag \\ 
  &\qquad\quad \times e^{-E_k(\vec{p}_f)(\tsep-\tins)} e^{-E_l(\vec{p}_i) \tins}  \,, 
 \label{eq:3pt_pspace}
\end{align}
where the source-sink separation $\tsep = t_f - t_i$ and the short-hand $\tins=t_\mathcal{O}-t_i$ for the insertion time have been defined and again $t_i=0$ has been assumed. The extraction of the ground state matrix element $\bra{0,\vec{p}_f}  \mathcal{O}^X_{\mu_1...\mu_n} \ket{0,\vec{p}_i}$ requires cancellation of the unknown overlap factors $A_k(\vec{p}_f)$ and $A_l(\vec{p}_i)$. This can be achieved by taking an appropriate ratio of two- and three-point functions. A common choice is the ratio \cite{Alexandrou:2006ru}
\begin{align}
 & R^X(\vec{p}_f, \vec{p}_i, \tins, \tsep) = \frac{C^X_{\mu_1...\mu_n}(\vec{p}_f, \vec{p}_i, \tins, \tsep)}{C^\mathrm{2pt}(\vec{p}_f, \tsep)} \notag \\ 
 &\quad \times \sqrt{\frac{C^\mathrm{2pt}(\vec{p}_i,\tsep-\tins) C^\mathrm{2pt}(\vec{p}_f, \tins) C^\mathrm{2pt}(\vec{p}_f, \tsep)}{C^\mathrm{2pt}(\vec{p}_f, \tsep-\tins) C^\mathrm{2pt}(\vec{p}_i, \tins) C^\mathrm{2pt}(\vec{p}_i, \tsep)}} \,, 
 \label{eq:ratio}
\end{align}
which has been shown to be particularly beneficial with respect to statistical errors in Ref.~\cite{Alexandrou:2008rp}. For asymptotically large Euclidean time separations ground state dominance is achieved 
\begin{equation}
 \lim_{\tins\rightarrow\infty} \lim_{\tsep\rightarrow\infty} R^X(\vec{p}_f, \vec{p}_i, \tins, \tsep) = \bra{0, \vec{p}_f} \mathcal{O}^X_{\mu_1...\mu_n} \ket{0, \vec{p}_i} \,.
 \label{eq:ratio_limit}
\end{equation}
giving access to the matrix element from a plateau of the effective form factor, similar to the extraction of the ground state energy from eq.~(\ref{eq:eff_mass}). For this reason the ratio method for matrix elements is also sometimes referred to as plateau method. \par

\subsection{Lattice techniques}
The computation of NMEs from the ratio in eq.~(\ref{eq:ratio}) requires three-point functions at several source-sink separations $\tsep$ to allow for a controlled study of excited state effects. The {\it{de facto}} standard approach is to use sequential inversion through a fixed sink, which gives access to all values of the insertion time $\tins$ for any given value of $\tsep$. However, in this setup each source-sink separation needs additional sequential inversions. Besides, the number of inversion per source-sink separation increases if non-local operator insertions are used, e.g. for point-split currents or the derivative operators in eqs.~(\ref{eq:OvD})--(\ref{eq:OtD}). \par

Another possibility is to perform sequential inversions through a fixed operator insertion at $\tins=\mathrm{const}$ \cite{Martinelli:1988rr}. This approach is complementary to the previous one as it gives access to all values of $\tsep$ without the need for additional inversions while allowing for also freely choosing momentum projection and operator at sink without incurring extra computational cost. The latter makes this an attractive choice for certain setups used for employing a dedicated variational analysis that will be discuss in sect.~\ref{sec:variational_techniques}. However, additional inversions are needed for each value of $\tins$ as well as every operator insertion which usually outweighs the aforementioned advantages in terms of computational cost. \par

A third method that effectively eliminates $\tins$ is obtained by summation over the operator insertion. This approach can be considered a particular case of the summation method~\cite{Maiani:1987by} and will be discussed in sect.~\ref{subsec:Feynmann_Hellmann}. In this setup it is possible to change the operator at sink without the need for additional inversion, but again not the operator insertion. It has only been employed in a few studies in recent years \cite{Bouchard:2016heu,Savage:2016kon,Chang:2018uxx}. \par

Most calculations of connected contributions use point-like sources in combination with smearing of the quark fields $q(x)$ to increase the ground state overlap, hence improving suppression of excited state contamination. This setup gives also full flexibility with respect to momentum projection at the source unlike \eg timeslice sources that would require new inversion for each momentum. A common choice for the smearing of quark fields is Gaussian smearing \cite{Gusken:1989qx,Alexandrou:1992ti}
\begin{equation}
 q(x) \rightarrow \tilde{q}(x) = \l( 1 + \kappa_G \Delta \r)^{N_\mathrm{smear}}
 \label{eq:gaussian_smearing}
\end{equation}
together with spatial APE smearing \cite{Albanese:1987ds} of the gauge fields that enter through the three-dimensional Laplacian $\Delta$. In general, the smearing parameter $\kappa_G$ and the number of smearing steps $N_\mathrm{smear}$ need to be tuned to give optimal results. Besides, smearing can be used as a comparably cheap way to define additional interpolating operators which will be discussed in the context of variational techniques in sect.~\ref{sec:variational_techniques}. For the fixed sink and fixed insertion method the same point-to-all forward propagators can be used for two- and three-point functions.\par

Since control over excited states depends on large source-sink separations, methods to reduce the computational cost are of particular importance for nucleon structure calculations as this requires high statistics. Commonly used approaches for the sequential method are the truncated solver method \cite{Bali:2009hu} or all-mode averaging (AMA) \cite{Blum:2012uh}, that allow to increase statistics at reduced cost compared to performing only exact inversions. Further methods that have been investigated include replacing the conventional sequential propagators by a stochastic estimator \cite{Alexandrou:2013xon,Yang:2015zja,Bali:2017mft} which facilitates the reuse of the propagator at the expense of introducing stochastic noise, the coherent sink method \cite{Bratt:2010jn,Bali:2019yiy} that allows to invert simultaneously on multiple, temporally separated sequential sources, and finally distillation \cite{Peardon:2009gh} which is suitable to systematically build a basis of interpolating operators and that has been employed in a recent study in ref.~\cite{Egerer:2018xgu}.  \par

Furthermore, a lot of progress has been made in the last few years regarding the inclusion of disconnected diagrams which has become feasible through the development and application of variance reduction techniques such as hierarchical probing \cite{Stathopoulos:2013aci}, low-mode deflation \cite{Gambhir:2016uwp}, the one-end trick \cite{McNeile:2006bz}, the hopping parameter expansion \cite{Thron:1997iy,Michael:1999rs} and combinations thereof \cite{Alexandrou:2018sjm,Giusti:2019kff}. This lead to a substantial reduction in computational cost compared to \eg using naive stochastic all-to-all estimators. However, the relative statistical error of quark-disconnected contributions remains inherently larger than the one of purely quark-connected diagrams even if gauge noise is reached. \par

\subsection{Observables} \label{subsec:observables}
There exists a large variety of nucleon structure related observables that are computed in lattice QCD. However, for the purpose of this review mainly a subset of observables will be considered for which uncontrolled excited state contributions have been shown to produce a bias at the current level of precision. As a first example, the insertion of a local vector current for a generic quark flavor
\begin{equation}
 \mathcal{O}^{V}_{\mu}(x)=\bar{q}(x) \g{\mu}\g{5} q(x) \,,
 \label{eq:O_V}
\end{equation}
or the corresponding point-split current gives rise to the electromagnetic form factors $F_1(Q^2)$, $F_2(Q^2)$ through
\begin{equation}
 W^V_\mu = \gamma_\mu F_1(Q^2) + \frac{\sigma_{\mu\nu} Q_\nu}{2N_N} F_2Q^2) \,.
 \label{eq:F1F2}
\end{equation}
Usually, mean square radii $\langle r^2 \rangle$ are defined from the expansion of a form factor in the Euclidean four-momentum transfer $Q^2$ around zero 
\begin{equation}
 F(Q^2) = F(0) \l(1 + \frac{Q^2}{6} \langle r^2 \rangle +\mathcal{O}(Q^2) \r) \,.
 \label{eq:FF_expansion}
\end{equation}
For example the electric and magnetic mean square charge radii of the nucleon are given by
\begin{equation}
 \langle r_{E,M}^2 \rangle = -6 \l.\frac{dG_{E,M}(Q^2)}{dQ^2} \r|_{Q^2=0} \,,
 \label{eq:EM_radii}
\end{equation}
where $F_1(Q^2)$, $F_2(Q^2)$ enter through the Sachs form factors
\begin{align}
 G_E =& F_1(Q^2) - \frac{Q^2}{4M_N^2} F_2(Q^2) \,, \label{eq:G_E} \\
 G_M =& F_1(Q^2) + F_2(Q^2)  \,. \label{eq:G_M}
\end{align}
Note that for the neutron $G_E^n(0)=F_1^n(Q^2)=0$ and the factor $1/F_1^n(0)$ is dropped from eq.~(\ref{eq:EM_radii}) to obtain a finite definition. Since radii are not explicitly momentum dependent they are more readily compared to other lattice calculations than the full momentum dependence of the form factors themselves. Moreover, radii can be directly compared to experimental results as well. \par

\begin{figure*}
 \includegraphics[totalheight=0.2\textheight]{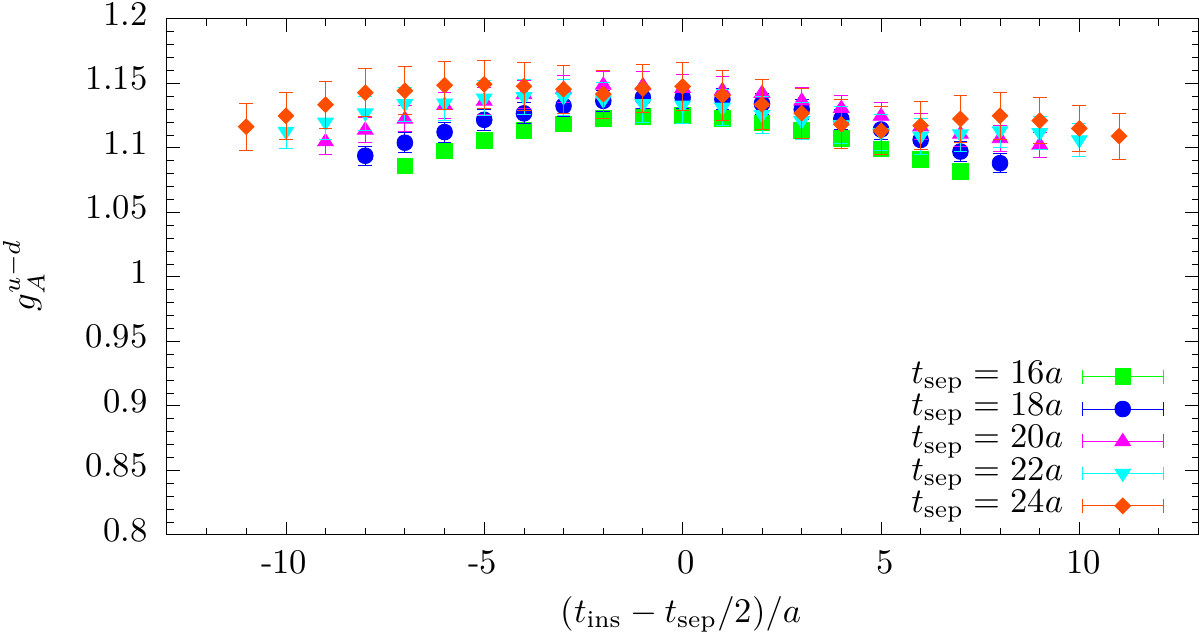}
 \includegraphics[totalheight=0.2\textheight]{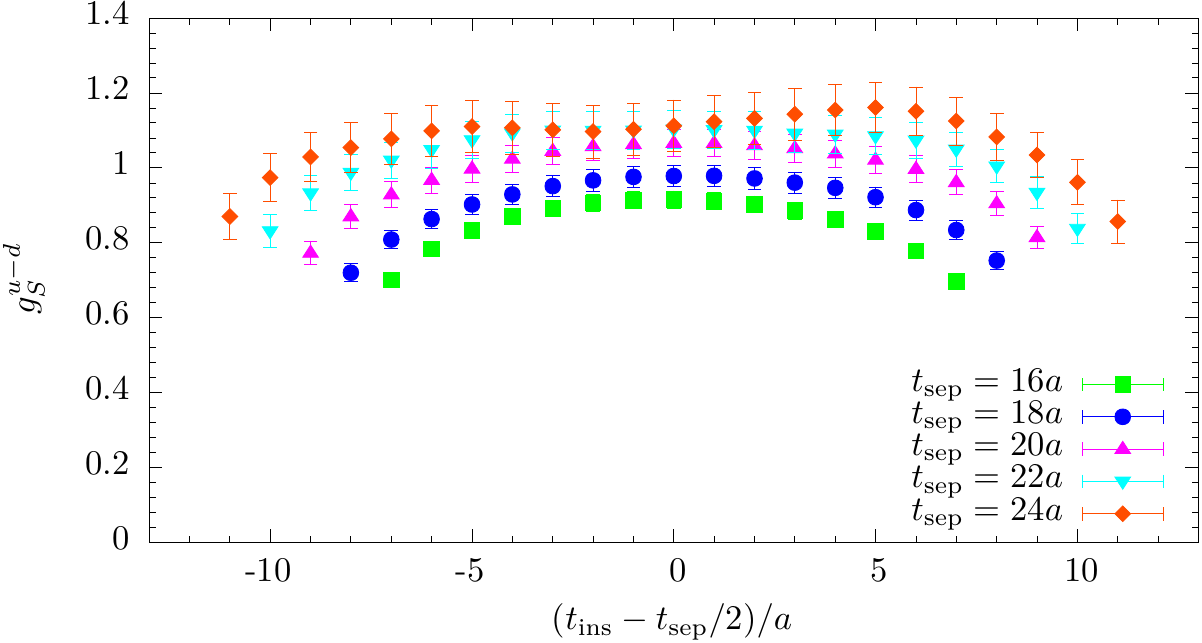} \\
 \includegraphics[totalheight=0.2\textheight]{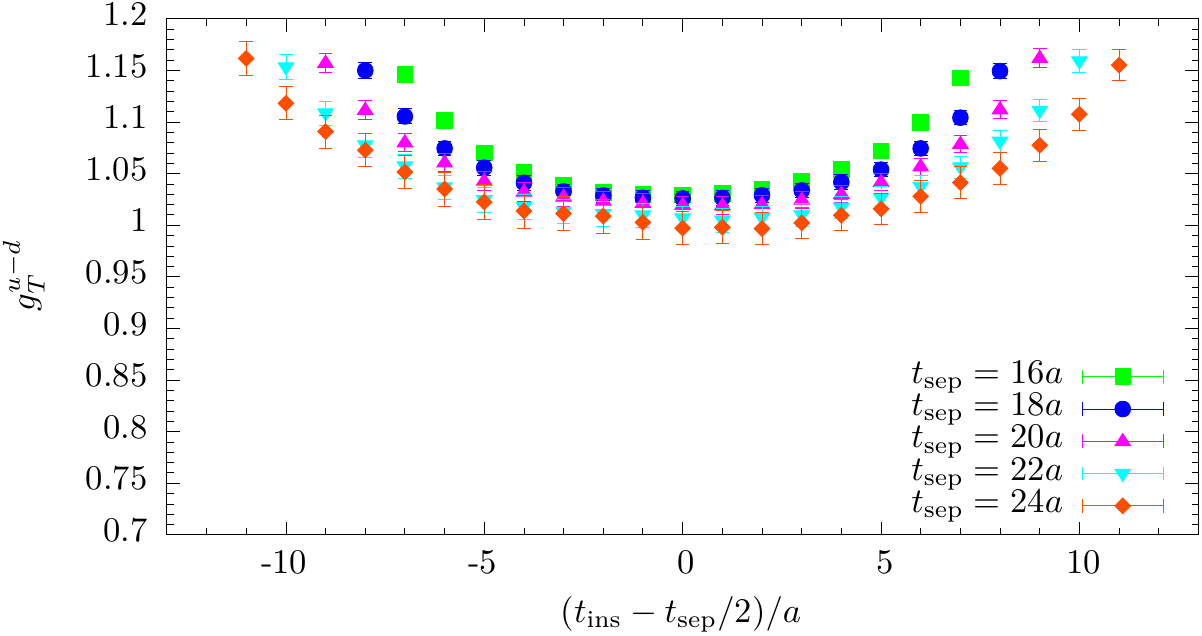}
 \includegraphics[totalheight=0.2\textheight]{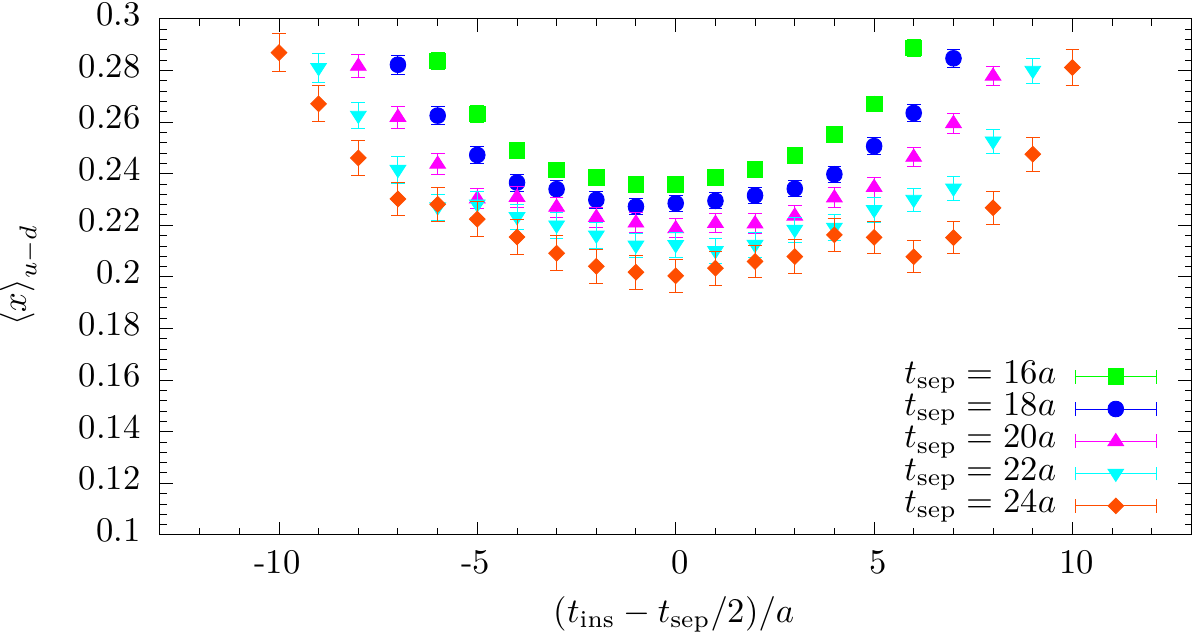}
 \caption{Examples for the effective form factors of the isovector charges $g_{A,S,T}^{u-d}$ and the average quark momentum fraction $\langle x \rangle_{u-d}$ from the ratio method as a function of the insertion time $\tins$ and for five values of the source-sink separation corresponding to $\tsep\in\l[1.03\fm, 1.54\fm\r]$. Results have been computed on 1540 configurations of a CLS ensemble (labeled N203; c.f. ref.~\cite{Bruno:2014jqa}) with $M_\pi\approx 350\mev$, $a=0.06426\fm$ and $T\times L^3 = 128a \times (48a)^3$ as part of a study published in Ref.~\cite{Harris:2019bih} for a fixed number of measurements on each configuration independent of the value of $\tsep$.}
 \label{fig:N203_g_AST_and_avg_x}
\end{figure*}

Another set of observables for which excited state effects are particularly relevant are nucleon charges defined at zero momentum transfer. However, for the vector current the electric charge $G_E(0)$ is conserved and the computation of the magnetic moment $\mu_M = G_M(0)$ is hindered by a momentum prefactor in the form factor decomposition which vanishes at $Q^2=0$. The latter introduces additional systematic effects as the extraction of $\mu_M$ requires either an extrapolation in $Q^2$ or position space methods, see \eg refs.~\cite{deDivitiis:2012vs,Alexandrou:2016rbj}. The situation is different for NMEs from other local bilinear operators, \ie the axial, scalar and tensor currents
\begin{align}
 \mathcal{O}^{A}_{\mu}(x)    &= \bar{q}(x) \g{\mu}\g{5} q(x) \,, \label{eq:OA} \\
 \mathcal{O}^{S}_{}(x)       &= \bar{q}(x) q(x) \,, \label{eq:OS} \\ 
 \mathcal{O}^{T}_{\mu\nu}(x) &= \bar{q}(x) \sigma_{\mu\nu} q(x) \,. \label{eq:OT}
\end{align}
for which the associated charges $g_{A,S,T}$ are not conserved. While radii require knowledge of the momentum dependence of the form factor, the axial, scalar and tensor charges are readily computed at zero-momentum transfer (up to renormalization) from a simple ratio of a three- and a single two-point function
\begin{equation}
 R^X_{\mu_1...\mu_n}(\vec{p}, \vec{p}, \tins, \tsep) = \frac{C^X_{\mu_1...\mu_n}(\vec{p}, \vec{p}, \tins, \tsep)}{\Ctwopt{\tsep}{p}} \,,
 \label{eq:ratio_zero_momentum}
\end{equation}
for equal initial and final state momentum $\vec{p}\equiv\vec{p}_i=\vec{p}_f$. For example, the isovector nucleon axial charge $g_A^{u-d}$ that is defined as
\begin{equation}
 \bra{P(p,s_f)} \bar{u} \g{\mu}\g{5} d \ket{N(p, s_i)} = g_A^{u-d} \bar{u}(p,s_f) \g{\mu} \g{5} u(p, s_i) \,,
 \label{eq:definition_isov_g_A}
\end{equation} 
for $p_f=p_i\equiv p$ where $P(p,s_f)$ and $N(p, s_i)$ refer to the final proton and initial neutron state, is obtained from
\begin{equation}
  \lim_{\tins\rightarrow\infty} \lim_{\tsep\rightarrow\infty} \frac{C^{A}_\mu(\vec{p}, \vec{p}, \tins, \tsep)}{\Ctwopt{\tsep}{\vec{p}}} \rightarrow g_A \,.
  \label{eq:g_A_from ratio}
\end{equation}
Some example data for the ratio as a function of $\tsep$ for the three charges are shown in the first three panels of fig.~\ref{fig:N203_g_AST_and_avg_x}. Since radii are less straightforward to compute, it is the experimentally well-known isovector axial charge $g_A^{u-d}$ that is considered a benchmark quantity for lattice QCD, although the statistical quality for the signal of a vector current insertion is superior to any of the operators in eqs.~(\ref{eq:OA})--(\ref{eq:OT}). Still, the axial charge has rather good signal quality compared to \eg the scalar charge shown in the upper right panel of fig.~\ref{fig:N203_g_AST_and_avg_x}, which makes statistical errors of order $\mathcal{O}(1\%)$ achievable in modern lattice simulations. Beyond charges also the axial and (induced) pseudoscalar form factors $G_A(Q^2)$ and $G_P(Q^2)$ have very recently received attention with focus on excited states from both, the theory \cite{Bar:2018xyi,Bar:2019gfx,Bar:2019igf} and the lattice side \cite{Bali:2018qus,Bali:2019yiy,Jang:2019vkm}. This will be discussed in more detail in subsects.~\ref{subsec:multi_particle_states}~and~\ref{subsec:ratio_fits}. \par

In addition to nucleon charges, the related form factors at non-zero $Q^2$ and radii, NMEs have been studied for operators involving derivatives, \eg the twist-2 operators
\begin{align}
 \mathcal{O}^{vD}_{\mu\nu}(x)     &= \bar{q}(x) \g{\l\{\mu\r.} \DBF{\l.\nu\r\}} q(x) \,, \label{eq:OvD} \\
 \mathcal{O}^{aD}_{\mu\nu}(x)     &= \bar{q}(x) \g{\l\{\mu\r.} \g{5} \DBF{\l.\nu\r\}} q(x) \,, \label{eq:OaD} \\
 \mathcal{O}^{tD}_{\mu\nu\rho}(x) &= \bar{q}(x) \sigma_{\l[\mu\l\{\nu\r.\r]} \DBF{\l.\rho\r\}} q(x) \,, \label{eq:OtD}
\end{align}
where $\left\{...\right\}$ and $\left[...\right]$ denote symmetrization with subtraction of the trace and antisymmetrization, respectively, and the symmetric derivative is defined as $\DBF{\mu}=\frac{1}{2} (\DF{\mu}-\DB{\mu})$. Matrix elements of these operators occur for example in the computation of the average quark momentum fraction of the nucleon $\langle x \rangle_q$, helicity and transversity moments $\langle x \rangle_{\Delta q}$ and $\langle x \rangle_{\delta q}$ or for the generalized parton distribution functions contributing in the computation of the spin decomposition of the proton, see \eg refs.~\cite{Alexandrou:2017oeh,Alexandrou:2020sml}. The latter involves also a gluonic operator insertion. In principle, any such NME may exhibit sizable excited state effects, as will be discussed in the next section. A rather well-studied example is the average quark momentum fraction of the nucleon for which example data is shown in the lower right panel of fig.~\ref{fig:N203_g_AST_and_avg_x} for the isovector combination. However, for many of these observables a systematic treatment of excited states is more difficult due to the signal quality which especially becomes an issue if quark-disconnected contributions are involved, see \eg refs.~\cite{Gupta:2018lvp,Djukanovic:2019jtp,Alexandrou:2019olr}. It is for this reason that methods to treat excited states are currently of particular importance compared for simple observables like $g_{A,S,T}^{u-d}$, $\langle x \rangle_{u-d}$ or in precision calculations of the nucleon radius because the achievable statistical error for these quantities does not allow to conceal the systematic effect of residual excited state contamination in final results. \par

\section{Excited states} \label{sec:excited_states}
While excited states are present in essentially any lattice calculation, they become especially an issue in calculations of nucleon structure. In the following the reasons for this are discussed in more detail and some theoretical expectations from chiral perturbation theory ($\chi$PT) and modeling excited states are reviewed. \par

\subsection{The signal-to-noise problem} \label{subsec:signal2noise}
\begin{figure}
 \includegraphics[totalheight=0.25\textheight]{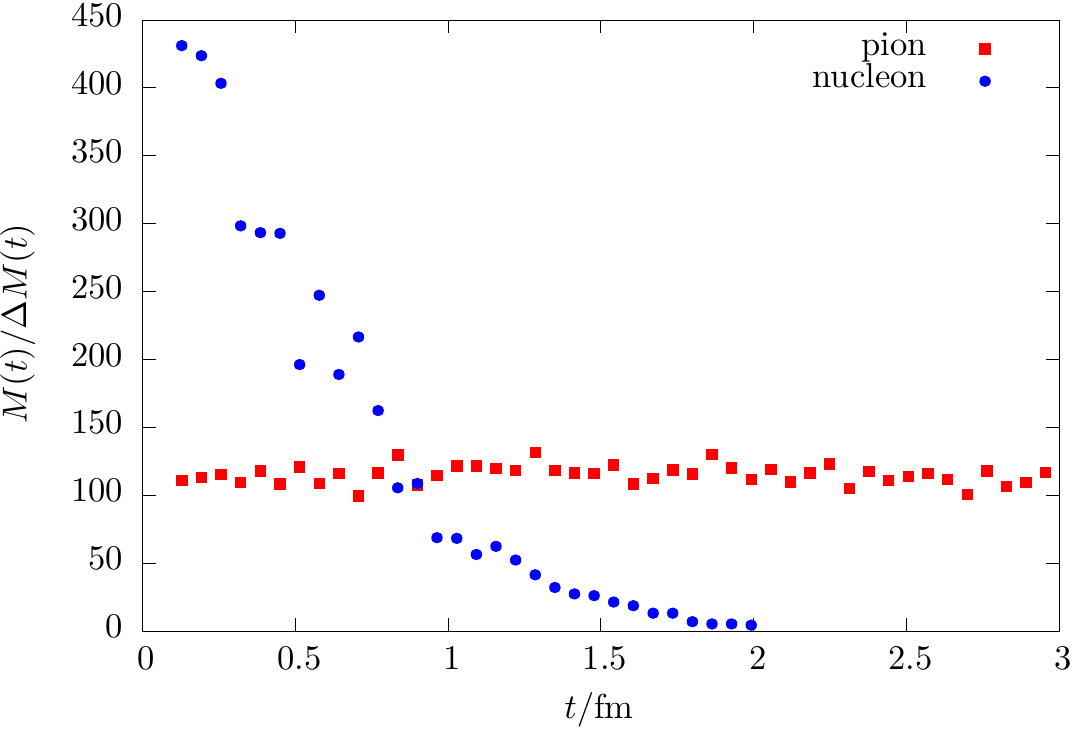}
 \caption{Signal-to-noise ratio for the effective mass $M(t)/\Delta M(t)$ of a pion and a nucleon two-point correlator computed on 250 configurations of a CLS ensemble (E250) with physical quark mass ($T\times L^3 = 192a\times (96a)^3$, $a=0.06426\fm$).}
 \label{fig:M_N_vs_Mpi_signal2noise}
\end{figure}

The primary cause why excited states are a persistent issue in nucleon structure calculation is an exponential signal-to-noise problem that prevents computation of n-point functions at large Euclidean time separations as required for \eg obtaining a plateau in the effective mass in eq.~(\ref{eq:eff_mass}) or for the ratio method in eq.~(\ref{eq:ratio_limit}). For the nucleon this problem has been formulated a long time ago \cite{Parisi:1983ae,Lepage:1989hd} from a field theoretical computation of the variance $\sigma_\stat^2(t)$ of the nucleon two-point function, showing that the dominating contribution at large values of $t$ is a three pion state, i.e.
\begin{equation}
 \sigma_\stat^2(t) \sim e^{-3M_\pi t} \,.
 \label{eq:variance_nucleon_2pt}
\end{equation}
For the signal-to-noise behavior of a nucleon two-point function this implies
\begin{equation}
 \frac{\Ctwopt{t}{0}}{\Delta \Ctwopt{t}{0}} \sim e^{-(M_N-\frac{3}{2} M_\pi) t} \,,
 \label{eq:nucleon_2pt_signal_to_noise}
\end{equation}
unlike the pion, for which a constant signal-to-noise ratio is expected. An example is shown in fig.~\ref{fig:M_N_vs_Mpi_signal2noise} for the effective mass of the nucleon and the pion computed on an ensemble at physical quark mass. Clearly, the signal for the nucleon is lost between $1.5\fm$ and $2\fm$, while for the pion at rest the ratio $M_\pi(t) /\Delta M_\pi(t)$ remains constant as expected. \par 

\begin{figure}
 \includegraphics[totalheight=0.25\textheight]{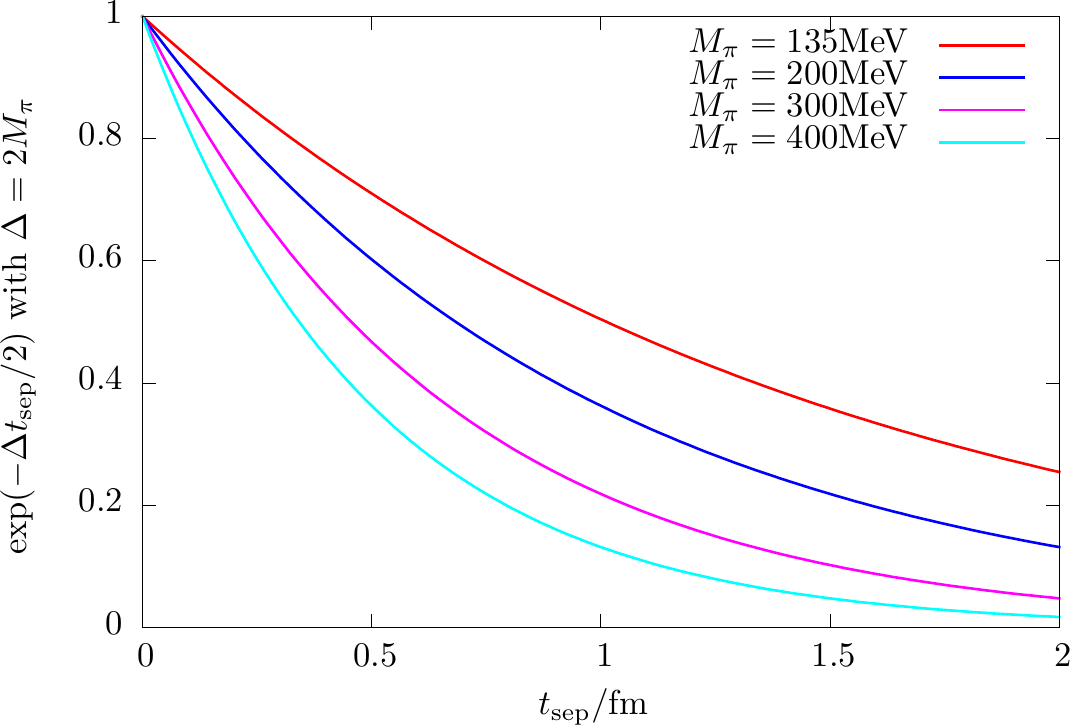}
 \caption{Expected scaling of the leading excited state contamination for the midpoint in the ratio method assuming a mass gap of $\Delta = 2 M_\pi$ as a function of $\tsep$ for different values of the pion mass.}
 \label{fig:ratio_midpoint_scaling}
\end{figure}

Regarding the behavior of excited states and the signal-to-noise problem in the ratio method, it is instructive to consider the two-state truncation for the zero-momentum case (initial and final state at rest) of the ratio in eq.~(\ref{eq:ratio_zero_momentum})
\begin{align}
  R^X(\tins, \tsep) =& A_{00} + A_{01} e^{-\Delta \tins} + A_{10} e^{-\Delta (\tsep-\tins)} \notag \\
  & + A_{11} e^{-\Delta\tsep} + ... \,,
  \label{eq:ratio_zero_momemtum_twostate}
\end{align}
where momentum arguments as well as Dirac indices have been suppressed and $\Delta = E_1-E_0$ denotes the energy gap between the first excited state and the ground state. Overlap factors and matrix elements from the spectral decomposition of the two- and three-point functions in eqs.~(\ref{eq:2pt_pspace},\ref{eq:3pt_pspace}) have been collected in the $A_{kl}$ factors. In particular, the first term $A_{00}$ is proportional to the ground state matrix element, e.g. the nucleon charges $g_{S,A,T}$ if the generic operator $\mathcal{O}_{\mu_1,...,\mu_n}^X$ in eq.~(\ref{eq:ratio_zero_momentum}) is chosen appropriately and $A_{01}=A_{10}$ holds for the zero-momentum case. Using the midpoint of the ratio as an estimate at any given source-sink separation, the leading excited state contamination $\sigma_\mathrm{esc}$ is therefore expected to scale as
\begin{equation}
 \sigma_\mathrm{esc} \sim e^{-\Delta \tsep / 2} \,.
 \label{eq:excited_state_scaling}
\end{equation}
Note that at non-zero momentum transfer the behavior of the ratio midpoint is similar, however, the time dependence in eq.~(\ref{eq:ratio_zero_momemtum_twostate}) is no longer symmetric if initial and final state differ by momentum. Considering the forward scattering case and neglecting the $N\pi$ interaction one may assume at least for a qualitative analysis that the leading gap is close to $2M_\pi$, \ie the scaling is approximately $\sim \exp(-M_\pi \tsep)$. In fig.~\ref{fig:ratio_midpoint_scaling} this behavior is plotted for several pion masses as a function of $\tsep$ and it is obvious that the issue of residual excited states for the ratio method becomes more severe towards the physical quark masses. For example, at a source-sink separation of $1.5\fm$ that is typically reached in NME lattice calculations, $\sigma_\mathrm{esc}$ is roughly one order of magnitude larger at physical light quark mass compared to the situation at $M_\pi=400\mev$. Assuming that the matrix element is of $\mathcal{O}(1)$ this implies corrections of up to $\sim 40\%$ for $\tsep=1.5\fm$ at physical pion mass. Therefore, it is doubtful if the values of $\tsep$ that can be reached in lattice calculations are sufficiently large to ensure control over excited states in the ratio method particularly at small light quark mass. Furthermore, excited-state effects can be strongly operator-dependent which is also observed empirically in lattice calculations; see \eg fig.~\ref{fig:N203_g_AST_and_avg_x}. Consequently, it cannot be concluded from observing ground state dominance for an observable in the ratio method that similar values of $\tsep$ will be sufficient for a different observable. \par

\begin{figure}
 \includegraphics[totalheight=0.25\textheight]{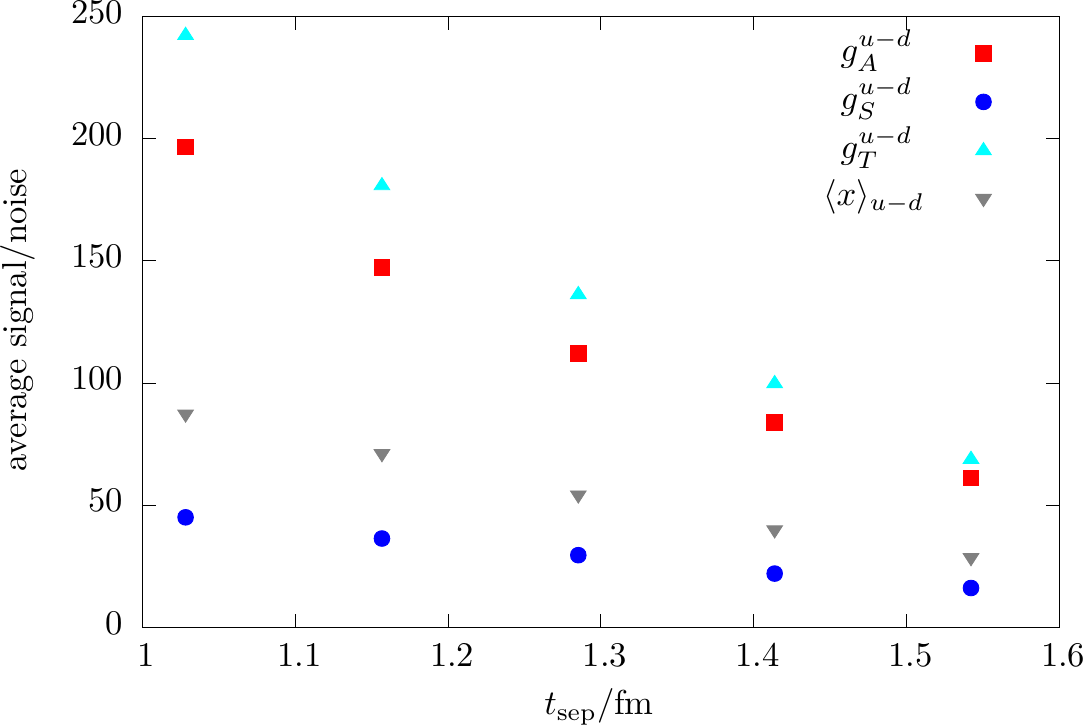}
 \caption{Signal-to-noise ratio averaged over individual timeslices $\tins$ at given value of $\tsep$ obtained from the effective form factor in eq.~(\ref{eq:ratio_zero_momentum}) for isovector $g_{A,S,T}^{u-d}$ and $\langle x\rangle_{u-d}$. Data are shown for the same ensemble as used for fig.~\ref{fig:N203_g_AST_and_avg_x}.}
 \label{fig:signal2noise_gAST_avgx}
\end{figure}

On the other hand, the statistical error of the effective form factor behaves as 
\begin{equation}
 \sigma_\stat \sim \frac{e^{\l(M_N-\frac{3}{2}M_\pi\r)\tsep}}{\sqrt{N}}  \,,
 \label{eq:stat_error_scaling}
\end{equation}
\ie it exponentially grows with $\tsep$. Even at moderately heavy quark mass the effect is sizable as shown in fig.~\ref{fig:signal2noise_gAST_avgx} for a selection of zero-momentum NMEs measured with fixed statistics at several values of $\tsep$. In fact, going from $\tsep=1\fm$ to $\tsep=1.5\fm$ the signal-to-noise ratio becomes roughly four times smaller. This means that any desired reduction in residual excited state contamination by increasing $\tsep$ competes with the statistical precision of the final result. Moreover, lowering the light quark mass increases both $\sigma_\stat$ and $\sigma_\mathrm{esc}$ individually. If one demands that the statistical error $\sigma_\stat$ and the systematic uncertainty $\sigma_\mathrm{esc}$ scale in the same way, \ie $\sigma_\mathrm{esc} = \mathrm{const} \cdot \sigma_\stat \equiv \sigma$, the required statistics $N$ for a target uncertainty $\sigma$ can be inferred in the asymptotic regime from eqs.~(\ref{eq:excited_state_scaling},\ref{eq:stat_error_scaling}) leading to \cite{Green:2018vxw} 
\begin{equation}
 N \sim \sigma^{-\l(2+\frac{2M_N - 3M_\pi}{\Delta}\r)} \,.
 \label{eq:common_scaling}
\end{equation}
In the review in ref.~\cite{Green:2018vxw} it has been further pointed out that at physical quark mass and again assuming $\Delta=2M_\pi$ the exponent is approximately $-13$ which must be compared to the naive factor of $-2$ obtained when neglecting the effects of excited states. It is in this sense that the presence of excited states strongly enhances the issue of the signal-to-noise problem in lattice calculations of NMEs, particularly when approaching physical quark mass. \par

\subsection{Multi-particle states and theory predictions} \label{subsec:multi_particle_states}
\begin{figure*}
\includegraphics[totalheight=0.276\textheight]{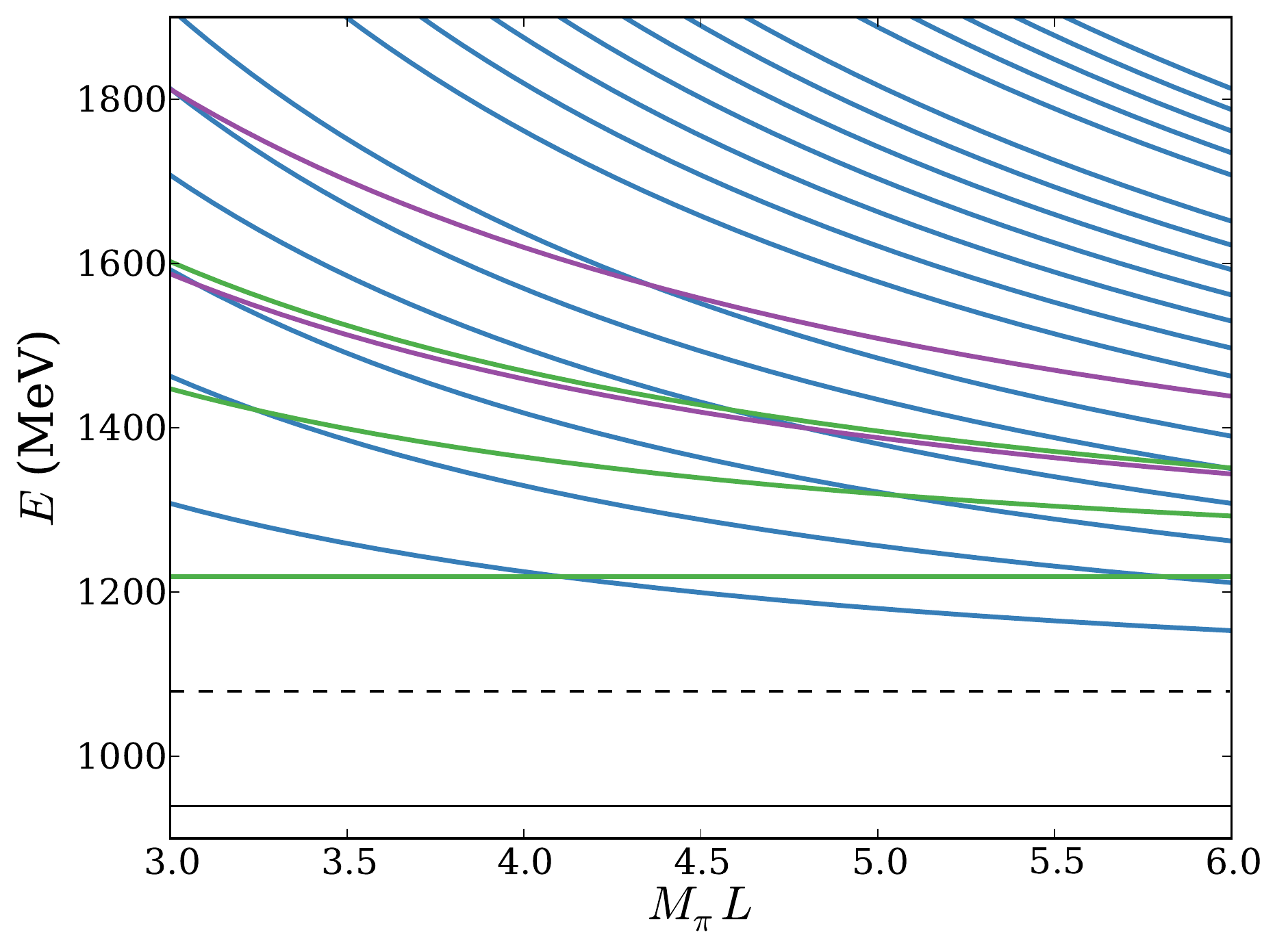}
\includegraphics[totalheight=0.3\textheight]{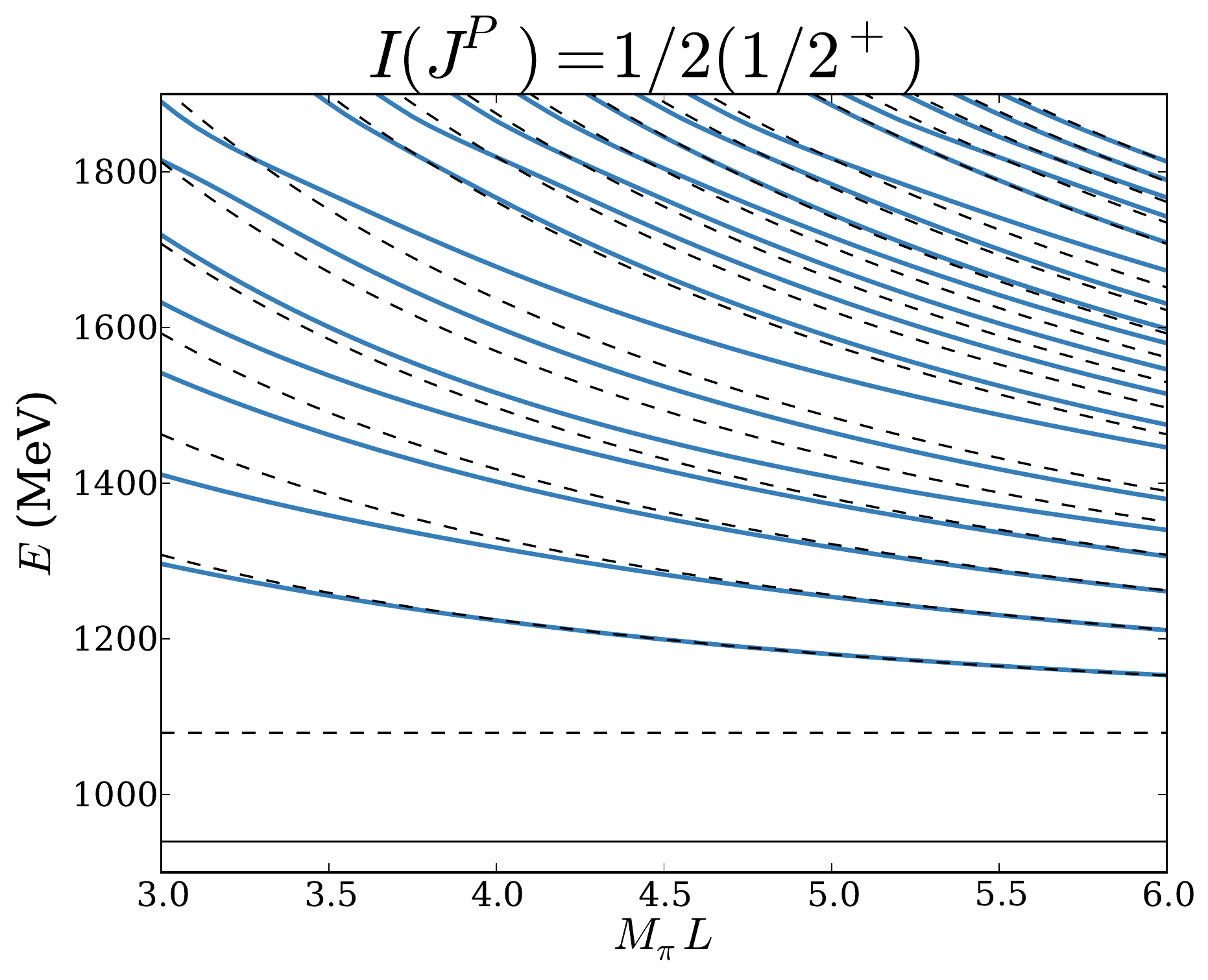}
 \caption{Multi-particle energy spectra of finite volume states as a function of $M_\pi L$. Left panel: Non-interacting states with quantum numbers of a single nucleon at rest. The black solid line corresponds to the nucleon mass and the dashed line to the threshold for the $N\pi$ state with opposite parity. The blue, green and magenta lines correspond to $N\pi$ states with back-to-back momentum, first few $N\pi\pi$  states with one pion at rest and first few $N\pi\pi$ states with the nucleon at rest, respectively. Right panel: Interacting $N\pi$ states with quantum numbers $I(J^P) = 1/2(1/2^+)$ (blue, solid curves) and corresponding non-interacting levels (black, dashed curves). Both figures are reproduced from ref.~\cite{Hansen:2016qoz} under the Creative Commons CC-BY license.}
 \label{fig:multi_particle_spectrum}
\end{figure*}

A further complication arises from the particular excited state pattern expected in nucleon structure calculations, \ie a dense spectrum of multi-particle states in addition to resonances, which limits the efficacy of a variational treatment or fits. In finite volume only discrete momenta $\vec{p}=\frac{2\pi}{L} \vec{n}$ with $\vec{n}=(n_1,n_2,n_3)^T$ and integer $n_i$ are allowed and any combination of a single nucleon with an arbitrary number of pions that is compatible with symmetries and momentum conservation will occur in the spectrum. Increasing the (spatial) box size $L$ while keeping the remaining physics parameter fixed, will reduce the size of a momentum unit $2\pi/L$, thus leading to a denser spectrum. Neglecting the interactions between the nucleon and the pions, an approximation of the multi-particle spectrum can be obtained that is shown in the left panel of fig.~\ref{fig:multi_particle_spectrum} for physical pion mass. Up until $M_\pi L=4$ the lowest lying excited state is a $N\pi\pi$ state, which justifies the previously made assumption of $\Delta=2M_\pi$ for the lowest gap. Note that at heavier pion mass this remains true for even larger values of $M_\pi L$. Moreover, corrections to the lower, non-interacting $N\pi$ levels are rather small as can be seen in the right panel of fig.~\ref{fig:multi_particle_spectrum}. The results shown in this figure have been obtained in ref.~\cite{Hansen:2016qoz} using the $\chi$PT in infinite volume together with the Lellouch-L\"uscher formalism. For higher $N\pi$ states $E \gtrsim 1400\mev$ deviations become more apparent, although the general pattern is not affected. \par

\begin{figure}
 \includegraphics[totalheight=0.25\textheight]{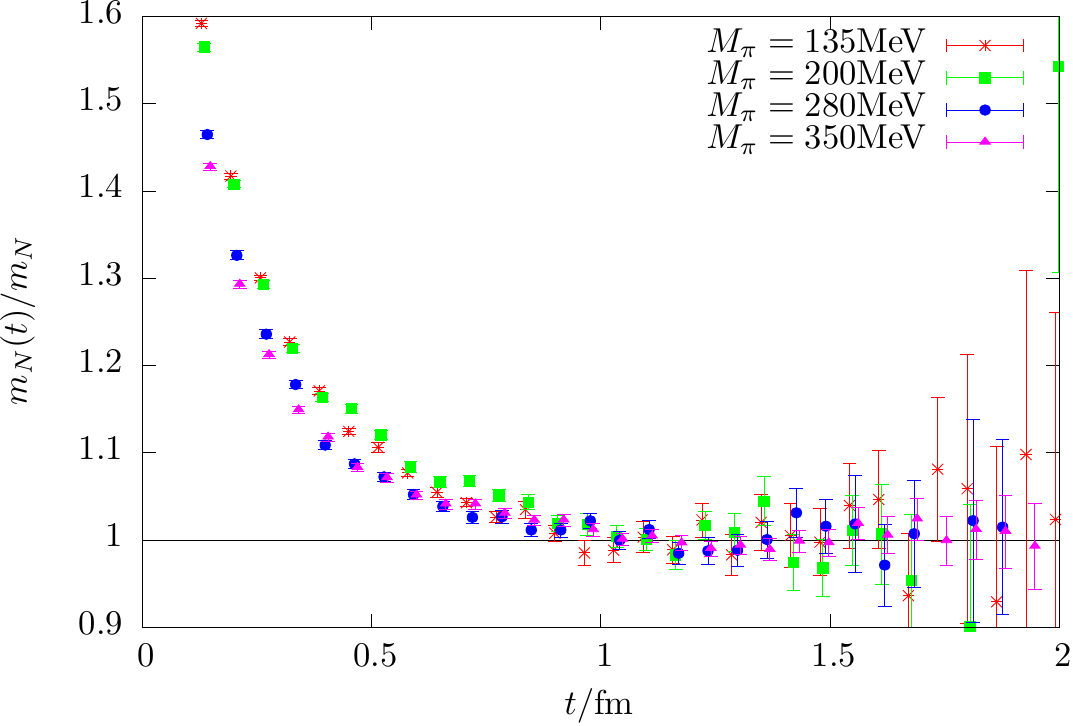}
 \caption{Examples for the ground state convergence in the nucleon effective masses. The relative excited state contamination in the nucleon effective mass $M_N(t)$ defined as the ratio of $M_N(t)$ and its fitted (asymptotic) value $M_N$ is shown as a function of $t$ computed on four CLS ensembles (E250, D200, N200, N203) at a common value of the lattice spacing $a=0.06426\fm$. The pion masses on these ensembles cover a range of roughly $\l[135\mev, 350\mev \r]$ corresponding to $M_\pi L=4.2, 4.2, 4.4, 5.4 $.}
 \label{fig:M_N_excited_states_different_Mpi}
\end{figure}

At heavier-than-physical quark mass the spectrum of multi-particle states is thinned out while keeping $M_\pi L $ fixed. Therefore, systematics related to excited states are again expected to become more severe at lighter quark masses. This is demonstrated in fig.~\ref{fig:M_N_excited_states_different_Mpi}, where the ratio $M_N(t)/M_N$ of the effective mass and the fitted ground state value is shown for four ensembles with quark masses covering a range from $\sim 135\mev$ up to $350\mev$. At small values of $t$ a clear hierarchy is observed before the signal becomes to noisy, \ie a plateau is approached more rapidly at heavier quark masses. \par

In the last few years there have been several studies \cite{Tiburzi:2009zp,Tiburzi:2015tta,Bar:2015zwa,Bar:2016uoj,Bar:2017kxh,Bar:2018wco,Bar:2018xyi,Bar:2019gfx,Bar:2019igf} carried out using $\chi$PT to predict effects of excited states in nucleon structure calculations. While recent lattice calculations are performed near or at physical quark mass which already removes a major systematic effect, other restrictions with respect to the applicability of $\chi$PT results to lattice data remain. This concerns \eg the size of the available source-sink separations in case of three-point functions. Regarding the use of smeared interpolating operators it has been pointed out in ref.~\cite{Bar:2015zwa} that they are mapped onto the point-like nucleon field in the effective theory, provided that the smearing radius is small compared to the Compton wavelength of the pion. The resulting effective operators containing the pion-nucleon coupling in the second term
\begin{equation}
 \chi_\mathrm{eff}(x) = \tilde{\alpha}  \l( \psi(x) + \frac{i}{2f} \pi(x) \g{5} \psi(x)\r) \,,
 \label{eq:eff_operator}
\end{equation}
then only differ by the value of a low energy constant (LEC) $\tilde{\alpha}$ for different smearings. Furthermore, at leading order this LEC is canceled in ratios, hence $\chi$PT predictions for excited state corrections are independent of the actual choice of smearing at leading order. At the very least, these studies provide qualitative insight into the behavior of excited state contamination, but in more recent work $\chi$PT predictions have also been used to systematically remove excited state contamination from lattice data; see \eg ref.~\cite{Bar:2018xyi}. \par

Excited states in nucleon two-point functions have been studied in refs.~\cite{Tiburzi:2009zp,Tiburzi:2015tta,Bar:2015zwa,Bar:2018wco}. For the nucleon effective mass it was found that the excited state correction due to $N\pi$ contributions are expected to be below $2\%$ at $t\geq 0.5\fm$ and to become a sub-percent effect for $t\geq1\fm$. This is roughly consistent with empirical findings in lattice studies \eg considering the behavior of the (total) relative excited state contribution in fig.~\ref{fig:M_N_excited_states_different_Mpi} as a function of $t$. In the most recent ref.~\cite{Bar:2018wco} the study has been extended to three-particle ($N\pi\pi$) states which where found to contribute at most at the permille level and thus considered to be negligible for all practical purposes in the foreseeable future. \par

For three-point functions and the resulting matrix elements the situation is more complicated. In the past, the main focus has been on $N\pi$ contributions in the three-point function with an axial vector insertion relevant for $g_A^{u-d}$ \cite{Tiburzi:2009zp,Tiburzi:2015tta,Bar:2016uoj,Bar:2017kxh}. The predicted effect on $g_A^{u-d}$ is an overestimation of at least several percent at typical values of $\tsep\lesssim 1.5\fm$ that are accessible in lattice simulations. An effect of similar size has been predicted for $g_T^{u-d}$, while for $g_S^{u-d}$ the overestimation was determined to be $\sim 50\%$ larger \cite{Bar:2016uoj}. In ref.~\cite{Bar:2017kxh} the first moments of parton distribution functions, $\langle x \rangle_{u-d}$, $\langle x \rangle_{\Delta u-\Delta d}$ and $\langle x \rangle_{\delta u-\delta d}$ have been analyzed in addition to charges and it has been found that the contribution of the $N\pi$ state lead to an overestimation of $5$--$10\%$ at $\tsep=2\fm$ in all of these observables. Furthermore, it has been pointed out that the source-sink separations that are currently accessible in lattice simulations are too small for a direct application of $\chi$PT and hence these predictions are of limited applicability. In particular, resonances that are not included in $\chi$PT might give non-negligible contributions. This may explain why the predicted effect for $g_A^{u-d}$ is opposite to what is actually observed in lattice simulations, \ie excited states lead to smaller values for $g_A^{u-d}$. An example is shown in the first panel of fig.~\ref{fig:N203_g_AST_and_avg_x} where the effective form factor for $g_A^{u-d}$ appears to approach the asymptotic value from below. \par

\begin{figure}
\includegraphics[totalheight=0.29\textheight]{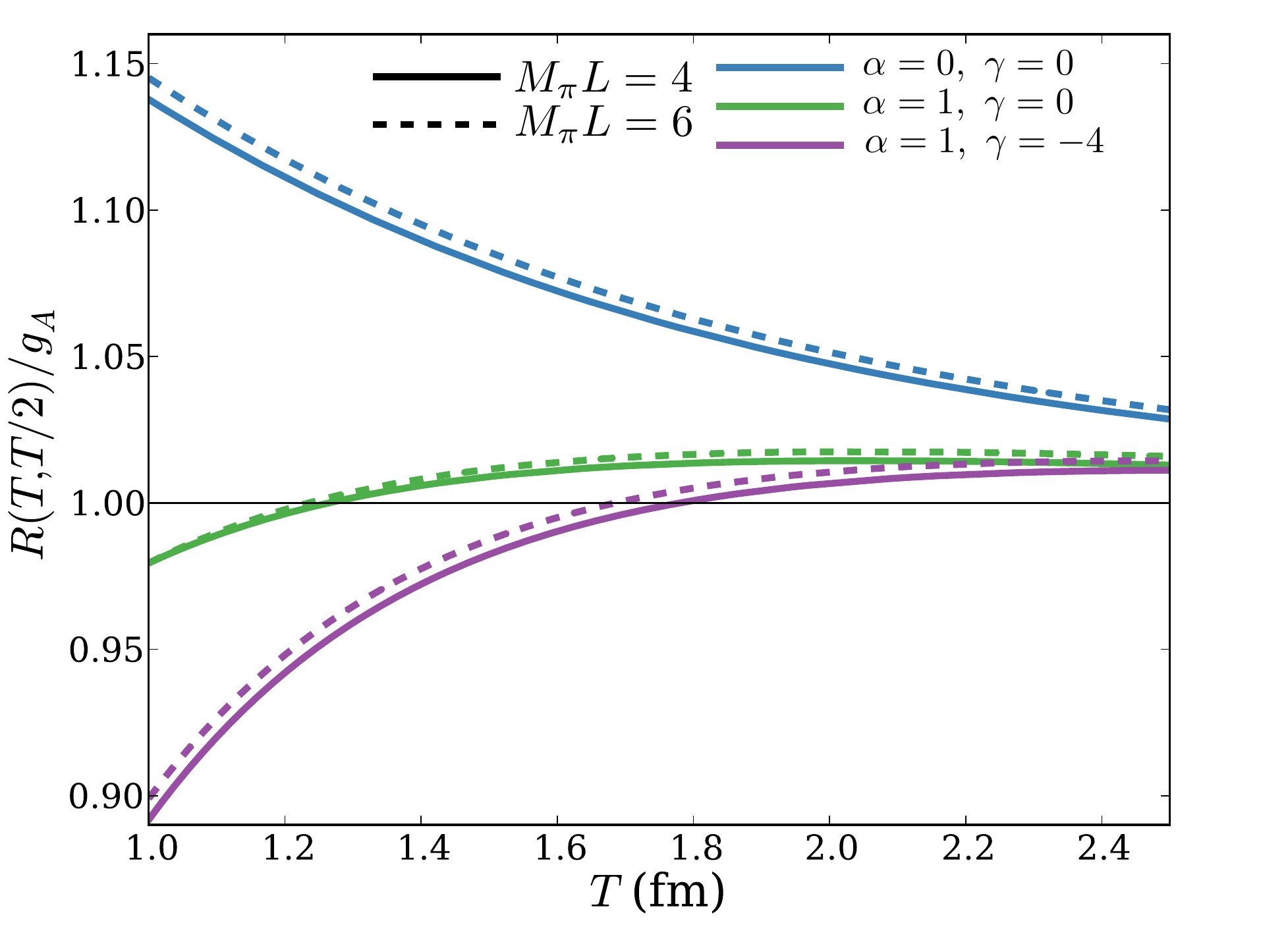}
 \caption{Prediction of the excited state contamination in the ratio method for $g_A$ from ref.~\cite{Hansen:2016qoz}. The two upper curves represent the leading order $\chi$PT prediction, but taking into account interacting values of the Lellouch-L\"uscher factors. The remaining two pairs of curves correspond to certain variations of model assumptions for the overlap factor and axial-vector transition amplitude. The lower pair of curves qualitatively reproduces the behavior observed in lattice calculations and the two values of $M_\pi L$ roughly cover the range found in modern lattice calculations. The figure has been originally published in ref.~\cite{Hansen:2016qoz} and is reproduced under the Creative Commons CC-BY license.}
 \label{fig:gA_excited_state_prediction}
\end{figure}

In ref.~\cite{Hansen:2016qoz} the issue of excited states in lattice calculations of $g_A$ has been revisited using the Lellouch-L\"uscher formalism and experimental results for the $N\pi$ scattering phases. While the inclusion of information on the spectrum has been found to only give a small correction to the $\chi$PT prediction for the excited state contamination in $g_A$, it has been argued that plausible model assumptions for the overlap factor and particularly postulating a sign change in the infinite volume axial-vector transition amplitude can qualitatively reproduce the observed behavior of lattice data which is shown in fig.~\ref{fig:gA_excited_state_prediction}. The mechanism behind this indeed turned out to be a cancellation between the (positive) contribution of lower lying states and the (negative) contribution of higher excited states to the overall excited state contamination. \par

\begin{figure}
 \includegraphics[totalheight=0.33\textheight]{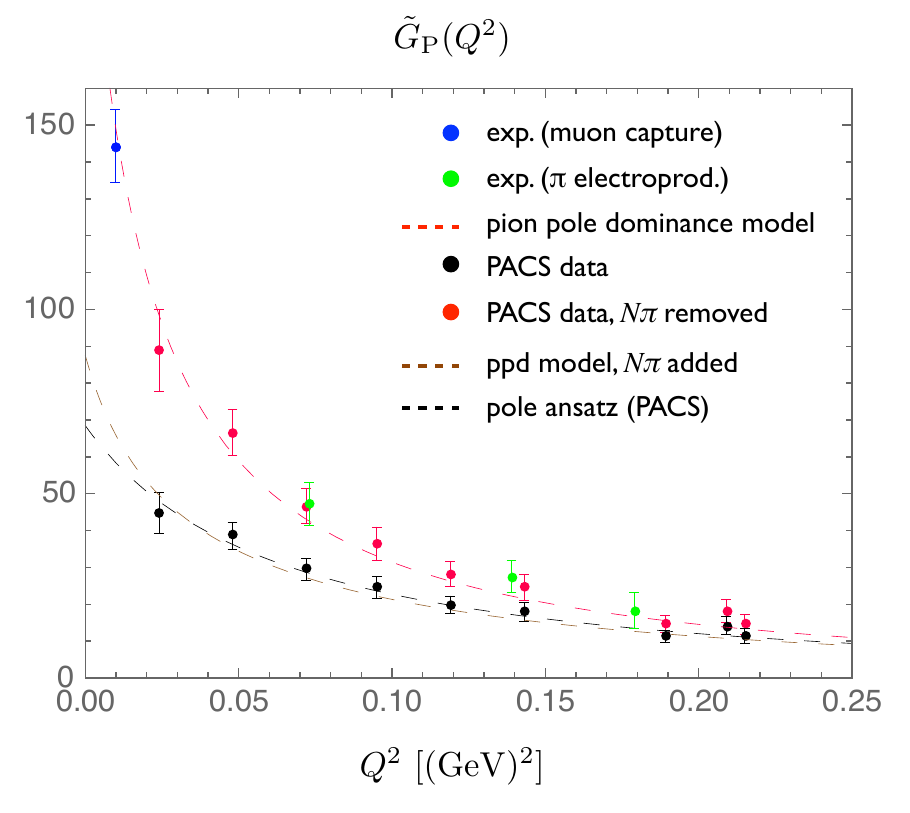}
 \caption{Lattice data on a PACS ensemble ($T\times L^3=(96a)^4$, $M_\pi\approx 146\mev$, $a\approx0.085\fm$) for isovector $\tilde{G}_P(Q^2)$ from ref.~\cite{Ishikawa:2018rew} with and without correction for $N\pi$ states. The lattice data is generally well described by the pion pole dominance (ppd) model before and after correction for the $N\pi$ states and is found to be in agreement with experimental results after the correction. The figure has been originally published in ref.~\cite{Bar:2018xyi} and is reproduced under the Creative Commons Attribution 4.0 International license.}
 \label{fig:tildeG_P_Npi_correction}
\end{figure}

Recently, $\chi$PT studies of $N\pi$ state contamination have been extended beyond zero-momentum for the axial form factor $G_A(Q^2)$ and the related, induced pseudoscalar form factor $\tilde{G}_P(Q^2)$ in ref.~\cite{Bar:2018xyi}. For $G_A(Q^2)$ it has been found that the leading order $\chi$PT prediction for the $N\pi$ contribution remains at the $5\%$ level at $\tsep=2\fm$ and is basically independent of $Q^2$. This is in stark contrast to the findings for $\tilde{G}_P(Q^2)$ for which an underestimation of $10$ to $40\%$ due to $N\pi$ states has been predicted that is strongly dependent on $Q^2$. This is in good agreement with numerical findings of lattice calculations, unlike the prediction of an overestimation in case of the axial form factor. This very different behavior of the two axial form factors has been attributed to how $N\pi$ states contribute in the two observables: For $G_A(Q^2)$ the entire tower of $N\pi$ state contributes, while for $\tilde{G}_P(Q^2)$ the $N\pi$ excited state contamination consists of only a single $N\pi$ state associated with a pion with its spatial momentum determined by the momentum transfer $Q^2$. Consequently, the prediction for $\tilde{G}_P(Q^2)$ remains valid for much smaller source-sink separations than for $G_A(Q^2)$ and it has been demonstrated that this can used to reliably correct lattice data at source-sink separations well below $1.5\fm$. An example for this correction of lattice data from ref.~\cite{Ishikawa:2018rew} is shown in fig.~\ref{fig:tildeG_P_Npi_correction} together with experimental data and predictions from the pion pole dominance model. \par

\begin{figure}
 \includegraphics[totalheight=0.33\textheight]{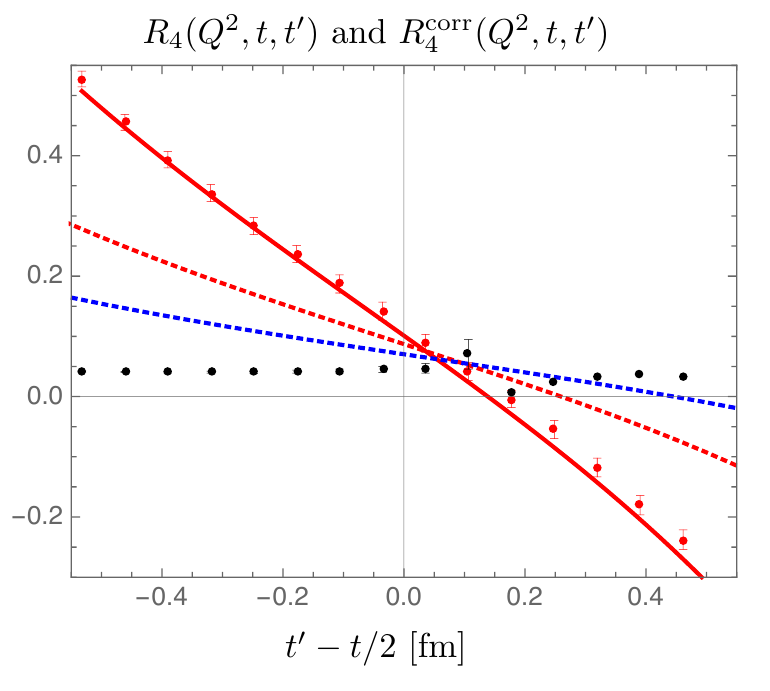}
 \caption{Raw (red circles) and corrected (black circles) lattice data for the effective form factor from the insertion of the time component of the axial vector current at $\tsep\approx1.07\fm$ and $Q^2\approx 0.073\gev^2$. The lattice data have been produced in a study published in ref.~\cite{Bali:2018qus} on an ensemble with $T \times L^3 = (64a)^4$, $M_\pi\approx 150\mev$ and $a\approx 0.071\fm$. The solid red line represents the $\chi$PT prediction; not a fit to the data. In addition, results from $\chi$PT for $\tsep=2.1\fm$ and $\tsep=3.2\fm$ are shown as red and blue dotted lines. The figure has been originally published in ref.~\cite{Bar:2018xyi} and is reproduced under the Creative Commons Attribution 4.0 International license.}
 \label{fig:A0_Npi_correction}
\end{figure}

In the same study in ref.~\cite{Bar:2018xyi} $N\pi$ states have also been found to be a likely cause for the strong excited state contamination observed in the effective form factor obtained from an operator insertion of the time component of the axial vector current, see e.g. refs.~\cite{Bali:2018qus,Meyer:2018twz}. While the excellent agreement of lattice data with the $\chi$PT prediction demonstrated in fig~\ref{fig:A0_Npi_correction} may be accidental to some extent, it can be concluded that the $N\pi$ contribution to the excited state contamination is large and dominant in this case and source-sink separations even beyond $3\fm$ might be required to sufficiently suppress them. Note that this is also the reason why lattice computations of the axial charge and $G_A(Q^2)$ commonly employ three-point functions from spatial components of the axial vector insertion, which exhibit much milder excited state contamination. \par

\begin{figure*}
 \includegraphics[totalheight=0.245\textheight]{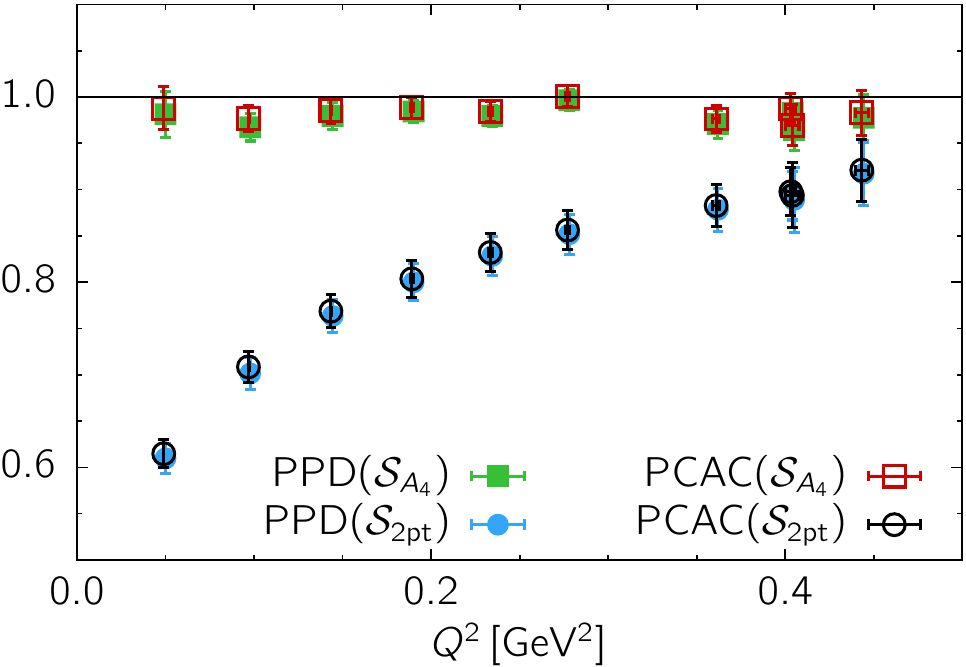} \ 
 \includegraphics[totalheight=0.245\textheight]{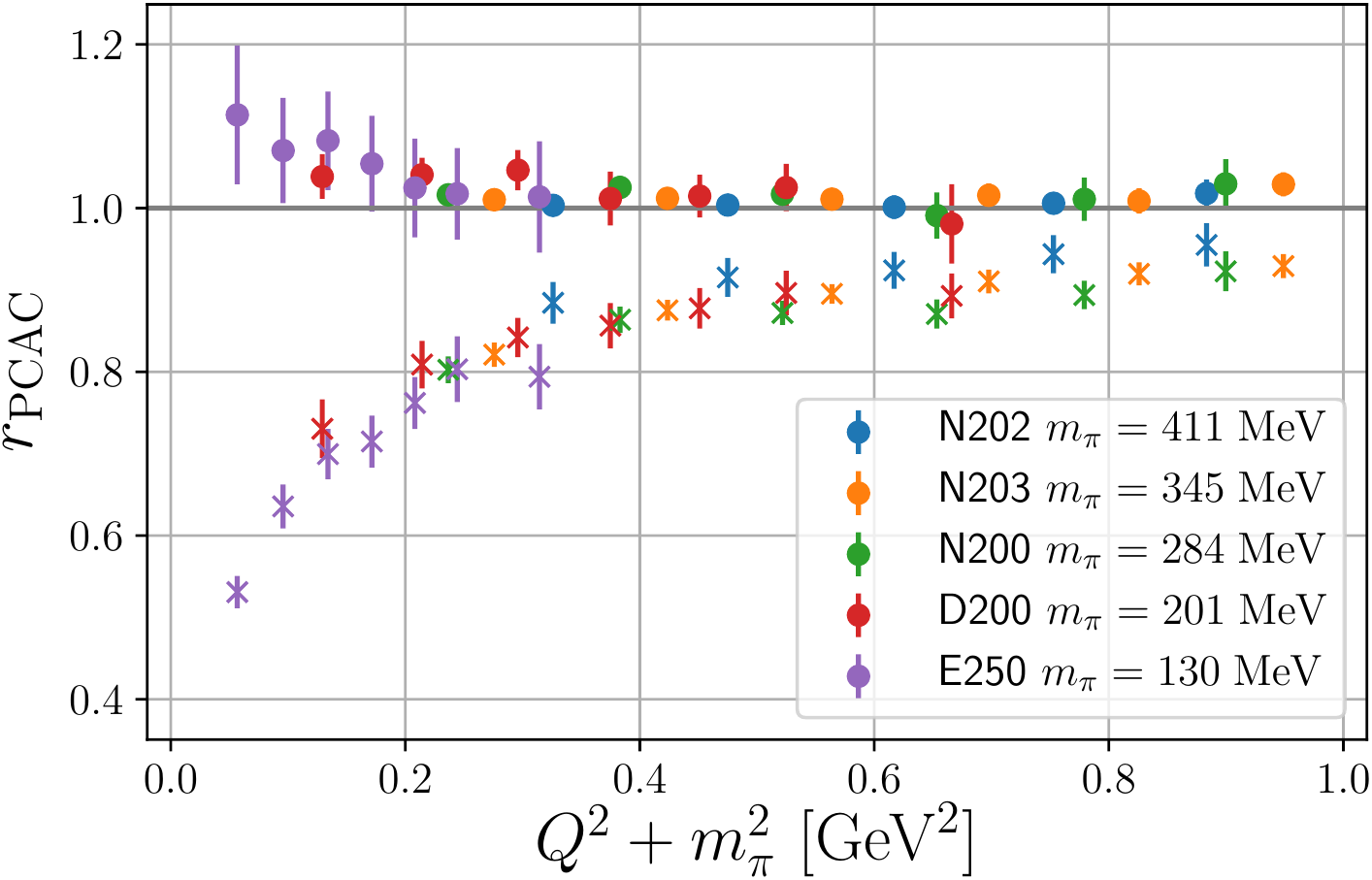}
 \caption{Lattice results for the violation of the generalized Goldberger-Treiman relation. Left panel: Results from ref.~\cite{Jang:2019vkm} on an ensemble with physical pion mass for $r_\mathrm{PCAC}$ (open symbols) and $r_\mathrm{PPD}$ (filled symbols) obtained from the two analysis strategies $S_{A_4}$ and $S_\mathrm{2pt}$ in this study. Right panel: Results from ref.~\cite{Bali:2019yiy} for $r_\mathrm{PCAC}$ on a set of CLS ensembles at fixed lattice spacing $a\approx0.064\fm$. Data denoted by filled symbol are obtained from a modified fit ansatz accounting for excited states enhanced by the presence of the pion pole, while open symbols refer to data obtained from a conventional fit. The two figures have been originally published in refs.~\cite{Jang:2019vkm} and \cite{Bali:2019yiy}, respectively, and are reproduced under the Creative Commons Attribution 4.0 International license.}
 \label{fig:PCAC_and_PPD_test}
\end{figure*}

Furthermore, in ref.~\cite{Bar:2019gfx} an analysis of the $N\pi$-state contamination has been performed for the pseudoscalar form factor $G_P(Q^2)$. Similar to the induced form factor $\tilde{G}_P(Q^2)$ a large contamination of $-20\%$ up to $50\%$ depending on $Q^2$ has been predicted at $\tsep=2\fm$. Moreover, it has been argued that a $N\pi$-state contamination is the most relevant source for the large violations of the generalized Goldberger-Treiman relation that have been observed in lattice simulations \cite{Ishikawa:2018rew,Rajan:2017lxk,Alexandrou:2017hac,Jang:2018lup,Liang:2018pis}. This relation establishes a connection between the form factors $G_A(Q^2)$ and $\tilde{G}_P(Q^2)$ associated with the partially conserved axial-vector current (PCAC) and the form factor $G_P(Q^2)$ of the pseudoscalar density. A quantitative measure of its violation is given by the deviation from unity for the following expression \cite{Rajan:2017lxk,Bali:2018qus}
\begin{equation}
 r_\mathrm{PCAC} = \frac{\hat{m}}{M_N} \frac{G_P(Q^2)}{G_A(Q^2)} + \frac{Q^2}{4M_N^2} \frac{\tilde{G}_P(Q^2)}{G_A(Q^2)} \,,
 \label{eq:PCAC_ratio}
\end{equation}
where $\hat{m}$ denotes the average bare PCAC quark mass. A second, closely related ratio that can be directly tested from lattice data is the pion-pole dominance (PPD) hypothesis
\begin{equation}
 r_\mathrm{PPD} = \frac{Q^2+M_\pi^2}{4M_N^2} \frac{\tilde{G}_P(Q^2)}{G_A(Q^2)} \,,
 \label{eq:PPD_ratio}
\end{equation}
relating $G_A(Q^2)$ and $\tilde{G}_P(Q^2)$. An example for the violation of these relations by lattice results are the lower sets of data points in fig.~\ref{fig:PCAC_and_PPD_test} that have been obtained using conventional fit ans\"atze to two- and three-point functions in two recent studies in refs.~\cite{Jang:2019vkm,Bali:2019yiy}. Again, the theoretical findings in ref.~\cite{Bar:2019gfx} were corroborated by lattice data, even at source-sink separations as small as $\tsep=1.3\fm$, although no definite conclusion has been drawn in this study. This has motivated further lattice investigations of the issue in refs.~\cite{Bali:2018qus,Jang:2019vkm,Bali:2019yiy}. In ref.~\cite{Bali:2018qus} a projection method has been introduced to remove excited state contamination in nucleon form factor calculations, however, this only lead to improvement for $r_\mathrm{PCAC}$ but not for $r_\mathrm{PPD}$. Subsequently, it has been argued in ref.~\cite{Bar:2019igf} using chiral perturbation theory that this projection method in fact enhances the $N\pi$ contamination in $G_P(Q^2)$ and the improvement observed for the PCAC relation in eq.~(\ref{eq:PCAC_ratio}) is caused by the enhanced $N\pi$ contribution in $G_P(Q^2)$ compensating the underestimation of $\tilde{G}_P(Q^2)$. However, no such cancellation takes place for $r_\mathrm{PPD}$ which does not depend on $G_P(Q^2)$. \par

Finally, in ref.~\cite{Jang:2019vkm} it has been demonstrated that the deviations from unity for eqs.~(\ref{eq:PCAC_ratio})~and~(\ref{eq:PPD_ratio}) observed in previous lattice studies can indeed be attributed to a low-lying excited state which is missed in commonly used fits, \eg fit strategy $\mathcal{S}_\mathrm{2pt}$ in the left panel of fig.~\ref{fig:PCAC_and_PPD_test}. This result and the correspondingly modified fit procedure $\mathcal{S}_{A_4}$ determining the energy of this state from a fit to the three-point function will be discussed in some more detail in subsect.~\ref{subsec:ratio_fits} in the context of multi-state fits. A different ansatz explicitly modeling the excited state contamination in presence of the pion pole has been introduced in ref.~\cite{Bali:2019yiy} and has been found to lead to very similar results. Results from the modified analysis strategies in both studies are visible in fig.~\ref{fig:PCAC_and_PPD_test}, \ie the upper sets of data points in both panels, which are compatible with one. \par

\section{Summed operator insertions}
\label{sec:summation_method}
The ratio method introduced in subsect.~\ref{subsec:NME} depends on two Euclidean time separations to become large such that excited states are sufficiently suppressed. As discussed in the previous section this can hardly be achieved in the presence of the nucleon signal to noise problem. While it may be possible to explicitly remove specific excited state contamination for certain matrix elements like e.g. $N\pi$ states in axial and pseudoscalar form factors, there is no general way how this can be achieved using additional, theoretical input only. Therefore, methods are needed that improve the suppression of excited states using data from the available source-sink separations and that are equally applicable to a broad class of nucleon structure observables. A simple approach that satisfies this requirement is the summation method that has been originally published in Ref.~\cite{Maiani:1987by}. It is based on summing the operator insertion over $\tins$, leaving $\tsep$ as the only time dependence. \par

\subsection{Summation method} \label{subsec:standard_summation_method}
\begin{figure}
 \includegraphics[totalheight=0.24\textheight]{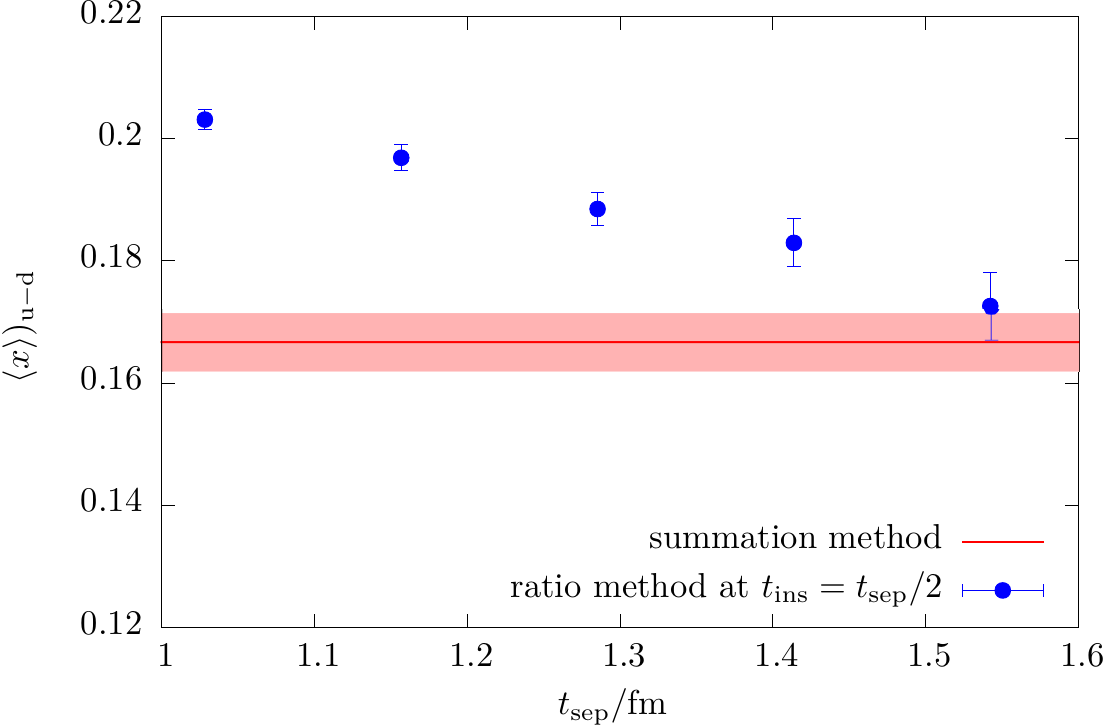}
 \caption{Comparison of results for $\langle x\rangle_{u-d}$ from the summation method and from the midpoint ($\tins=\tsep/2$) in the ratio method. Example data shown for the same ensemble as in fig.~\ref{fig:N203_g_AST_and_avg_x}.}
 \label{fig:ratio_vs_summation_method}
\end{figure}

In its commonly used version \cite{Dong:1997xr,Capitani:2012gj} the summation is performed at the level of the ratio in eq.~(\ref{eq:ratio}) running only over timeslices between source and sink, i.e. $\tins\in\l[t_\mathrm{ex}, \tsep - t_\mathrm{ex}\r]$ with some additional freedom of leaving out further timeslices parameterized by $t_\mathrm{ex}$
\begin{align}
 S^X_{\mu_1 ... \mu_n}(\tsep) &= \sum_{\tins=t_\mathrm{ex}}^{\tsep-t_\mathrm{ex}} R^X_{\mu_1 ... \mu_n}(\tsep, \tins) \notag \\
 &= \mathrm{const} + \bra{0} \mathcal{O}^X_{\mu_1 ... \mu_n} \ket{0} \tsep + \mathcal{O}(e^{-\Delta \tsep}) \,.
 \label{eq:summation_method}
\end{align}
The leading excited state contamination related to the gap $\Delta$ is still present but more strongly suppressed compared to the excited state contamination at the midpoint of the effective form factor in the ratio method that scales with $\mathcal{O}(e^{-\Delta\tsep/2})$. It should be noted, that only at zero-momentum transfer the gap $\Delta$ is actually the same as in eq.~(\ref{eq:ratio_zero_momemtum_twostate}). However, the resulting suppression of the leading excited state contamination is always of $\mathcal{O}(e^{-\Delta\tsep})$, where $\Delta$ is the smallest gap to the ground state, see also refs.~\cite{Brandt:2011sj,Capitani:2015sba}. Figure~\ref{fig:ratio_vs_summation_method} shows an example for the improved suppression of excited states compared to the ratio method for $\langle x \rangle_{u-d}$. While the midpoint of the effective form factor approaches the result from the summation method, the errors overlap only at the largest source-sink separation of $\tsep=1.54\fm$ and the central value from the ratio method is still larger. \par

Key advantages of the summation method are that it is trivial to implement, does not introduce model dependence or additional parameters (apart from the choice of $t_\mathrm{ex}$) and that it is often possible to reuse data generated for other methods \eg the ratio method or multi-state fits. Therefore, it has seen very widespread use in lattice calculations, often for crosschecking results obtained from other methods, but also for obtaining final results. On the other hand, the summation method tends to produce larger statistical errors in the matrix element of interest compared to the ratio method; \cf fig.~\ref{fig:ratio_vs_summation_method}. Besides, the excited state suppression in the summation method still depends on the values of $\tsep$ used. In actual lattice calculations the results is often dominated by the smallest source-sink separations that enter the linear fit because the effective statistics rapidly deteriorates with increasing $\tsep$ if the number of measurements is kept constant independent of $\tsep$. Since a similar issue arises with other methods, it may be desirable to spend extra computational effort to keep effective statistics constant, \ie scale the number of measurements at different source-sink separations so as to achieve (approximately) constant statistical error. This has been done in some recent lattice studies, see \eg refs.~\cite{Alexandrou:2018sjm,Alexandrou:2020aja}. Note that even for the ratio method it is not sufficient to check convergence in the presence of exponential error growth by inspecting the behavior of the observable as a function of $\tsep$ because the signal is typically lost before the asymptotic value is reached. \par

\subsection{Feynmann-Hellmann inspired approach} \label{subsec:Feynmann_Hellmann}
Another realization of the summation over the operator insertion that appeared in the original paper for the summation method \cite{Maiani:1987by} has been revisited and used a few times in the last years \cite{Bulava:2011yz,deDivitiis:2012vs,Chambers:2014qaa,Bouchard:2016heu,Savage:2016kon}. Most recently it has been employed in a dedicated calculation of the axial charge \cite{Chang:2018uxx,Walker-Loud:2019cif}, leading to a precise estimate of $g_A^{u-d}=1.271(13)$ in agreement with the experimental result. The method is inspired by the Feynman-Hellmann theorem that relates the energy shift due to a perturbation in the action 
\begin{equation}
 \mathcal{S} \rightarrow \mathcal{S} + \lambda \int d^4x \mathcal{O}_\Gamma(x) \,,
 \label{eq:S_shift}
\end{equation}
for a local bilinear operator $\mathcal{O}(x)$ with Dirac structure $\Gamma$ and some parameter $\lambda$ to the matrix element of a state $\ket{k}$ 
\begin{equation}
 \l.\frac{\partial}{\partial\lambda} E_k^\lambda\r|_{\lambda=0} = \frac{1}{2E_k} \bra{k} \mathcal{O}_\Gamma \ket{k} \,.
 \label{eq:FHT}
\end{equation}
Following the presentation in Refs.~\cite{Bouchard:2016heu,Walker-Loud:2019cif} the spectral decomposition of the nucleon two-point function in eq.~(\ref{eq:2pt_pspace}) at takes the following form
\begin{align}
 C^\mathrm{2pt}_\lambda(\vec{p},t) = \sum_k \l|\bra{k} \chi \ket{\Omega}_\lambda \r|^2 e^{-E_k^\lambda(\vec{p}) t} \,,
 \label{eq:2pt_pspace_lambda}
\end{align}
where we have again assumed the source to be at $t_i=0$ and $\ket{\Omega}_\lambda$ denotes the modified vacuum in the presence of the perturbation. The effective energy in eq.~(\ref{eq:eff_mass}) is then related to the ground state matrix element $\bra{0} \mathcal{O}_\Gamma \ket{0}$ by virtue of eq.~(\ref{eq:FHT}). Considering the zero-momentum case as required for \eg the computation of $g_A^{u-d}$ one finds the following relation for the effective mass
\begin{align}
 &\l.\frac{\partial }{\partial\lambda} m_\mathrm{eff}^\lambda(t, \tau)\r|_{\lambda=0} \notag \\
 &\quad = \l.\frac{1}{\tau} \l( \frac{\partial_\lambda C^\mathrm{2pt}_\lambda(t)}{C^\mathrm{2pt}(t)} - \frac{\partial_\lambda C^\mathrm{2pt}_\lambda(t+\tau)}{C^\mathrm{2pt}(t+\tau)} \r) \r|_{\lambda=0} \notag \\
 &\quad = \frac{1}{2E_0} \bra{0} \mathcal{O}_\Gamma \ket{0} + \mathcal{O}\l(\frac{e^{-\Delta(t+\tau)} - e^{-\Delta t}}{\tau}\r) \,,
 \label{eq:CalLat_method}
\end{align}
where the leading excited state behavior from the smallest energy gap $\Delta$ has been indicated by the second term in the last line. The correlation function $\partial_\lambda C^\mathrm{2pt}_\lambda(\vec{p},t) = \frac{\partial}{\partial\lambda} C^\mathrm{2pt}_\lambda(\vec{p},t)$ can be straightforwardly computed w.r.t. the original vacuum at $\lambda=0$ by replacing one of the propagators in a standard two-point function by a so-called Feynman-Hellmann propagator \cite{Bouchard:2016heu}
\begin{equation}
  S(y,x) = \sum_{z=(t_z, \vec{z})} S(y,z) \Gamma S(z,x) \,,
  \label{eq:FH_propagator}
\end{equation}
\ie a sequential propagator that is summed over the insertion time. Note that unlike in the previously discussed version of the summation method, here the summation over the operator insertion time is performed over the entire lattice, which is also the result originally derived in ref.~\cite{Maiani:1987by}. \par

An advantage of this approach over the sequential method is that it requires only the computation of two-point functions which depend on the source-sink separation $t=\tsep$ but not on the insertion time. Therefore, it does not require new inversions for each value of $\tsep$, and the computation of the usual sequential propagators for three-point functions is treated for the computation of the sequential propagator in eq.~(\ref{eq:FH_propagator}). This mitigates some of the signal-to-noise problem when performing a dedicated calculation for a single observable. However, it allows only to compute results for a single operator insertion and a single momentum transfer at a time, which changes the cost comparison in favor of the standard method when computing multiple observables or especially for computing form factors at non-zero $Q^2$. \par

\begin{figure*}
 \includegraphics[totalheight=0.27\textheight]{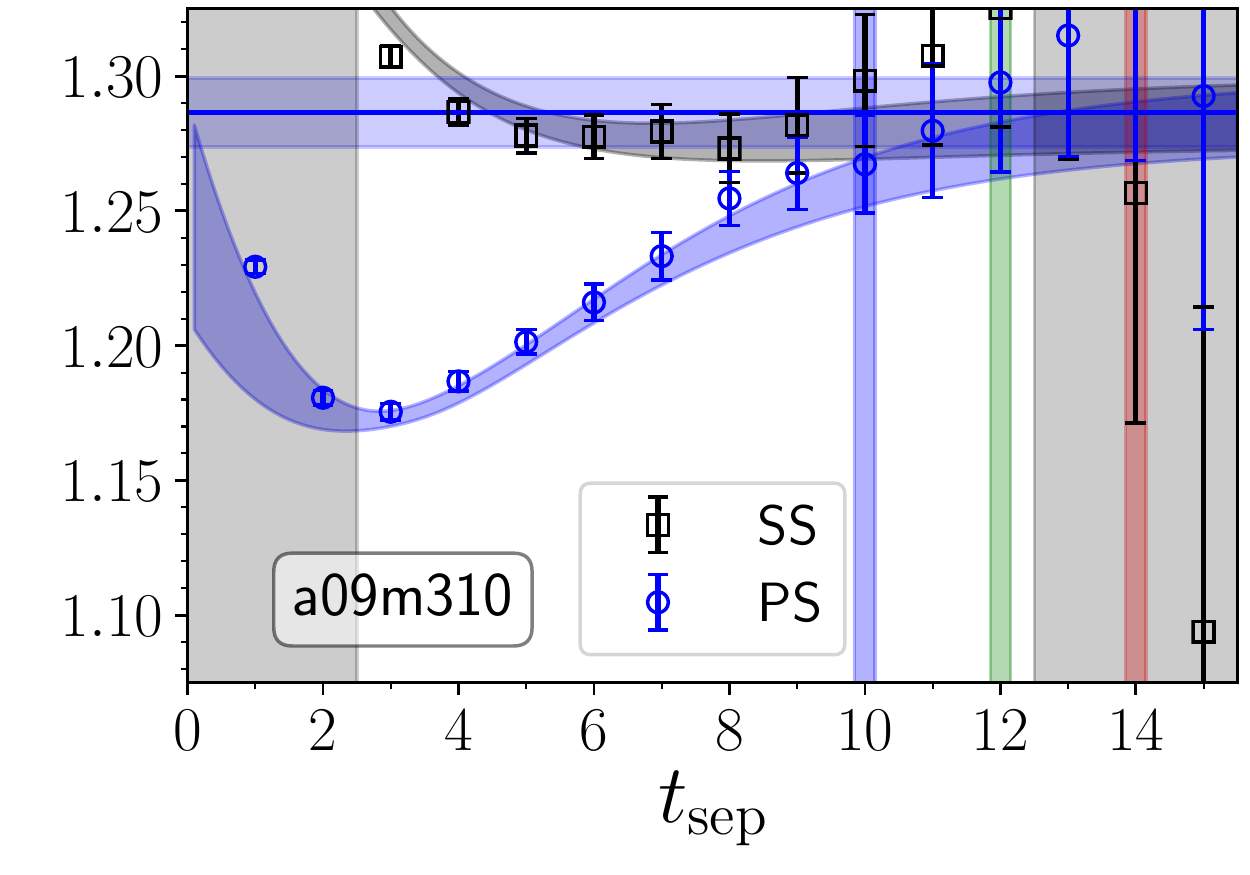}
 \includegraphics[totalheight=0.27\textheight]{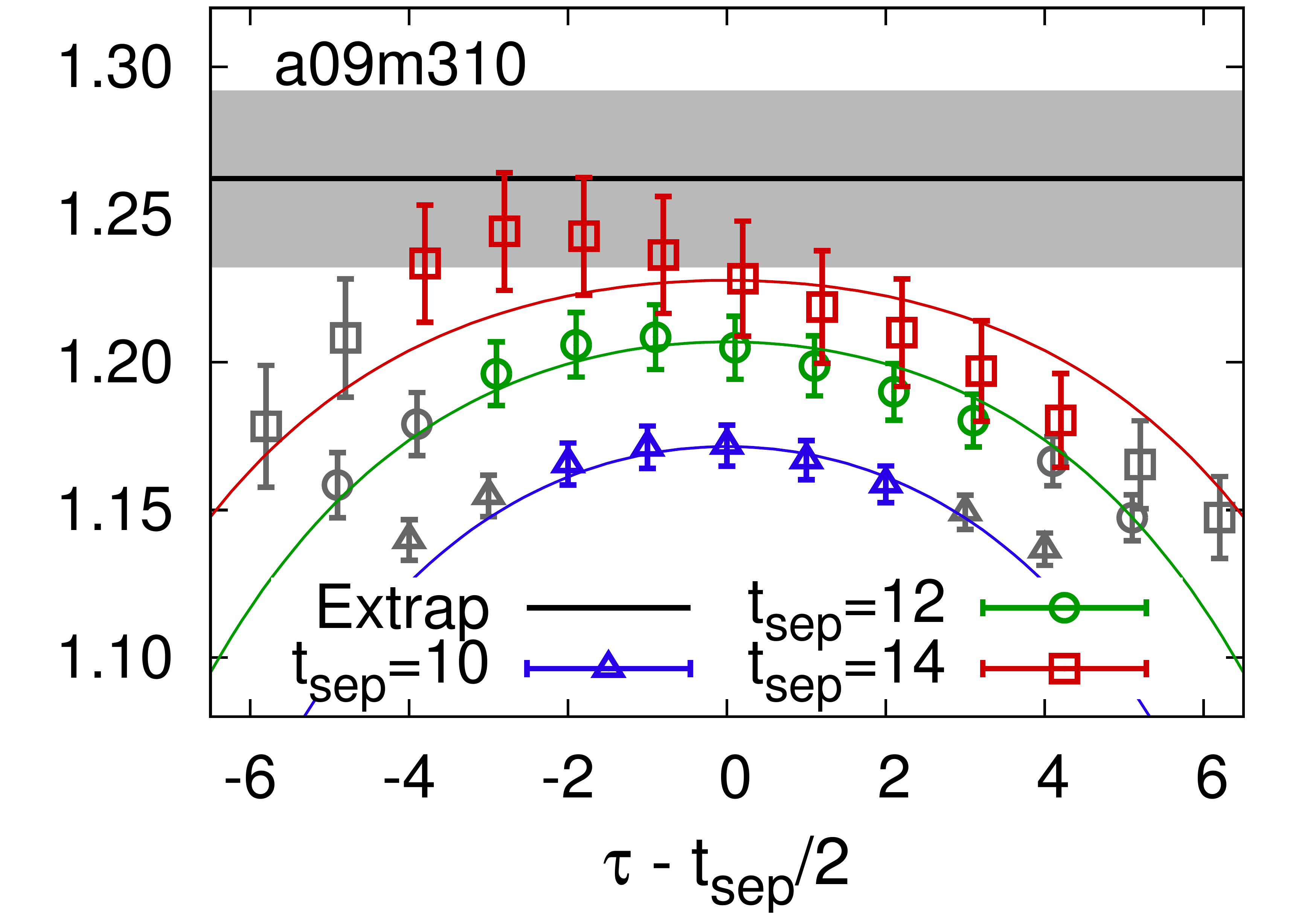}
 \caption{Comparison of results from the Feynmann-Hellmann inspired approach (left panel) and the standard sequential method (right panel) for the nucleon axial charge on a HISQ ensemble ($M_\pi\approx 310\mev$, $a\approx 0.09\fm$) generated by MILC \cite{Follana:2006rc,Bazavov:2012xda}. The data in the left panel have been produced in the study published in ref.~\cite{Chang:2018uxx} with a point (P) and smeared (S) sink and the source always smeared. The blue and back curves have been obtained from simultaneous two-state fits to SS and SP two-point functions and correlation functions for $g_A^{u-d}$ and $g_V^{u-d}$. The vertical gray bands mark data excluded from the fit and the blue, green and red bands highlight the values of $\tsep/a=10,12,14$ that have been used for the data from the sequential method shown in the right panel. The curves in the right panel correspond to a standard two-state fit ansatz; for further details see the original analysis published in ref.~\cite{Bhattacharya:2016zcn}. In both plots the horizontal line indicates the result for the ground state matrix element. The figure is reproduced with kind permission of the authors of ref.~\cite{Walker-Loud:2019cif}.}
 \label{fig:FH_vs_sequential_method}
\end{figure*}

In fig.~\ref{fig:FH_vs_sequential_method} a comparison of results for $g_A^{u-d}$ is shown for the Feynmann-Hellmann inspired approach (left panel) used in ref.~\cite{Chang:2018uxx} and the sequential method (right panel) from an analysis in ref.~\cite{Bhattacharya:2016zcn} on a common ensemble, but using a different action in the valence sector. In both cases the final result has been obtained from a two-state fit to the data; see also sect.~\ref{sec:multi_state_fits}. While the results are compatible, the statistical error in the left panel is smaller by roughly a factor of two. In refs.~\cite{Chang:2018uxx,Walker-Loud:2019cif} it has been argued that a key advantage of the Feynmann-Hellmann inspired approach is that the leading excited state contamination is not just of $\mathcal{O}\l(e^{-\Delta \tsep}\r)$ as for the commonly used version of the summation method but that it is further suppressed by the difference between contributions separated by $\tau$ as given in eq.~(\ref{eq:CalLat_method}). This is used to justify the inclusion of data at much smaller values of $\tsep$ in these fits than for the sequential method, which causes the final statistical errors to be smaller. In fact, the fit in the left panel of fig.~\ref{fig:FH_vs_sequential_method} includes values of  $\tsep\gtrsim0.3\fm$, while for the sequential method in the right panel only data for $\tsep/a\gtrsim1.0\fm$ has been used; see also the discussion in the 2019 FLAG review in ref.~\cite{Aoki:2019cca}. Note that in principle it is possible to perform a similar fit for the summation method by including the leading excited state contamination in eq.~\ref{eq:summation_method}, which would allow to include of data at smaller values of $\tsep$ as well; see the discussion in subsect.~\ref{subsec:two_state_truncation}. However, the analysis and the resulting error estimate for $g_A^{u-d}$ published in ref.~\cite{Chang:2018uxx} has been subject to criticism in ref.~\cite{Gupta:2018qil} by the PNDME collaboration (re-)analyzing an extended set of ensembles and in ref.~\cite{Green:2018vxw} because the distribution of fit qualities fails the Kolmogorov-Smirnov test, unlike the updated results from the sequential method published in ref.~\cite{Gupta:2018qil}. Note that there has also been disagreement regarding the central value of $g_A^{u-d}$ obtained in ref.~\cite{Chang:2018uxx} and the result from the sequential method in refs.~\cite{Bhattacharya:2016zcn,Gupta:2018qil}, which has been attributed to the chiral and continuum extrapolation as discussed in ref.~\cite{Gupta:2018qil} leading to the claim that an overall error of $\sim5\%$ for $g_A^{u-d}$ is realistic. A more detailed assessment of this claim is beyond the scope of this review. \par

\section{Multi-state fits} \label{sec:multi_state_fits}
A widespread approach to deal with excited states in nucleon structure calculations are fit models that explicitly include the effects of some of the excited states in the determination of the ground state NME. Since fits can be applied to the same data as the ratio and the summation method they are straightforward to implement and are currently used in most nucleon structure calculations. Furthermore, fits are quite flexible with respect to the actual fit ansatz, the choice of fit ranges and priors as well as \eg the choice of simultaneously fitted observables. Therefore, basically every major lattice collaboration performing nucleon structure calculations has adopted their own favored method in the last years. In this section several approaches will be discussed that have been used in modern lattice calculations. \par

\subsection{Two-state truncation} \label{subsec:two_state_truncation}
Fit models for excited state contamination in NME calculations are based on truncations of the spectral decomposition of the individual two- and three-point functions in eqs.~(\ref{eq:2pt_pspace})~and~(\ref{eq:3pt_pspace}). A popular choice are the respective two-state truncations that are given by
\begin{equation}
 \Ctwopt{t}{p} = \l|A_0(\vec{p})\r|^2 e^{-E_0(\vec{p})t} + \l|A_1(\vec{p})\r|^2 e^{-E_1(\vec{p})t} + ... \,,
 \label{eq:2pt_two_state_truncation}
\end{equation}
and
\begin{align}
 C&^X_{\mu_1...\mu_n}(\vec{p}_f, \vec{p}_i, \tins, \tsep) \notag \\
 &=       A_0^f {A_0^i}^* \bra{0,\vec{p}_f} \mathcal{O}^X_{\mu_1 ... \mu_n} \ket{0, \vec{p}_i} e^{-E_0^f(\tsep-\tins)} e^{-E_0^i\tins} \notag \\
 &\quad + A_0^f {A_1^i}^* \bra{0,\vec{p}_f} \mathcal{O}^X_{\mu_1 ... \mu_n} \ket{1, \vec{p}_i} e^{-E_0^f(\tsep-\tins)} e^{-E_1^i\tins} \notag \\
 &\quad + A_1^f {A_0^i}^* \bra{1,\vec{p}_f} \mathcal{O}^X_{\mu_1 ... \mu_n} \ket{0, \vec{p}_i} e^{-E_1^f(\tsep-\tins)} e^{-E_0^i\tins} \notag \\
 &\quad + A_1^f {A_1^i}^* \bra{1,\vec{p}_f} \mathcal{O}^X_{\mu_1 ... \mu_n} \ket{1, \vec{p}_i} e^{-E_1^f(\tsep-\tins)} e^{-E_1^i\tins} \notag \\ 
 &\quad + ... \,,
 \label{eq:3pt_two_state_truncation}
\end{align}
where the superscripts $i$ and $f$ indicate the dependence of the overlap factors and energies on the initial and final state momenta $\vec{p}_i$ and $\vec{p}_f$, \eg $A^f_k=A_k(\vec{p}_f)$. In principle, all the overlap factors, matrix elements and energies in this expression are free parameters that need to be determined by the fit. Correlated fits are performed to the lattice data by minimizing
\begin{equation}
 \chi^2_\mathrm{corr} = \vec{\chi}^T \tens{C} \vec{\chi} \,,
 \label{eq:corr_chisq}
\end{equation}
where $\tens{C}$ denotes the covariance matrix and $\vec{\chi}=\vec{Y} - \vec{f}(X_0,...,X_n)$ the difference between the vector of lattice data $\vec{Y}$ entering the fit and the fit model $\vec{f}(X_1,...,X_n)$ depending on a set of fit parameters $X_1,...,X_n$ that are to be determined by the fit. Although it is possible to perform a direct, simultaneous fit to the model given by eqs.~(\ref{eq:2pt_two_state_truncation})~and~(\ref{eq:3pt_two_state_truncation}) in its most general form, see \eg the nucleon form factor calculations in refs.~\cite{Alexandrou:2017ypw,Alexandrou:2017hac}, in practice it can be difficult to obtain reliable results in this way considering the finite statistical precision of lattice data. The reason for this are the fairly large number of parameters even in case of the two-state truncation and possibly the size of the covariance matrix. This is particularly true at non-zero momentum transfer because the number of independent fit parameter is larger than for the zero-momentum case while the signal quality deteriorates with increasing $Q^2$. Therefore, additional assumptions and knowledge about parameters are often used to stabilize fits. \par

A common way to increase stability of the fit are assumptions on the energy gaps between the first excited state and the ground state. A rather basic ansatz has been explored several years ago by the Mainz group for a calculation of isovector electromagnetic form factors in ref.~\cite{Capitani:2015sba}. Instead of fitting two- and three-point functions separately, the corresponding two-state truncations have been used to parametrize the leading excited state contribution to the effective form factors obtained from the ratio eq.~(\ref{eq:ratio}) as
\begin{align}
 G_{E,M}^\mathrm{eff}(Q^2, \tins, \tsep) =& G_{E,M}(Q^2) + c_{E,M}^{(1)}(Q^2) e^{-M_\pi \tins} \notag \\ 
 &+ c_{E,M}^{(2)}(Q^2) e^{-2M_\pi(\tsep-\tins)} \,.
\end{align}
In this expression the energy gaps have been fixed to $M_\pi$ and $2M_\pi$ assuming that multi-particle states are responsible for the leading excited state contamination while neglecting nucleon-pion interactions. The choice of $M_\pi$ for the first gap is motivated by the initial state nucleon having momentum $\vec{p}_i$ allowing for a corresponding $N\pi$ state with a moving nucleon and a pion at rest. On the other hand the final state nucleon is produced at rest, \ie $\vec{p}_f=\vec{0}$ which implies that the lowest state for the second gap is a $N\pi\pi$ state with two pions in an S-wave, motivating the choice of $2M_\pi$ for this gap. While this approach helps to stabilize the fit to lattice data, it comes at the price of introducing additional model dependence. \par

\begin{figure*}
\includegraphics[totalheight=0.195\textheight]{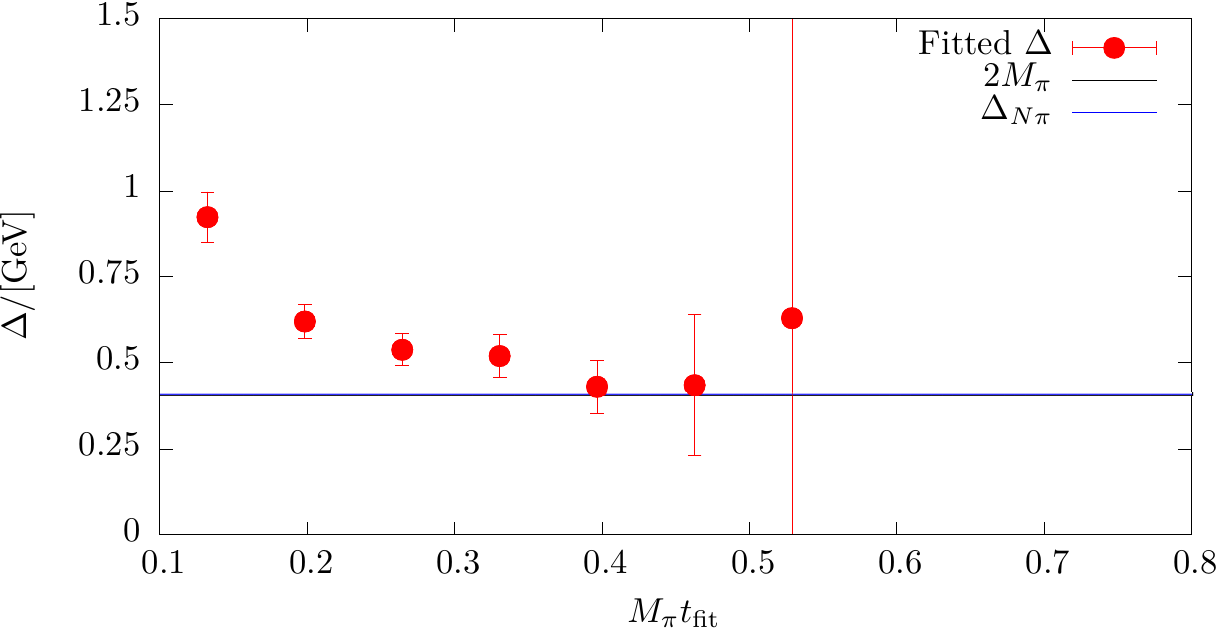}
\includegraphics[totalheight=0.195\textheight]{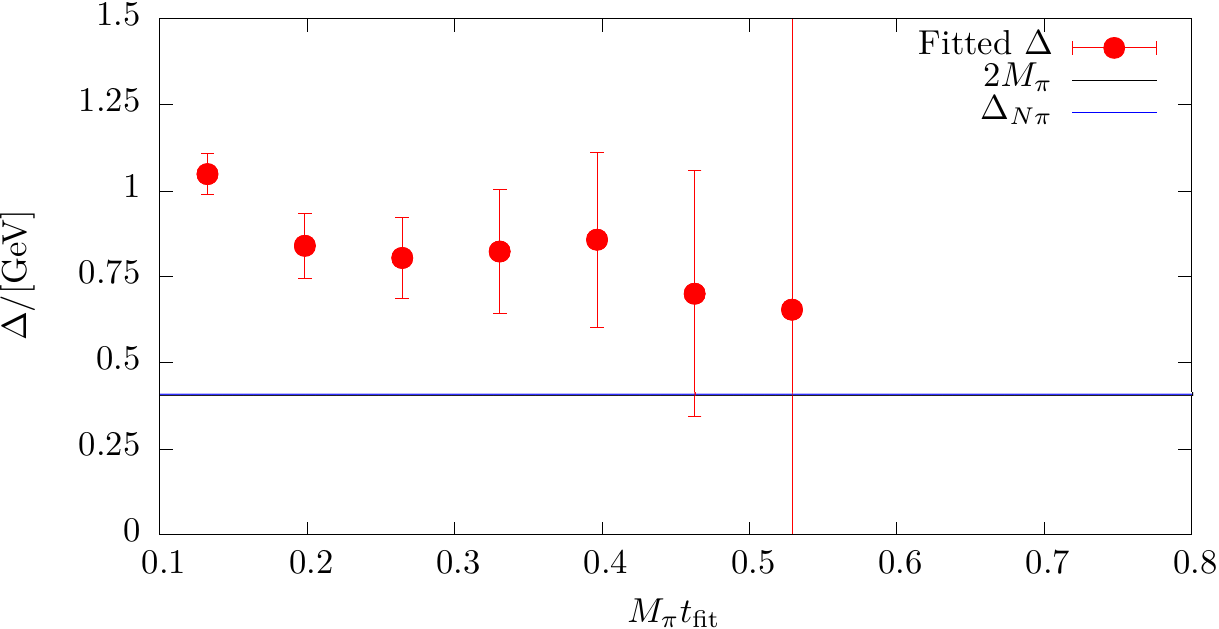}
\caption{Results for the energy gap $\Delta$ from a simultaneous fit to the two-state truncation of the ratio in eq.~(\ref{eq:ratio_zero_momemtum_twostate}) for six observables as a function of the lower bound of the fit range $t_\mathrm{fit}$ in units of $M_\pi$ computed on a CLS ensemble (D200) with $M_\pi\approx 200\mev$, $a=0.06426\fm$ and $T\times L^3 = 128a \times (64a)^3$. The blue, horizontal line marks the gap $\Delta_{N\pi}$ between the ground state and the lowest (noninteracting) $N\pi$ state. Left panel: Results without including the last term $\sim \exp(-\Delta \tsep)$. Right panel: Results for including all terms in eq.~(\ref{eq:ratio_zero_momemtum_twostate}). The data for both figures has been generated in the study in Ref.~\cite{Harris:2019bih}.}
\label{fig:multi-state_fit_gaps}
\end{figure*}

In a more recent study of nucleon charges and moments of twist-2 operators in ref.~\cite{Harris:2019bih} similar fits to ratio data have been carried out, which amounts to fitting the expression in eq.~(\ref{eq:ratio_zero_momemtum_twostate}). Although the number of fit parameters is reduced for the analysis of data at $Q^2=0$, this alone turned out insufficient to yield stable results when fitting the data for a single observable. The same problem has been observed in another calculation of nucleon charges in ref.~\cite{Hasan:2019noy} that used single observable fits to the two-state truncated ratio among other methods. In order to remedy this issue, simultaneous fits to all six observables, \ie $g_{A,S,T}^{u-d}$, $\langle x\rangle_{u-d}$, $\langle x\rangle_{\Delta u-\Delta d}$ and $\langle x\rangle_{\delta u-\delta d}$ have been introduced in ref.~\cite{Harris:2019bih}. This exploits correlations in the data which was found to further stabilize the fits and improve statistical precision without the need resorting to further assumptions or priors. However, one must keep in mind that the covariance matrix entering eq.~(\ref{eq:corr_chisq}) can become a limiting factor for such simultaneous fits, as it may be poorly estimated for the available number of independent measurements if too many observables (or timelices) are included. Still, for the two-state truncation fitting several observables was sufficient to track the convergence of the gap as a function of the lower value of the fit range $\l[t_\mathrm{fit}, \tsep/2\r]$ on plateau data symmetric around $\tins=\tsep/a$. Results from this procedure are shown in fig.~\ref{fig:multi-state_fit_gaps} on an ensemble with a pion mass of $M_\pi=200\mev$. In the left panel the last term in eq.~(\ref{eq:ratio_zero_momemtum_twostate}) has been neglected and clear convergence towards the expected gap of $\Delta\approx2M_\pi$ is observed for $M_\pi t_\mathrm{fit}\gtrsim 0.4$. Unlike approaches that use information from \eg fitting the two-point function to fix the gap in the fit of the ratio or two- and three-point functions, this gives additional confidence that the fit model is applied in a regime for which ground state dominance is reached and higher states beyond the leading contamination become sufficiently suppressed. \par

The right panel of fig.~\ref{fig:multi-state_fit_gaps} shows results for $\Delta$ from a fit including the excited-to-excited state contribution in the last term of eq.~(\ref{eq:ratio_zero_momemtum_twostate}). This leads to significantly larger errors on $\Delta$ and to larger values of $\Delta$ itself although at least the trend towards $2M_\pi$ remains visible\footnote{Note that the matrix elements of interest and their errors are less affected than $\Delta$ by the additional term in the fit.}. The primary reason for this is again related to the signal-to-noise problem, \ie the additional parameter $A_{11}$ in the fit is rather poorly constrained by the lattice data due to the exponential error growth with $\tsep$ and the limited number of different $\tsep$ values that can be included in the fit. Similar to what has been discussed before for the linear fits in the summation method that tend to be dominated by data at the smallest values of $\tsep$, the situation could be improved by increasing the number of measurement for every step in $\tsep$ such that the effective statistics remains approximately constant across the available values of $\tsep$. At any rate, going beyond zero-momentum transfer with this method without additional assumptions would remain difficult due to the increasing number of parameters and less precise data. \par

\begin{figure}
 \includegraphics[totalheight=0.275\textheight]{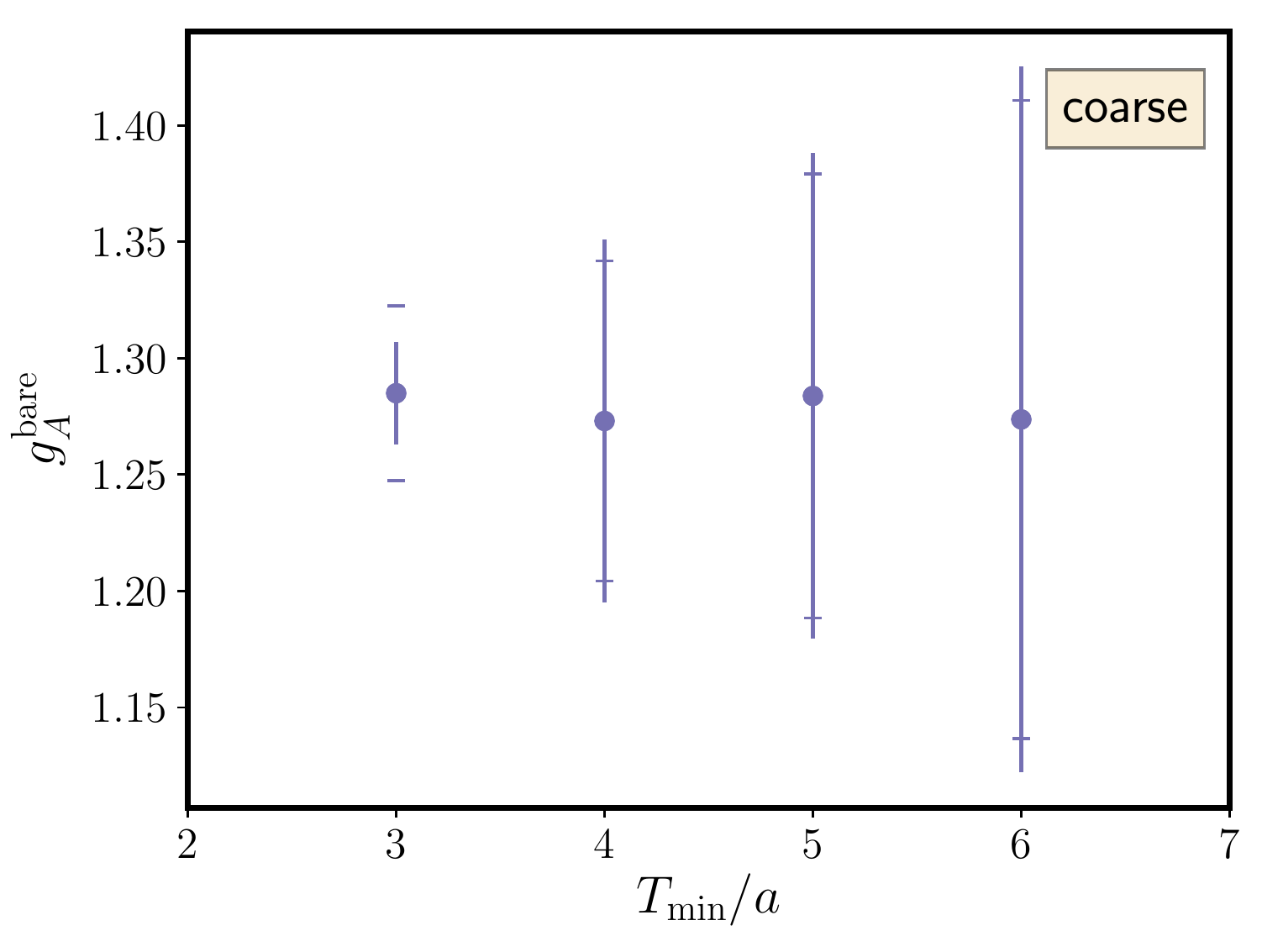}
 \caption{Unrenormalized $g_A^{u-d}$ from a fit to eq.~\ref{eq:summation_method_with_corrections} as a function of the minimal source-sink separation denoted by $T_\mathrm{min}/a$. Results have been obtained on an $N_f=2+1+1$ 2-HEX-smeared Wilson-clover ensemble with $M_\pi=137(2)\mev$, $a=0.1163(4)\fm$ and $T\times L^3=(48a)^4$. The vertical lines indicate the usual statistical error from the fit, while the end caps of the error bars represent the errors obtained when fixing the gap $\Delta$ to its central value. The label ``coarse" refers to the coarser of the two ensembles used in ref.~\cite{Hasan:2019noy}. The figure has been originally published in ref.~\cite{Hasan:2019noy} and is reproduced under the Creative Commons Attribution 4.0 International license.}
 \label{fig:g_A_summation_method_with_corrections}
\end{figure}

A different approach that employs a two-state truncation on the summation method \cite{Hippel:2014kta} has been explored in refs.~\cite{Hasan:2019noy} to complement the usual ratio and summation methods as well as the aforementioned (single observable) two-state fits to the ratio. The method corresponds to an extension of the usual summation method by supplementing the linear behavior of the summed ratio in $\tsep$ as given in eq.~(\ref{eq:summation_method}) with the first excited state correction. This leads to the following fit form
\begin{align}
 S^X_{\mu_1 ... \mu_n}(\tsep) =&  c^{X,0}_{\mu_1 ... \mu_n} + \bra{0} \mathcal{O}^X_{\mu_1 ... \mu_n}\ket{0} \tsep  \notag \\
                               &+ c^{X,1}_{\mu_1 ... \mu_n} \tsep e^{-\Delta \tsep} + c^{X,2}_{\mu_1 ... \mu_n} e^{-\Delta \tsep} \,,
 \label{eq:summation_method_with_corrections}
\end{align}
where the operator-dependent coefficients $c^{X,i}_{\mu_1 ... \mu_n}$ and the operator-independent gap $\Delta$ are left as free parameters of the fit. Exemplary results for the unrenormalized, axial charge as a function of the minimal source-sink separation entering the fit are shown in fig.~\ref{fig:g_A_summation_method_with_corrections}. Regarding the central value there is no dependence observed on the minimal value of $\tsep$ within the rapidly growing error. This kind of fit is similar to the ones performed in the determination of $g_A^{u-d}$ using the Feynman-Hellmann inspired approach as discussed in subsect.~\ref{subsec:Feynmann_Hellmann}, \ie from a purely technical point of view the two calculations mainly differ in how the summation over the operator insertion has been implemented for the data entering the fit. However, the gauge ensembles used and most likely the resulting computational cost are very different, which makes a conclusive comparison impossible, even if one were to ignore the fact that in the sequential method multiple observables and momenta are computed simultaneously. It is interesting to note though that in both studies similarly small statistical errors are obtained for including data starting from $\tsep\gtrsim0.3\fm$. Still, the preferred value quoted in ref.~\cite{Hasan:2019noy} from this method has been chosen such that only data with $\tsep\gtrsim0.58\fm$ have been included due to concerns regarding the statistical quality of the fit including data at smaller values of $\tsep$. In a comprehensive study of results obtained from different methods, this lead to a less favorable statistical signal quality for this method compared to other approaches. Since the gap $\Delta$ in eq.~(\ref{eq:summation_method_with_corrections}) is independent of the observable it might be worth investigating if a simultaneous fit across multiple observables helps to improve stability and statistical precision of the fit as it was found for fits to the ratio in ref.~\cite{Harris:2019bih}. \par

\begin{figure*}
 \includegraphics[totalheight=0.205\textheight]{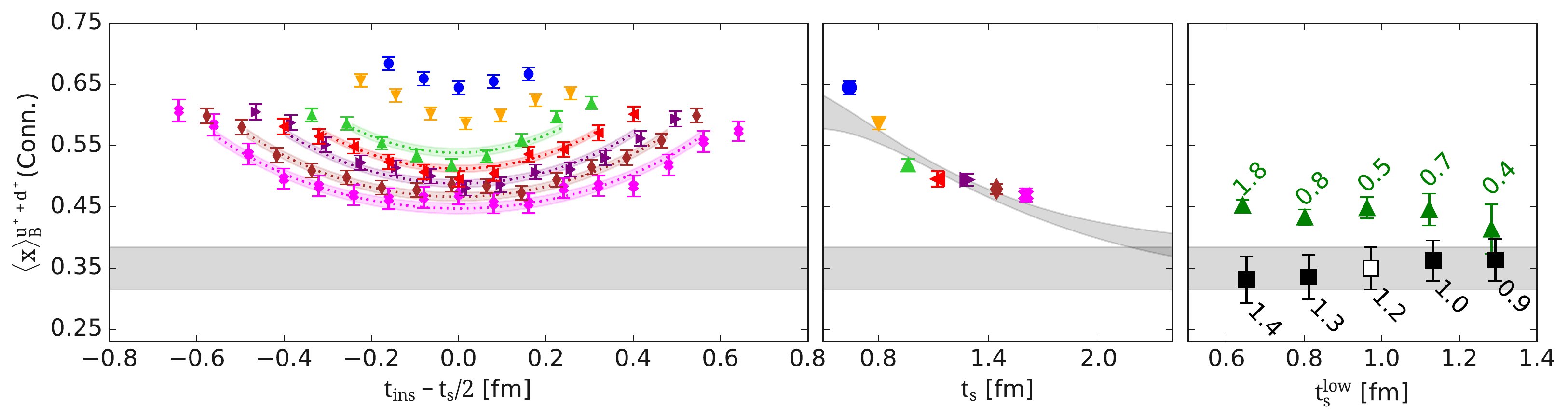}
 \caption{Excited state study for the quark-connected contribution to the isoscalar average quark momentum faction of the nucleon. In the left panel data for the effective form factor from the ratio method is shown for seven values of the source-sink separation $t_s$ with bands from the final two-state fit excluding the two smallest values of $t_s$. The ground state result is indicated by the horizontal gray band in all three panels. The central panel shows results from the ratio method and the predicted time-dependence using the parameters from the final fit with $t_s/a\geq12$ as a Gray band. In the right panel results from the summation method (green triangles) are compared to the two-state fit (black squares) as a function of the lowest source sink separation $t^\mathrm{low}_s$ for each analysis, together with the value of $\chi_\mathrm{corr}^2/\mathrm{d.o.f.}$ for each fit. Results have been obtained on an $N_f=2+1+1$ ETMC ensemble with $M_\pi=0.1393(7)\mev$, $a=0.0801(4)\fm$ and $T\times L^3 = 128a \times (64a)^3$. The figure has been originally published in ref.~\cite{Alexandrou:2020sml} and is reproduced under the Creative Commons Attribution 4.0 International license.}
 \label{fig:ETMC_two_state_fit_avg_x}
\end{figure*}

Instead of explicitly fixing an energy gap using model assumptions, or leaving it entirely as a free parameter as in the previously discussed approaches used in calculations of NMEs at zero-momentum transfer, it is possible to use information on energies from a separate analysis of the nucleon two-point function. This arguably leads to milder model dependence than \eg the explicit choice $\Delta=2M_\pi$ made in ref.~\cite{Capitani:2015sba} while still enabling two-state fits of NMEs at non-zero momentum transfer. A very recent account of this kind of approach is found in an analysis of the decomposition of the proton spin and momentum fraction by the Extended Twisted Mass Collaboration (ETMC) in ref.~\cite{Alexandrou:2020sml}. In their favored approach the final state is always produced at rest $\vec{p}_f=0$ and the fit is split up in several steps, starting with the two-point function at zero-momentum to extract the nucleon mass $E_0(\vec{0})=M_N$. The mass is used to fix all other ground state energies $E_0(\vec{p})$ from the dispersion relation $E_0(\vec{p})=\sqrt{M_N^2 + \vec{p}^2}$ removing the parameter from the fits of the two-state function at non-zero momentum. Since $M_N$ can be computed with rather good precision the values from the dispersion relation at higher momenta are much more precise than actual lattice data at the same momenta, thus one may assume that this helps to stabilize the fit, which is likely the motivation for this choice. Subsequently, fits to the two-point function at non-zero momentum are carried out to determine the energy gap $\Delta(\vec{p}) = E_1(\vec{p}) - E_0(\vec{p}) $ for all required values of $\vec{p}$, as well as the (ratio of) overlaps. Finally, the remaining four parameters (\ie the matrix elements) are extracted from a simultaneous fit to the ratio data for several values of $\tsep$. The fit ansatz is given by plugging eqs.~(\ref{eq:2pt_two_state_truncation})~and~(\ref{eq:3pt_two_state_truncation}) in eq.~(\ref{eq:ratio}) using the already known values of the remaining fit parameters to remove them from the final fit. Note that in another variation of this method used in \eg a calculation of electromagnetic form factors in ref~\cite{Alexandrou:2018sjm} the final fit as been applied directly to the three-point function instead of the ratio. This is in principle equivalent, but the ratio may be preferred as it is usually shown in figures instead of the actual three-point functions. \par 

Some results from the procedure in ref.~\cite{Alexandrou:2020sml} for the quark-connected contribution to the isoscalar average quark momentum fraction are shown in fig.~\ref{fig:ETMC_two_state_fit_avg_x} including a comparison with the summation method and the resulting values of $\chi^2_\mathrm{cor}/\mathrm{d.o.f.}$. While the fit ranges in $\tins$ have been fixed to $\tins\in\l[\tsep/2-5a, \tsep/2+5a\r]$, the lowest source-sink separation entering the fit $t_s^\mathrm{low}$ has been varied and for the two-state fit a slight upwards trend is observed. The fit qualities improve with increasing value of $t_s^\mathrm{low}$, indicating that at small $t_s^\mathrm{low}$ higher excited states likely are non-negligible. A similar, but slightly downwards pointing trend is observed for the summation method which yields higher values than the two-state fit at $t_s^\mathrm{low}\lesssim 1.4\fm$. For the final choice of $t_s^\mathrm{low}=0.96\fm$ in the two-state fit there appears to be some tension between the predicted dependence on the source-sink separation and data from the ratio method. Besides, the (correlated) errors of the summation method and the two-state fit overlap only at the largest value of $t_s^\mathrm{low}$, which gives a hint that results from the two methods may only converge at even larger values of $\tsep$. Since neither the fit range in $\tins$ has been varied nor the resulting energy gaps are given in ref.~\cite{Alexandrou:2020sml} it is difficult to judge if the fit is performed in the regime where contamination due to higher states are in fact already negligible, or a larger result as favored by the summation method and indicated by the trend in both methods would be more trustworthy. At any rate, the example highlights the importance of carefully crosschecking results from multi-state fits with other methods, apart from only demanding a good fit quality.\par

\begin{figure}
 \includegraphics[totalheight=0.245\textheight]{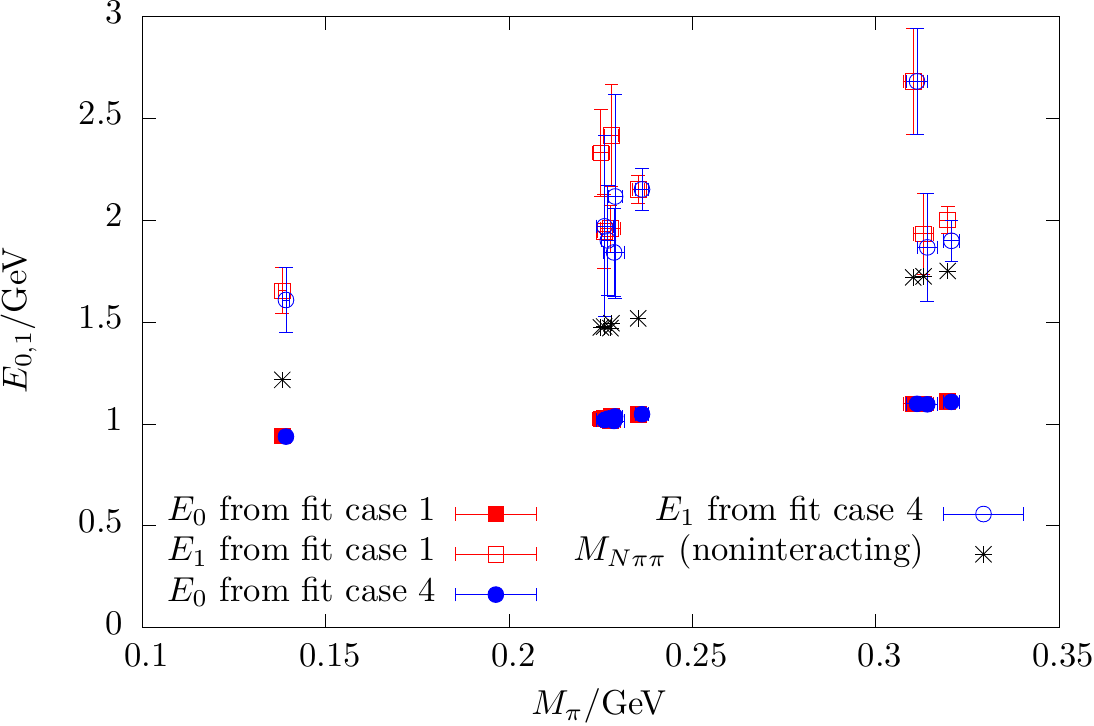}
 \caption{Energies from a simultaneous two-state fit of two- and three-point functions from a calculation of $g_{A,S,T}^{u-d}$ in ref.~\cite{Bhattacharya:2016zcn} as a function of (valence) $M_\pi$. The filled and open symbols refer to the ground state energy $E_0$ and the energy of the first excited state $E_1$, respectively. Results are shown for two different fits ``case 1'' and ``case 4'' which differ by the number of timeslices included when fitting two- and three-point functions as defined in table~V of ref.~\cite{Bhattacharya:2016zcn}. Whenever two different results have been quoted on the same ensemble in ref.~\cite{Bhattacharya:2016zcn}, \ie from using only high precision solves and the full AMA results, only the result with full statistics has been included.}
 \label{fig:PNDME_energies_multistate_fit}
\end{figure}

Ideally, also the energy gaps should be compared against theoretical expectations. In particular, the gap determined by the fit should converge towards the lowest gap in the spectrum for sufficiently large Euclidean time separations of the data in the fit. While corresponding results for the fitted gaps are rarely included in the literature, results from an older calculation of isovector nucleon charges by the PNDME collaboration in ref.~\cite{Bhattacharya:2016zcn} give a strong hint that convergence for the energy gap between ground and first excited state is difficult to achieve for this kind of two-state fits that use information on the energies from the two-point function. In this study a similar fit ansatz for the two-state truncations of two- and three-point functions has been used, albeit only at zero-momentum transfer. Results for the energies of the first two states from this approach are shown in fig.~\ref{fig:PNDME_energies_multistate_fit}. Clearly, the results for the energy $E_1$ of the first excited state lie systematically above the non-interacting $N\pi\pi$ energies. This is in line with what has been reported in ref.~\cite{Hasan:2019noy} for the leading gap obtained from a two-state fit to the two-point function on two ensembles at physical quark mass. Note that depending on the box size a $N\pi$ state might have even lower energy, however, for typical lattices the energies are usually close to each other. The authors of ref.~\cite{Bhattacharya:2016zcn} also performed some variations with respect to the fit ranges, \ie the fit labeled ``case 4'' starts at larger values of $t$ and $\tins$ for the two- and three-point functions, respectively. While this tends to increase the statistical errors as expected, there is only a minor shift (if any) towards smaller values of $E_1$ observed. Therefore, it must be considered doubtful whether such two-state fits correctly describe the contribution from the excited states that need to be subtracted to identify the ground state. \par

\subsection{Including additional states} \label{subsec:ratio_fits}
All of the fits discussed in the previous subsection share as a common feature that they only take into account the contribution of the lowest excited state as they are based on the two-state truncation in eqs.~(\ref{eq:2pt_two_state_truncation}) and~(\ref{eq:3pt_two_state_truncation}). In principle, this can be extended by including further states in the truncation. This is the approach favored by the PNDME collaboration which they have used in several recent NME calculations, \eg for isovector nucleon charges in ref.~\cite{Yoon:2016jzj,Gupta:2018qil}, moments of of twist-2 operator insertion in ref.~\cite{Mondal:2020cmt}, as well as electromagnetic and axial form factors in refs.~\cite{Jang:2019jkn,Rajan:2017lxk,Jang:2019vkm}. \par

\begin{figure}
 \includegraphics[totalheight=0.275\textheight]{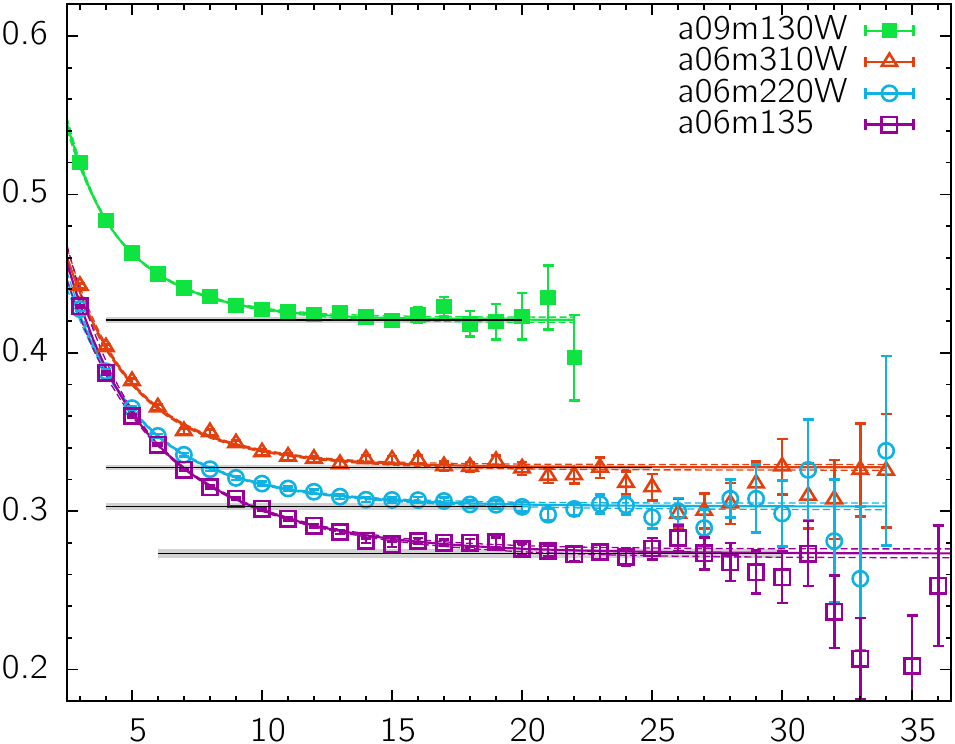}
 \caption{Nucleon effective mass $aM_N$ in lattice units as a function of $t/a$ and results from fits of the four-state truncation in eq.~(\ref{eq:2pt_four_state_truncation}) for some of the ensembles used in ref.~\cite{Gupta:2018qil}. The black, horizontal lines and gray bands indicate the ground state results and their errors, while the colored, solid and dashed curves indicate the full results and errors from the fit. The figure has been originally published in ref.~\cite{Gupta:2018qil} and is reproduced under the Creative Commons Attribution 4.0 International license.}
 \label{fig:PNDME_four_state_truncation}
\end{figure}

As outlined in refs.~\cite{Rajan:2017lxk,Gupta:2018qil} the two-point function is generally fitted to the four-state truncation
\begin{align}
\Ctwopt{t}{p} =& \l|A_0(\vec{p})\r|^2 e^{-E_0(\vec{p})t} + \l|A_1(\vec{p})\r|^2 e^{-E_1(\vec{p})t} \notag \\
               &+ \l|A_2(\vec{p})\r|^2 e^{-E_2(\vec{p})t} + \l|A_3(\vec{p})\r|^2 e^{-E_3(\vec{p})t} + ... \,.
\label{eq:2pt_four_state_truncation}
\end{align}
In order to stabilize these fits beyond the two-state truncation a (sequential) empirical Bayesian analysis with Gaussian priors \cite{Lepage:2001ym,Chen:2004gp} is carried out for the masses $E_{k\geq 1}$ and amplitudes $A_{k\geq 1}$ as described in ref.~\cite{Rajan:2017lxk}. Although the results for $E_{k>1}$ and amplitudes $A_{k>1}$ remain sensitive on both the priors and the lower bound $t_\mathrm{min}$ in the fit, the four-state truncation is found to describe the data sufficiently well as can be seen in fig.~\ref{fig:PNDME_four_state_truncation} for four ensembles covering several values of the pion mass and two lattice spacings. However, as discussed in ref.~\cite{Gupta:2018qil} the resulting energy gaps $E_1(\vec{0})-E_0(\vec{0})$ at zero-momentum transfer are again found to be mostly incompatible with the theoretical expectation of $E_1-E_0\approx 2M_\pi$ as it has been the case for the two-state truncation. This observation corroborates the conclusion that fits to a single two-point function are unable to fully resolve the excited state spectrum. Note that this behavior differs from what has been found in the previously mentioned simultaneous, direct fits to the ratio introduced in ref.~\cite{Harris:2019bih} for which convergence towards $2M_\pi$ is in fact observed. Therefore, while using masses and amplitudes obtained from the four-state truncation of the two point function in the fits of the three-point function can yield an effective description of the data, this procedure should not be expected to allow for a systematic elimination of the contribution from the lowest excited states. \par

For the truncation of the three-point function several ans\"atze have been used in recent calculations by PNDME that are referred to as $2^*$-, $2$- and $3^*$-fits. In their naming scheme, the $2^*$-fits include only the ground state contribution $\sim\bra{0,\vec{p}_f} \mathcal{O}^X_{\mu_1...\mu_n} \ket{0,\vec{p}_i}$ and the $0\rightarrow 1$ transition matrix elements, while the $2$-fit includes the full two-state truncation. Furthermore, the $3^*$ fits incorporate all terms involving the $1\rightarrow2$ and $0\rightarrow 2$ transition matrix elements. In a recent calculation of electromagnetic form factors and radii in ref.~\cite{Jang:2019jkn} final results are mostly obtained from the $3^*$ fits, however, in other cases these fits turn out unstable and the results from the two-state fit are preferred, see \eg ref.~\cite{Gupta:2018qil}. In fact, any matrix elements beyond the ones in the $2^*$-fits are found to be poorly constrained by these fits.\par

Including further states in the truncations of two- and three-point functions does still not guarantee the efficacy of multi-state fits for removal of excited states, which remains dependent on the observable, the available statistical precision and a careful choice of fit ranges and priors. In particular, for observables requiring non-zero momentum transfer like $r_{E,M}^{u-d}$ and $\mu_N^{u-d}$, the excited state contamination is $Q^2$-dependent For example, the result for the nucleon electric radius $r_E^{u-d}=0.769(27)(30)\fm$ in ref.~\cite{Jang:2019jkn} was found to be $16\%$ smaller than the experimental value. This may be explained by the fact that the excited state correction in $G_E^{u-d}(Q^2)$ increases with $Q^2$ and its convergence is from above. If the multi-state fit does not sufficiently remove these effects for data at larger $Q^2$ that enter the extraction of the radius, this may lead to an underestimation of the radius. On the other hand for $\mu_M^{u-d}$ the excited state contamination is large at small values of $Q^2$ and convergence is from below, which likely explains the low result of $\mu_N^{u-d}=3.939(86)(138)$ if the multi-state fits fails to fully remove the contamination. \par

Anyhow, the underestimation of electromagnetic radii and the magnetic moment is a common feature of several recent lattice calculations, see \eg refs.~\cite{Alexandrou:2017ypw,Alexandrou:2018sjm,Alexandrou:2020aja,Jang:2019jkn}, and it is difficult to judge if deviations from experimental results are dominated by residual excited state contamination alone. For example, for the electric form factor also finite size corrections are expected to be $Q^2$-dependent and can thus be difficult to disentangle from excited state effects. Two studies by the PACS collaboration in refs.~\cite{Ishikawa:2018rew,Shintani:2018ozy} give a hint that finite size corrections may indeed play a role, as their results obtained on two fairly large physical volumes ($M_\pi L\approx 6$ and $M_\pi L\approx 7.2$) are in better agreement with experimental results, albeit with larger errors. Moreover, final results for $r_{E,M}$ and $\mu_M$ may exhibit further dependence on the method used to extract them from lattice data, see \eg ref.~\cite{Alexandrou:2016rbj}, the available lattice momenta and the chiral, continuum and finite size extrapolations. \par

\begin{figure}
 \includegraphics[totalheight=0.24\textheight]{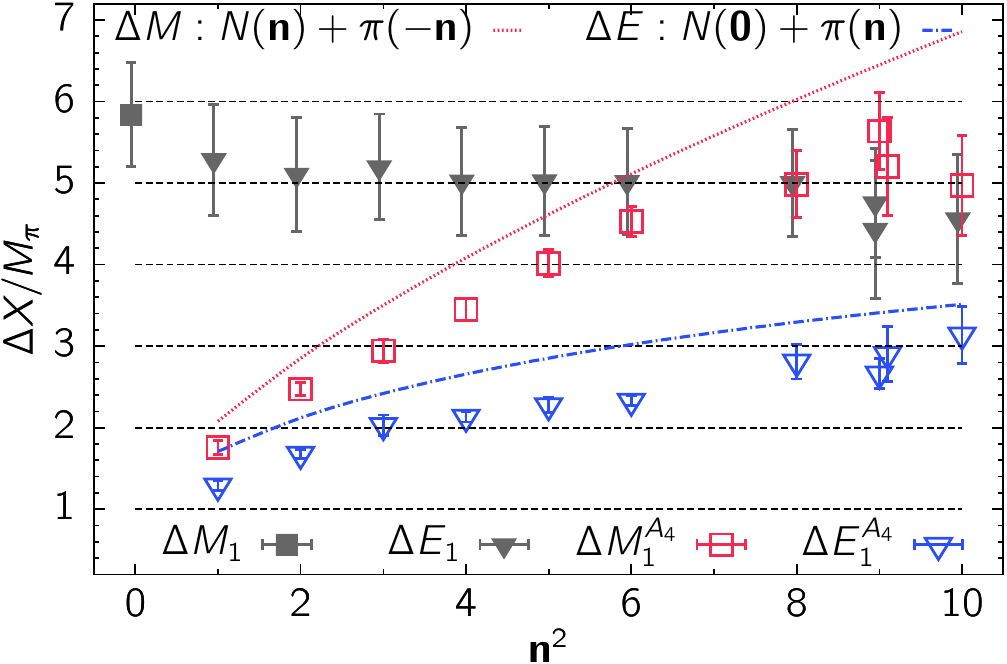}
 \caption{Energy gaps in units of $M_\pi$ determined from the two fit strategies $S_\mathrm{2pt}$ and $S_{A_4}$ in ref.~\cite{Jang:2019vkm}. The dotted lines represent estimates for noninteracting $N\pi$ states with back-to-back momentum (red line) and with the nucleon at rest (blue line). The figure has been originally published in ref.~\cite{Jang:2019vkm} and is reproduced under the Creative Commons Attribution 4.0 International license.}
 \label{fig:spec_splitting_A4}
\end{figure}
Another interesting application of this kind of fit ansatz is found in the recent study of axial form factors in ref.~\cite{Jang:2019vkm} that has already been mentioned in subsect.~\ref{subsec:multi_particle_states}. In this case the conventional multi-state fit strategy that uses information on the energy gaps from the two-point function, labeled by $S_\mathrm{2pt}$ in the notation of ref.~\cite{Jang:2019vkm}, fails badly to achieve ground state dominance. The reason for this is that the four-state fit applied to the two-point function does not resolve $N\pi$ states which are responsible for a large excited state contamination in the pseudoscalar form factor as discussed at the end of subsect.~\ref{subsec:multi_particle_states}. In ref.~\cite{Jang:2019vkm} an alternative fit strategy labeled $S_{A_4}$ has been adopted that uses information on this additional, low-lying energy gaps through a fit to the three-point function from an insertion of the temporal component of the axial vector current. The result for the relevant energy gaps obtained from both procedures as a function of the momentum transfer in integer units are shown in fig.~\ref{fig:spec_splitting_A4} together with the noninteracting estimates for $N\pi$ states with back-to-back momentum and with the nucleon at rest. The noninteracting estimates agree rather well with the behavior of the energy gaps extracted using strategy $S_{A_4}$. Using the fit strategy $S_{A_4}$ the lattice data for $r_\mathrm{PCAC}$ and $r_\mathrm{PPD}$ defined in eqs.~(\ref{eq:PCAC_ratio})~(\ref{eq:PPD_ratio}), respectively, are found to be compatible with unity as shown in the left panel of fig.~\ref{fig:PCAC_and_PPD_test}.\par

\begin{figure}
 \includegraphics[totalheight=0.275\textheight]{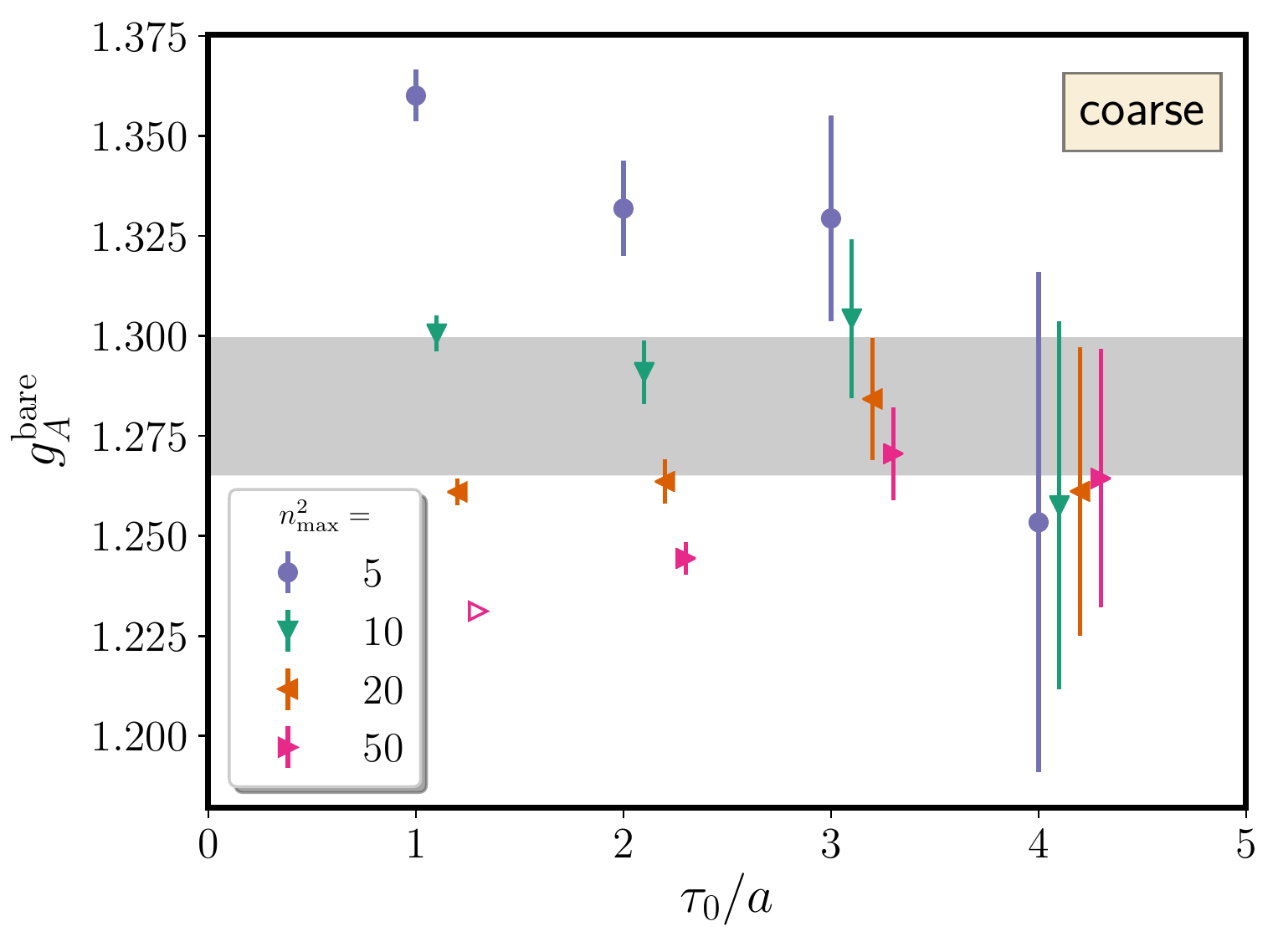}
 \caption{Results from ref.~\cite{Hasan:2019noy} for the unrenormalized isovector axial charge obtained from the many-state fit model in eq.~(\ref{eq:many_state_fit_model}) as a function of the starting value of the insertion time in the fit denoted by $\tau_0$. The open symbol indicates a $p$-value below $0.02$ and the gray band represents the final result from this study. The results are shown for the same ensemble as in fig.~\ref{fig:g_A_summation_method_with_corrections}. The figure has been originally published in ref.~\cite{Hasan:2019noy} and is reproduced under the Creative Commons Attribution 4.0 International license.}
 \label{fig:g_A_many_state_fit}
\end{figure}

A conceptually different approach that aims at modeling $N\pi$ state contributions in a many-state fit has been investigated in ref.~\cite{Hasan:2019noy} for the extraction of $g_{A,S,T}^{u-d}$. It relies on using the noninteracting energies for the first $N\pi$ states
\begin{equation}
  E_{\vec{n}} = \sqrt{\l(\frac{2\pi\vec{n}}{L}\r)^2 + M_N^2} + \sqrt{\l(\frac{2\pi\vec{n}}{L}\r)^2 + M_\pi^2} \,,
  \label{Eq:non_interacting_energies}
\end{equation}
to fix the energy gaps $\Delta_{\vec{n}}=E_{\vec{n}}-M_N$ for states with relative momentum $\vec{p}=2\pi\vec{n}/L$ up to some cutoff value $|\vec{n}|^2 \leq |\vec{n}_\mathrm{max}|^2 \equiv n_\mathrm{max}^2$. This approximation may be justified by the observation in ref.~\cite{Hansen:2016qoz} that the deviation between interacting and noninteracting energy levels is small compared to $\Delta_{\vec{n}}$ itself. The fit ansatz used in ref.~\cite{Hasan:2019noy} reads
\begin{align}
 R^X_{\mu_1...\mu_n}&(\tins, \tsep) = \bra{0} \mathcal{O}^X_{\mu_1...\mu_n} \ket{0} \notag \\
 &+ b^X_{\mu_1...\mu_n} \sum_{\substack{\vec{n}\neq 0 \\ |\vec{n}|^2 \leq |\vec{n}_\mathrm{max}|^2 }} \l( e^{-\Delta_{\vec{n}} \tins} + e^{-\Delta_{\vec{n}} (\tsep-\tins)}\r) \notag \\
 &+ c^X_{\mu_1...\mu_n} \sum_{\substack{\vec{n}\neq 0 \\ |\vec{n}|^2 \leq |\vec{n}_\mathrm{max}|^2 }} e^{-\Delta_{\vec{n}} \tsep}\,,
 \label{eq:many_state_fit_model}
\end{align}
assuming that all ground-to-excited state transitions enter with the same coefficient and that off-diagonal transitions between different excited states are negligible, while excited state contamination in the two-point function is considered important. In fig.~\ref{fig:g_A_many_state_fit} results from this procedure are shown for the unrenormalized axial charge. For small values of the lower bound $\tau_0$ in the insertion time a strong dependence on the choice of $n_\mathrm{max}$ is observed as expected in the presence of a tower of $N\pi$ states contributing to the signal. With increasing values of $\tau_0$ the values appear to converge across different $n_\mathrm{max}$ and the result is found to be consistent with other methods albeit with larger errors. However, in ref.~\cite{Hasan:2019noy} it has been concluded that the strong dependence on $n_\mathrm{max}^2$ in modeling the excited states at small $\tau_0$ likely make the approach unreliable. \par

\section{Variational techniques}
\label{sec:variational_techniques}
A conceptually different approach to tame excited state contamination is the variational method \cite{Michael:1985ne,Luscher:1990ck}, which has become a standard tool in spectroscopy calculations as it allows to systematically remove the lowest excited state contributions \cite{Blossier:2009kd}. The method is based on the computation of a matrix of correlations functions 
\begin{equation}
 \tens{C}^\mathrm{2pt}_{ij}(\vec{p},t) = \sum_{\vec{x}} e^{i\vec{p}\vec{x}}\langle \chi_i(\vec{x},t) \chi_j^\dag(\vec{0},0)\rangle \,,
 \label{eq:2pt_corr_matrix}
\end{equation}
for a basis of $N$ operators $\vec{\chi} = (\chi_1(t), ... , \chi_N(t))^T$ with suitable quantum numbers, and solving the generalized eigenvalue problem (GEVP)
\begin{equation}
 \tens{C}^\mathrm{2pt}(t) \vec{v}_k(t,t_0) = \lambda_k(t, t_0) \tens{C}^\mathrm{2pt}(t_0) \vec{v}_k(t,t_0) \,.,
\label{eq:GEVP}
\end{equation}
for $t>t_0$ and $k\in\l[0,...,N-1\r]$. Energy levels are obtained at large $t$ from the principal correlators $\lambda_k(t,t_0) \sim e^{-E_k(\vec{p})t}$, while the eigenvectors $\vec{v}_k(t,t_0)$ carry information on matrix elements. Note that solving the GEVP generally results in an unsorted set $\l\{s_k(t,t_0)|k\in\l[1,...,N\r]\r\}$ of states $s_k(t,t_0)=(\lambda_k(t,t_0), \vec{v}_k(t,t_0))$ on each timeslice and performing the state assignment going from timeslice $t$ to $t+1$ can be a non-trivial task particularly in the presence of an exponentially deteriorating signal-to-noise ratio; for a discussion of methods to sort the states see ref.~\cite{Fischer:2020bgv}. The most important feature of the variational approach is that ground state energies and matrix elements are improved with respect to the leading excited state contamination which now depends on the gap $E_N(\vec{p})-E_0(\vec{p})$ to the $N$-th state in the spectrum \cite{Blossier:2009kd} instead of the smallest gap in the spectral decomposition of a single two-point function $E_1(\vec{p})-E_0(\vec{p})$. Therefore, contamination from excited states is more strongly suppressed which can be systematically improved by adding more (independent) interpolating operators to the basis $\vec{\chi}$. \par

Beyond spectroscopy the variational approach can be applied in calculations of hadronic matrix elements \cite{Bulava:2011yz}. In this case the generalized eigenvectors that diagonalize the two-point correlation function matrix are used to construct optimized interpolating operators 
\begin{equation}
 \chi^\mathrm{opt}_k = \vec{\chi}\cdot \vec{v}_k \,,
 \label{eq:GEVP_optimized_interpolators}
\end{equation}
for the $k$-th state, that are used in the computation of optimized three-point functions. For nucleon structure calculations an issue arises again from the dense spectrum of multiparticle states, particularly towards physical quark masses and in large volumes. In principle, each state to be removed in the variational approach requires an additional operator, which may require a very large basis. Furthermore, it is known from spectroscopy calculations \cite{Dudek:2012xn} using distillation \cite{Peardon:2009gh} that multiparticle operators must be included to systematically resolve multiparticle states, which for NME calculations is also supported by chiral perturbation theory \cite{Bar:2015zwa}; see the discussion in subsect.~\ref{subsec:multi_particle_states}. However, these are numerically expensive to implement for the computation of nucleon three-point functions. In fact, there has so far only been a single, published study \cite{Egerer:2018xgu} using distillation in the context of NME calculations\footnote{A second study in ref.~\cite{Egerer:2020hnc} extending the work in ref.~\cite{Egerer:2018xgu} by combining momentum smearing with distillation appeared while writing this review.}, which will be discussed in some more detail in subsect.~\ref{subsec:hybrid_interpolators}. Although nonlocal operators have been used in this calculation, multiparticle operator have still not been included. \par

\subsection{Smeared interpolators} \label{subsec:smearing_GEVP}
Since building a basis that systematically accounts for the lowest-lying excited states has not been attempted for NME calculations due to the computational cost induced by the need for multiparticle operators, all existing NME studies using the variational method have aimed at constructing an improved interpolator from a basis of computationally affordable interpolators. A straightforward way is to use differently smeared operators, which has been explored in refs.~\cite{Owen:2012ts,Dragos:2016rtx,Yoon:2016dij}. However, it remains an open question if this approach is actually beneficial compared to the sequential method using a single interpolator with properly tuned smearing in combination with other methods for controlling excited states, \eg two-state fits or the summation method. This concerns the resulting suppression of excited states as well as the computational cost required to achieve a given target precision. \par

\begin{figure}
 \includegraphics[totalheight=0.275\textheight]{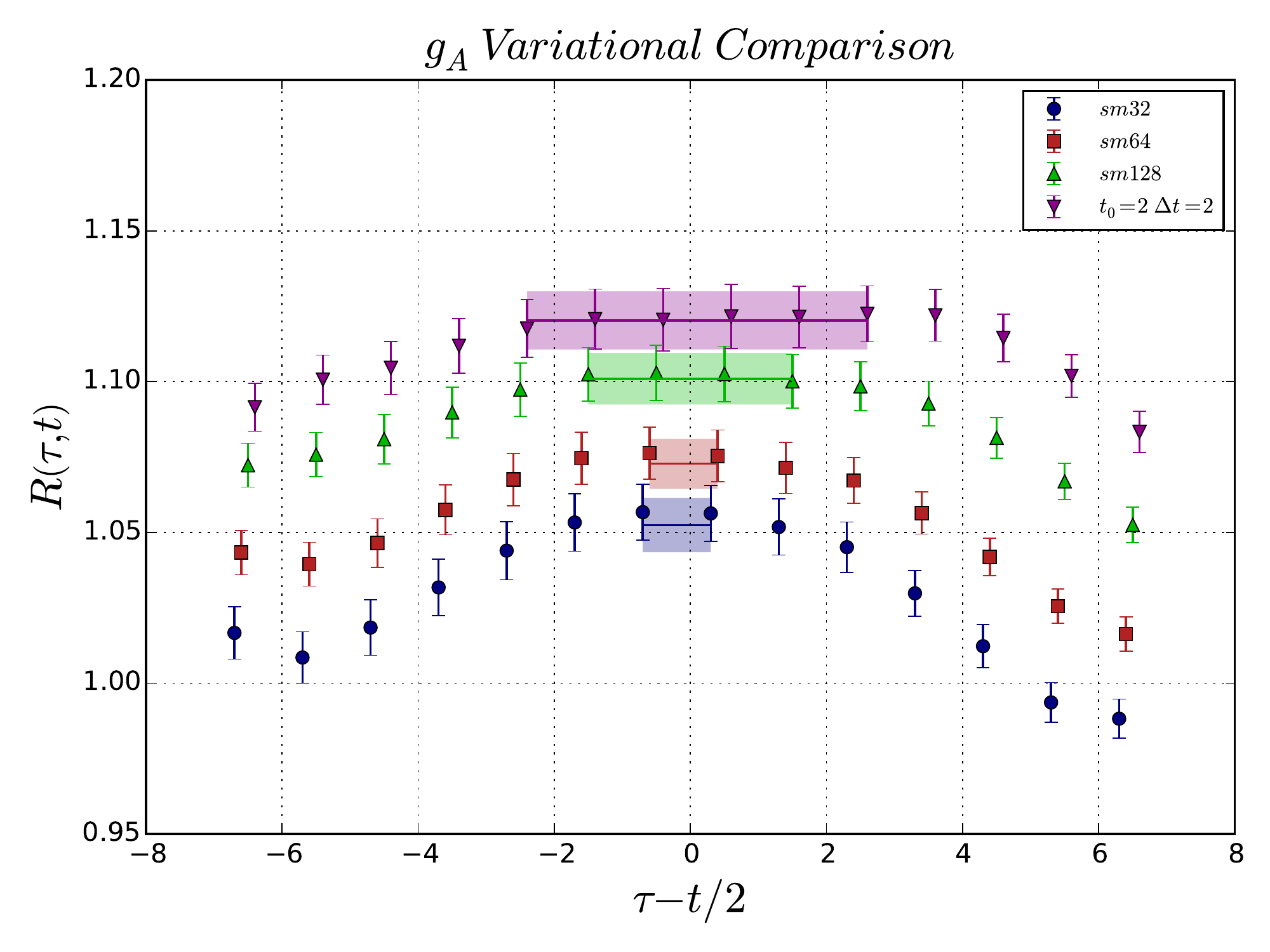}
 \caption{Effective form factor of the isovector axial charge for three different smearing levels and a variationally improved interpolator. Results from smeared interpolators are denoted by sm32, sm64 and sm128 corresponding to $N_\mathrm{smear}=32,63,128$. For the variationally optimized interpolator $t_0$ and $\Delta t$ refer to the sink times $t=t_0$ and $t=t_0+\Delta$ used in the construction of the optimized two-point correlator, see ref.~\cite{Dragos:2016rtx} for technical details. The figure has been originally published in ref.~\cite{Dragos:2016rtx} and is reproduced under the Creative Commons Attribution 4.0 International license.}
 \label{fig:gA_variational_comparison_smearing}
\end{figure}

Figure~\ref{fig:gA_variational_comparison_smearing} illustrates the first part of this issue for $g_A^{u-d}$: the choice of smearing steps $N_\mathrm{smear}$ has a large impact on the resulting excited state suppression for the single interpolator approach and careful tuning of the smearing is required for a meaningful comparison, because too small a value of $N_\mathrm{smear}$ leads to enhanced excited state contamination. Although the excited state contamination appears to be slightly more reduced for the variational method in this example from ref.~\cite{Dragos:2016rtx}, it is close to the one from the single interpolator approach for the largest value of $N_\mathrm{smear}$ and a further increase of $N_\mathrm{smear}$ might have resulted in a compatible value. \par 

While in refs.~\cite{Dragos:2015ocy,Dragos:2016rtx} some indication has been presented that the variational method is more robust with respect to excited state contamination than the two-state fit and summation method, this has been challenged by the study in ref.~\cite{Yoon:2016dij}. In this study it has been concluded that the efficacy of the variational method and a two-state fit applied to the single interpolator approach for reducing excited state contamination is similar. In particular, it has been stressed that the comparison to the two-state method in refs.~\cite{Dragos:2015ocy,Dragos:2016rtx} did not use the optimal smearing size and that the resulting statistical errors might be artificially large due to the choice of the values of $\tsep$ in the fit. Furthermore, it has been pointed out in ref.~\cite{Yoon:2016dij} that the computational cost depends on the setup and that lighter quark masses and larger temporal lattice sizes $T$ work in favor of a (properly tuned) two-state fit approach. At any rate, it appears fair to conclude that no clear advantage of the variational method using a basis of smeared interpolators over other commonly used methods has been found in existing studies. \par

\subsection{Generalized pencil of function (GPOF)} \label{subsec:GPOF}
Another, very cost-efficient approach to obtain a variational basis is the so-called generalized pencil of function method, that has been applied for NME calculations in refs.~\cite{Aubin:2010jc,Aubin:2011zz,Green:2014xba,Ottnad:2017mzd}. It offers an alternative way to analyze the data that are computed using the sequential method for several source-sink separations, as required for other methods such as two-state fits or the summation method. In the GPOF method a set of linearly independent interpolating operators \begin{equation}
 \chi_{\delta t}(t) \equiv \chi(t+\delta t) = e^{H \delta t} \chi(t) e^{-H\delta t} \,,
 \label{eq:GPOF_operator}
\end{equation}
is created by time-shifting an existing operator $\chi(t)$ and used to build a $(N+1)\times(N+1)$ matrix of two-point functions
\begin{align}
 \tens{C}^\mathrm{2pt}(t) &= \l(\begin{smallmatrix}
  \langle\chi_{0\cdot\delta t}(t_f) \chi^\dag(t_i)\rangle & \dots & \langle\chi_{0\cdot\delta t}(t_f) \chi^\dag_{N\cdot\delta t}(t_i)\rangle \\
  \vdots &  \ddots & \vdots \\
  \langle\chi_{N\cdot\delta t}(t_f) \chi^\dag(t_i)\rangle & \dots & \langle\chi_{n\cdot\delta t}(t_f) \chi^\dag_{N\cdot\delta t}(t_i)\rangle \\
 \end{smallmatrix}\r) \notag \\ 
  &= \l(\begin{smallmatrix}
  C^\mathrm{2pt}(t) & \dots & C^\mathrm{2pt}(t+N\cdot\delta t) \\
  \vdots & \ddots & \vdots \\
  C^\mathrm{2pt}(t+N\cdot\delta t) & \dots & C^\mathrm{2pt}(t+2N\cdot\delta t)
 \end{smallmatrix}\r) \,,
 \label{eq:GPOF_2pt}
\end{align}
where any momentum dependence has been suppressed in the notation. Typically, for NME calculations only a single, original operator is used, \eg the standard interpolator in eq.~(\ref{eq:standard_interpolator}). For the case of a nucleon three-point function a non-symmetric but otherwise similar correlation function matrix is built by applying the same procedure to its respective source and sink operators 
\begin{align}
 &\tens{C}^\mathrm{3pt}(\tins, \tsep) = \notag \\
 &\ \l( \begin{smallmatrix}
  C^\mathrm{3pt}(\tins, \tsep) & \dots  & C^\mathrm{3pt}(\tins+N\delta t, \tsep+N\delta t) \\
  \vdots         & \ddots & \vdots \\
  C^\mathrm{3pt}(\tins, \tsep +N\cdot\delta t) & \dots  & C^\mathrm{3pt}(\tins+N\delta t), \tsep+2N\delta t)
 \end{smallmatrix} \r) \,,
\end{align}
where all further indices and parameters besides the time-dependence have been suppressed. The matrix of eigenvectors obtained from the GEVP applied to $\tens{C}^\mathrm{2pt}(t)$ is then used to diagonalize the three-point function matrix
\begin{align}
 &\tens{C}^\mathrm{3pt}(\tins, \tsep) \rightarrow \vec{v}^T \tens{C}^\mathrm{3pt}(t_f, t, t_i) \vec{v} \notag \\
  &\quad=\diag(\Lambda_0(\tins,\tsep), ..., \Lambda_N(\tins,\tsep)) \,.
 \label{eq:GPOF_3pt_rotation}
\end{align}
The principal correlator $\Lambda_0(\tins,\tsep)$ for the ground state can be treated in the same way as the standard three-point functions in a subsequent analysis, \eg the ratio method. Anyhow, NME calculations are just a particular application of the GPOF method and it can in fact be applied to any problem in hadron spectroscopy and structure calculations whenever a variational basis is desired. Further details on GPOF based approaches and the relation between several families of methods, \ie the GEVP and GPOF, the Prony method \cite{Prony:1795,Fleming:2004hs,Beane:2009kya,Cushman:2019hfh} and the Gardner method \cite{doi:10.1063/1.1730560} have recently been discussed in ref.~\cite{Fischer:2020bgv}, where also a new combination of GEVP and GPOF has been proposed. \par

\begin{figure}
 \includegraphics[totalheight=0.195\textheight]{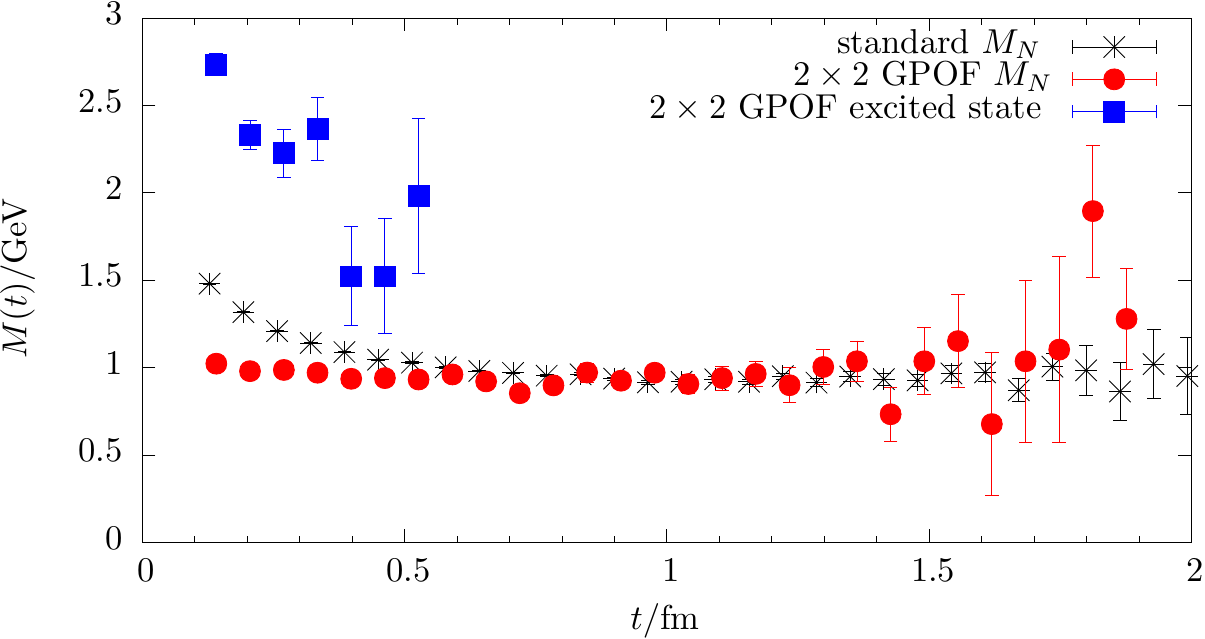}
 \caption{Example data for the nucleon effective mass from a $2\times2$ GPOF in comparison to the single correlator approach. Results from GPOF are horizontally displaced. Data have been generated on the same ensemble as used in fig.~\ref{fig:M_N_vs_Mpi_signal2noise}.}
 \label{fig:GPOF_E250_m_N_comparison}
\end{figure}

In practice, the signal-to-noise problem and availability of different values of $\tsep$ limits the application of the GPOF in NME calculations to a single additional interpolator and thus a $2\times2$ correlation function matrix. Figure~\ref{fig:GPOF_E250_m_N_comparison} shows corresponding results from a $2\times 2$ GPOF with $\delta t=2a$ for the nucleon effective mass on an ensemble at physical quark mass. The excited state contamination is found to be drastically reduced in the GPOF principal correlator at small values of $t/a$. However, the point errors for the GPOF ground state are larger than from the single. As pointed out in ref.~\cite{Green:2014xba} this behavior is expected, because for a system with exactly two-states with energies $E_{0,1}$ the two- and three-point ground state principal correlators read
\begin{equation}
 \lambda_0(t) = C^\mathrm{2pt}(t+2\delta t) - 2 e^{-E_1 \delta t} C^\mathrm{2pt}(t+\delta t)  + e^{-2E_1 \delta t} C^\mathrm{2pt}(t) \,,
\end{equation}
and
\begin{align}
 &\Lambda_0(\tins,\tsep) = C^\mathrm{3pt}(\tins+\delta t, \tsep + 2\delta t) \notag \\
                         &\ - e^{-E_1\delta t} \l( C^\mathrm{3pt}(\tins, \tsep + \delta t) + C^\mathrm{3pt}(\tins + \delta t, \tsep +\delta t) \r) \notag \\
                         &\  + e^{-2 E_1 \delta t} C^\mathrm{3pt}(\tins, \tsep) \,,
\end{align}
respectively, implying that the statistical uncertainties are dominated by the correlators at the largest value of $\tsep$. This strongly limits the statistical precision for the three-point case due to the signal-to-noise problem, at least if effective statistics are not kept constant for increasing values of $\tsep$. Still, for the nucleon mass this effect is compensated by the much earlier onset of the plateau at $\sim 0.4\fm$. \par

Furthermore it has been pointed out in ref.~\cite{Green:2014xba}, that the GPOF might not be efficient for contamination from transition matrix elements. This is because they are more strongly suppressed in the two-point function than the corresponding contributions to the three-point functions by transition to the ground state. However, within the statistical precision of that study it has been concluded that GPOF and summation method lead to comparable results. This has also been confirmed by a more recent study in ref.~\cite{Ottnad:2017mzd} of moments of twist-2 operator insertions, which are notorious for large excited state contamination. In the context of NME calculations the main advantage of the GPOF is clearly its simple implementation compared to \eg multi-state fits and that it can usually be applied on existing data without adding computational cost. However, despite its simplicity the method has not seen widespread use in recent NME calculations. Still it might be worthwhile exploring the GPOF approach for NME calculations further, particularly if the required data are readily available from other methods such that it can \eg serve as a crosscheck. \par

\subsection{Hybrid interpolators} \label{subsec:hybrid_interpolators}
\begin{figure}
 \includegraphics[totalheight=0.163\textheight]{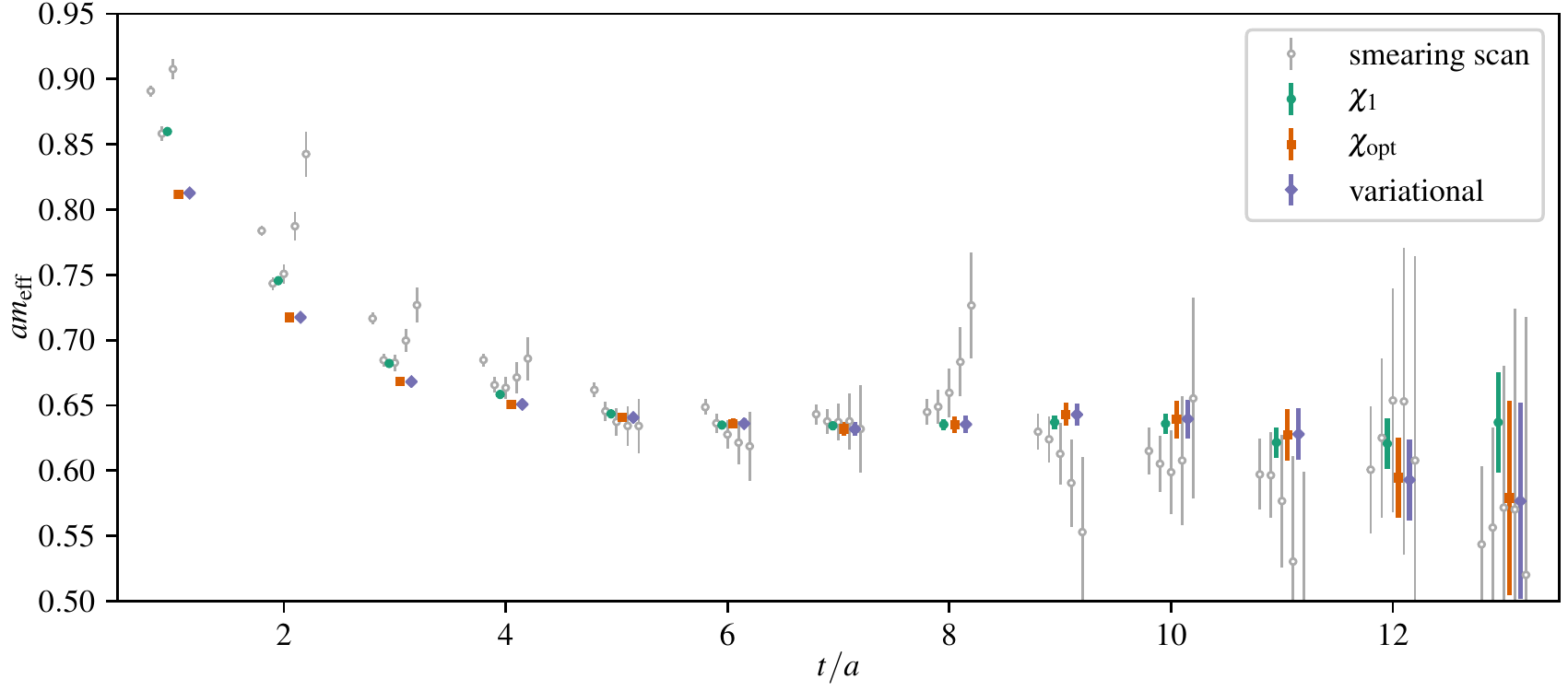} 
 \caption{Comparison of results for effective $M_N$ from a low-statistics smearing scan using the interpolator $\chi_1$ defined in eq.~(\ref{eq:hybrid_chi_1}) (open, gray circles), the variationally improved operator $\chi_\mathrm{opt}$ from a second tuning run and results for $\chi_1$ (filled, green circles) and $\chi_\mathrm{opt}$ (blue diamonds) from full statistics. The figure has been originally published in ref.~\cite{Green:2019zhh} and is reproduced under the Creative Commons Attribution 4.0 International license.}
 \label{fig:hybrid_m_N_eff_comparison}
\end{figure}

The search for an affordable basis of interpolators has recently been extended to so-called hybrid interpolators that include an insertion of a chromomagnetic field $B_i=-\frac{1}{2}\epsilon_{ijk}F_{jk}$ and that can be implemented at similar computational cost compared as the standard approach. In ref.~\cite{Green:2019zhh} the following basis of operators has been studied
\begin{align}
 \chi_1 =&\epsilon_{abc} \l( \tilde{u}^T_a C \g{5} P_+ \tilde{d}_b \r) \tilde{u}_c \,, \label{eq:hybrid_chi_1} \\
 \chi_2 =&\epsilon_{abc} \l( (B_i \tilde{u})^T_a C\g{j} P_+ \tilde{d}_b - \tilde{u}^T_a C \g{j} P_+ (B_i \tilde{d})_b \r) \g{i} \g{j} \tilde{u}_c \,, \label{eq:hybrid_chi_2} \\
 \chi_3 =&\epsilon_{abc} \l( (B_i \tilde{u})^T_a C\g{j} P_+ \tilde{d}_b - \tilde{u}^T_a C \g{j} P_+ (B_i \tilde{d})_b \r) P_{ij} \tilde{u}_c \,, \label{eq:hybrid_chi_3} 
\end{align}
where $\tilde{u}$, $\tilde{d}$ are smeared quark fields, $P_{ij}=\delta_{ij}-\g{i}\g{j}/3$, and $\chi_1$ is the standard interpolator in eq.~(\ref{eq:standard_interpolator}) up to an additional (positive) parity projector $P_+=(1+\g{0})/2$. Two tuning runs have been performed to optimize the smearing for $\chi_1$ and construct a variationally improved interpolator $\chi_\mathrm{opt}$ for the ground state as introduced in eq.~(\ref{eq:GEVP_optimized_interpolators}). Some results for the nucleon effective mass from ref.~\cite{Green:2019zhh} are shown in fig.~\ref{fig:hybrid_m_N_eff_comparison}. The variational basis is found to lead to a reduction of excited state contamination compared to the standard approach, albeit less than the reduction that is usually observed in the previously discussed GPOF approach. Furthermore, these results once more demonstrate the importance of tuning the smearing for optimal results and a meaningful comparison to other methods. \par

\begin{figure*}
 \includegraphics[totalheight=0.275\textheight]{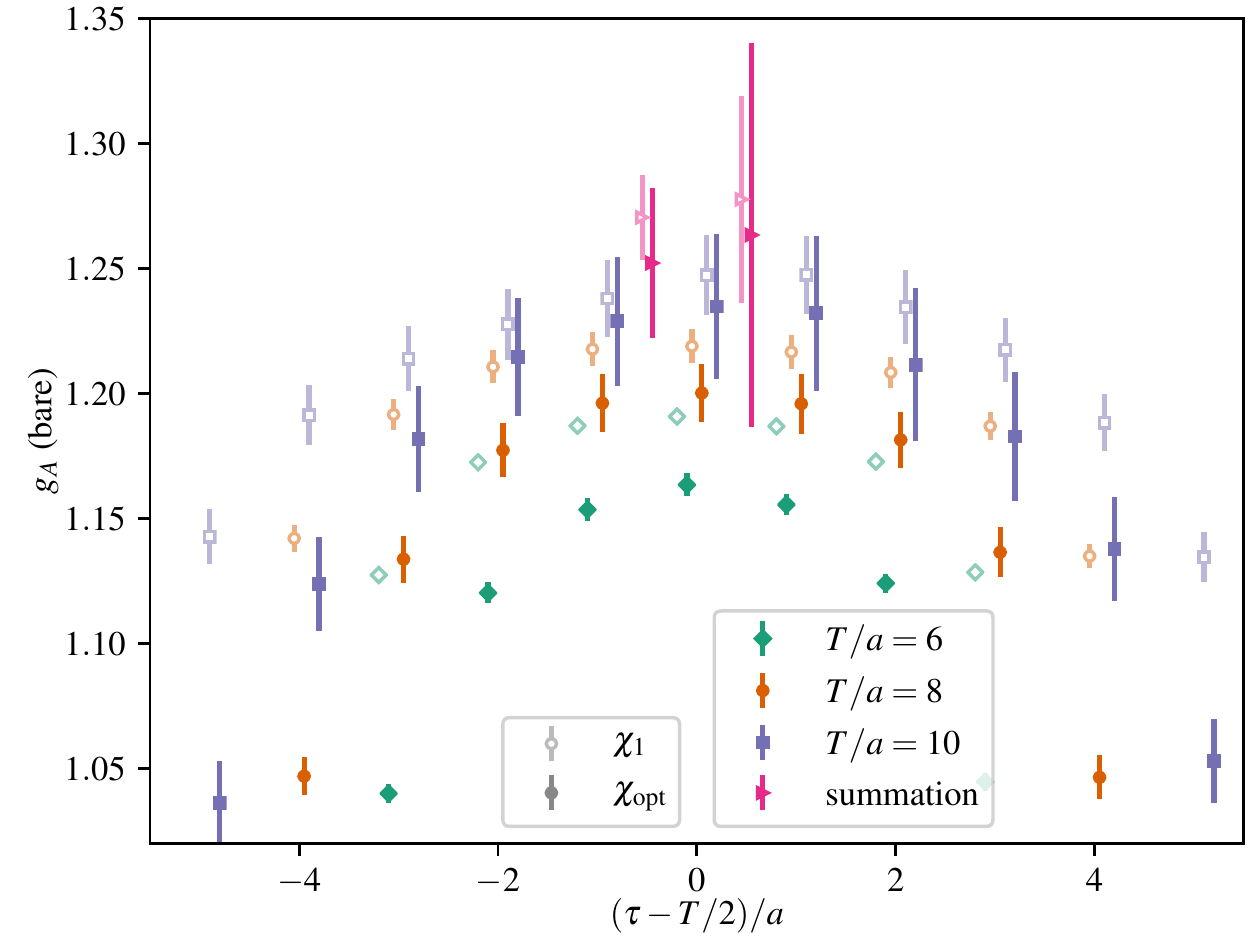} \ 
 \includegraphics[totalheight=0.275\textheight]{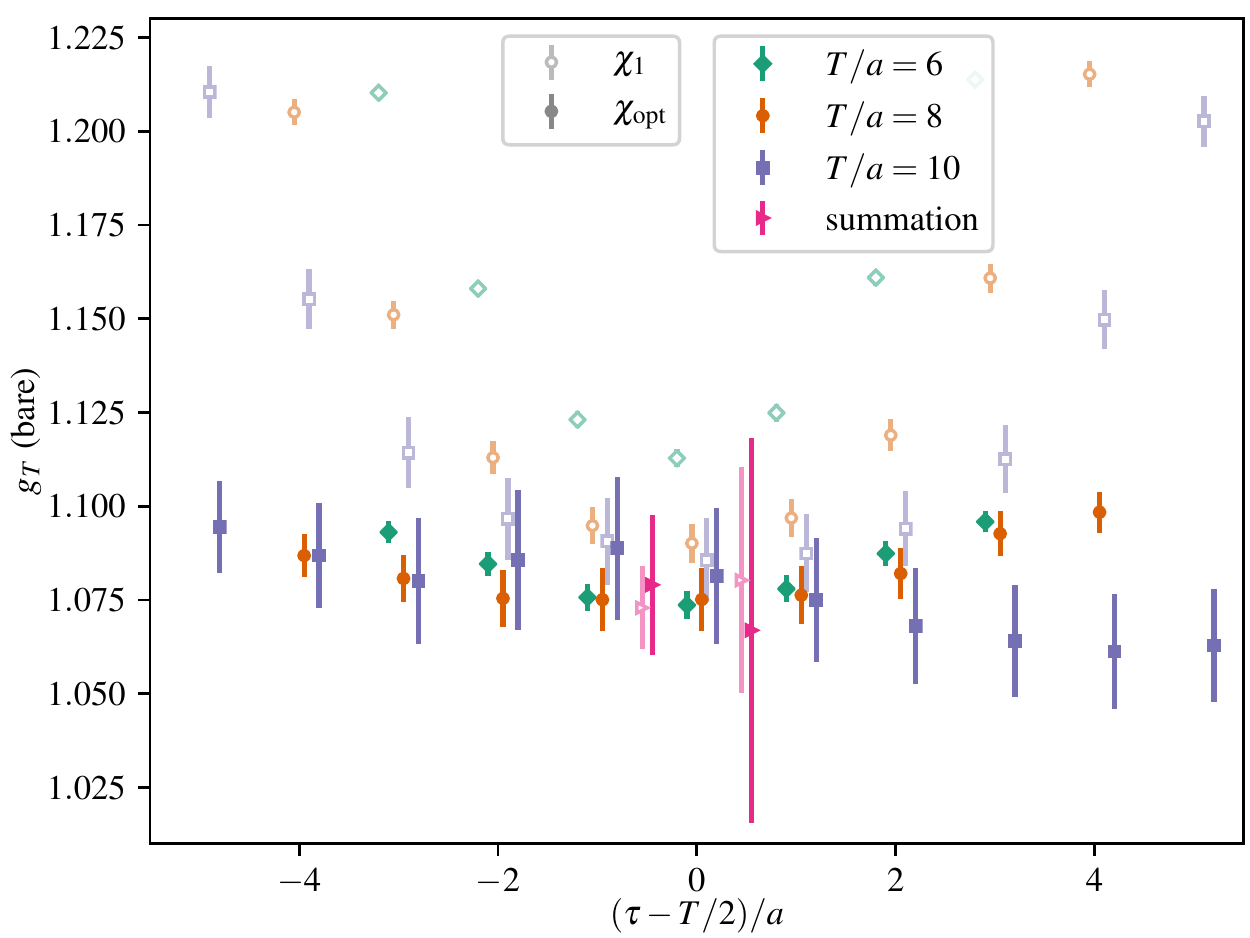}
 \caption{Effective form factors and results from summation method for $g_A^{u-d}$ (left panel) and $g_T^{u-d}$ (right panel) from the interpolator $\chi_1$ defined in eq.~(\ref{eq:hybrid_chi_1}) (open symbols) and variationally optimized operator $\chi_\mathrm{opt}$. The figures have been originally published in ref.~\cite{Green:2019zhh} and are reproduced under the Creative Commons Attribution 4.0 International license.}
 \label{fig:hybrid_g_A_and_g_T}
\end{figure*}

However, for the actual NME calculation the results found in ref.~\cite{Green:2019zhh} were mixed, as can be seen from the results for $g_{A,T}^{u-d}$ in fig.~\ref{fig:hybrid_g_A_and_g_T}. While the excited state contamination is significantly reduced for $g_T^{u-d}$ using the variationally improved interpolator $\chi_\mathrm{opt}$, it is actually increased for $g_A^{u-d}$. The latter result might be explained by a partial (accidental) cancellation of excited states present in the single interpolator approach, which is weakened or prevented in the variational approach. Besides, the statistical errors for $\chi_\mathrm{opt}$ are always larger than for $\chi_1$. These findings corroborate the conclusion that a small variational basis without multi-particle operators does not allow for a systematic treatment of excited states in nucleon structure calculations. \par

\begin{figure*}
 \includegraphics[totalheight=0.21\textheight]{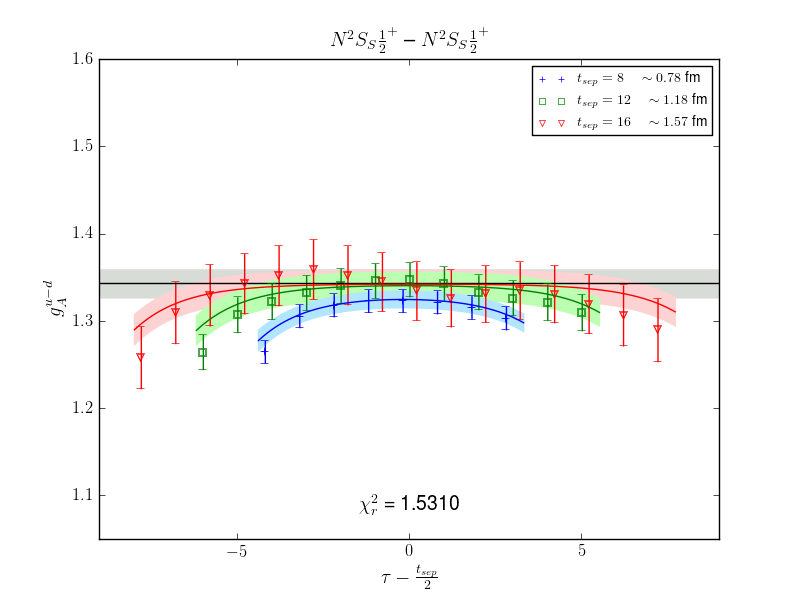} \hspace{-2.35em}
 \includegraphics[totalheight=0.21\textheight]{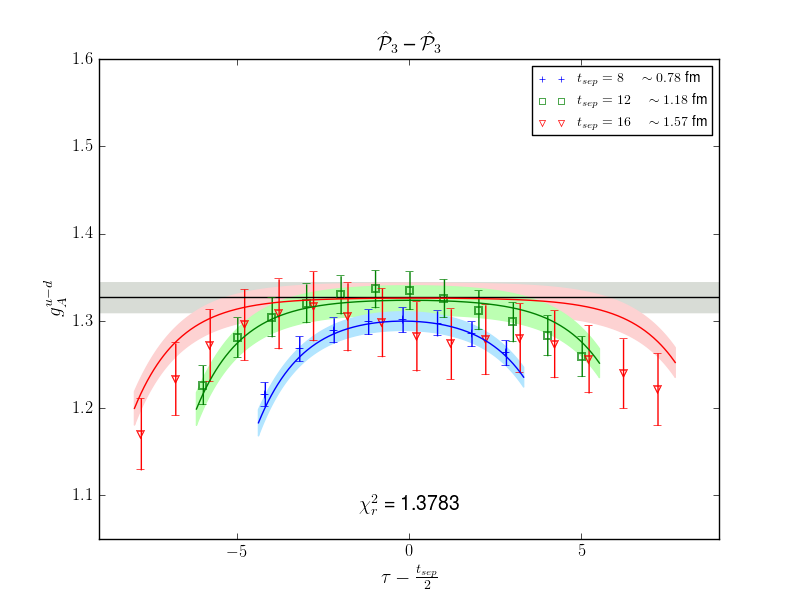} \hspace{-2.35em}
 \includegraphics[totalheight=0.21\textheight]{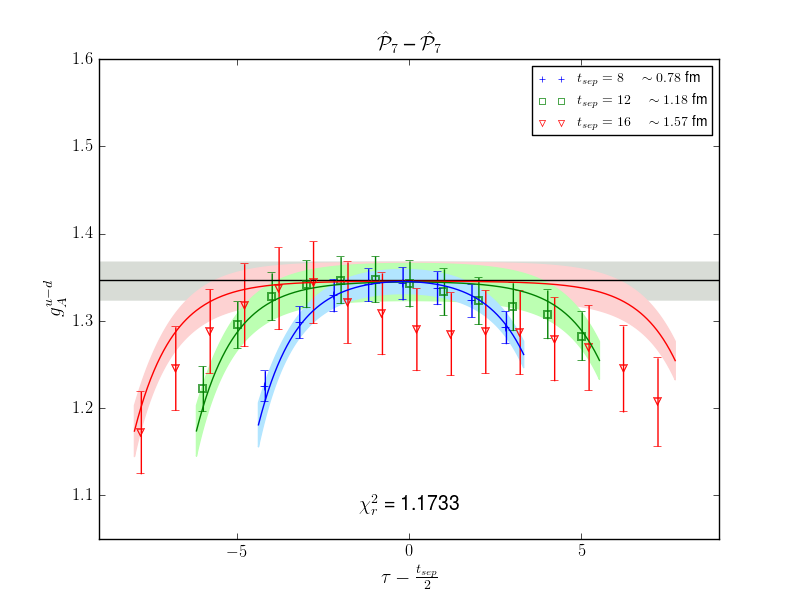}
 \caption{Effective form factor for $g_A^{u-d}$ from a single distilled operator (first panel), a variational improved operators from a basis of three (second panel) and from a basis with seven distilled operators (third panel). The horizontal gray band indicates the final result from a two-state fit and the colored, curved bands represent the fit results at any given value of $\tsep$. The figures have been originally published in ref.~\cite{Egerer:2018xgu} and are reproduced under the Creative Commons Attribution 4.0 International license.}
 \label{fig:distillation_g_A_comparison}
\end{figure*}

Similar results have been found previously in ref.~\cite{Egerer:2018xgu}. In addition to a standard, Jacobi smeared \cite{Allton:1993wc} interpolating field and a single distilled operator ${}^2S_S\frac{1}{2}^+$ that resembles the nucleon interpolator in eq.~(\ref{eq:hybrid_chi_1}), two bases of distilled operators $\mathcal{B}_3$, $\mathcal{B}_7$ have been constructed using covariant derivatives to obtain variationally improved interpolators denoted by $\hat{\mathcal{P}}_3$ and $\hat{\mathcal{P}}_7$, respectively. Basis $\mathcal{B}_3$ includes ${}^2S_S\frac{1}{2}^+$ and two hybrid interpolators which have been found to have large ground state overlap in ref.~\cite{Dudek:2012ag}, while basis $\mathcal{B}_7$ expands $\mathcal{B}_3$ by four additional operators that probe the radial structure of the nucleon. Note that $\mathcal{B}_3$ is similar to the basis used in ref.~\cite{Green:2019zhh} to obtain the variationally improved interpolator. Some results for $g_A^{u-d}$ are shown in the three panels of fig.~\ref{fig:distillation_g_A_comparison} for ${}^2S_S\frac{1}{2}^+$ and the variationally improved interpolators from bases $\mathcal{B}_3$ and  $\mathcal{B}_7$, respectively. For $g_A^{u-d}$ the variationally optimized interpolator $\hat{\mathcal{P}}_3$ shows enhanced excited state contamination compared to the single interpolator approach, similar to what is observed in ref.~\cite{Green:2019zhh}. A second observation is that further increasing the basis indeed improves the situation as one should expect. Again, for $g_T^{u-d}$ (not shown here) the excited state contamination has been found to be reduced by including hybrid interpolators when compared to the single interpolator approach. It should be further noted that the standard, smeared interpolator has been found to give results qualitatively consistent with using ${}^2S_S\frac{1}{2}^+$ as expected, albeit with much larger errors. However, it is not clear if the smearing has been tuned in ref.~\cite{Green:2019zhh}. Besides, it is not possible to directly compare computational cost in a meaningful way to distillation. Still, the findings of the two studies in refs.~\cite{Egerer:2018xgu,Green:2019zhh} using hybrid interpolators qualitatively agree in that a small variational basis without multiparticle operators is insufficient to reliably deal with excited state contamination in NME calculations across multiple observables. \par

\subsection{Parity-expanded variational analysis (PEVA)} \label{subsec:PEVA}
Yet another way to build a variational basis has been developed and applied in a series of papers~\cite{Menadue:2013kfi,Stokes:2018emx,Stokes:2019zdd} for studies of baryons at non-zero momentum. Originally introduced in ref.~\cite{Menadue:2013kfi} at the level of nucleon two-point functions, the method aims at resolving parity mixing between nucleon states at non-zero momentum, hence its name ``parity-expanded variational analysis''. Unlike other approaches that attempt to find a generally applicable variational basis to construct improved operators, the PEVA approach is designed to specifically deal with potential contamination caused by this mixing with states of opposite parity. \par

The basic idea of this method is to start from a basis of conventional interpolators denoted by $\{\chi_i\}$ and construct an extended basis
\begin{align}
 \chi_{i,\pm \vec{p}}  &= \Gamma_{\pm\vec{p}} \chi_i \,, \label{eq:PEVA_op1} \\
 \chi_{i,\pm \vec{p}}' &= \Gamma_{\pm\vec{p}} \g{5} \chi_i \,, \label{eq:PEVA_op2}
\end{align}
using a helicity projector
\begin{equation}
 \Gamma_{\pm\vec{p}}= \frac{1}{4} \l(1+\g{0}\r) \l(1\pm i\g{5}\g{k}\hat{p}_k\r) \,.
 \label{eq:PEVA_projector}
\end{equation}
At zero-momentum the two operators transform as eigenstates of parity, \ie $\chi_{i,\pm \vec{p}}\rightarrow +\chi_{i,\pm \vec{p}}$ and $\chi_{i,\pm \vec{p}}'\rightarrow -\chi_{i,\pm \vec{p}}'$, respectively, while at non-zero momentum they do not have definite parity. The authors of refs.~\cite{Menadue:2013kfi,Stokes:2018emx,Stokes:2019zdd} always consider an initial basis made up two interpolators
\begin{align}
 \chi_1 &= \epsilon_{abc} \l( \tilde{u}_a^T C \tilde{d}_b \r) \tilde{u}_c \,, \label{eq:PEVA_chi_1} \\
 \chi_2 &= \epsilon_{abc} \l( \tilde{u}_a^T C \g{5} \tilde{d}_b \r) \g{5} \tilde{u}_c \,, \label{eq:PEVA_chi_2}
\end{align}
with four different levels of Gaussian smearing to generate the smeared quark fields $\tilde{u}$, $\tilde{d}$, effectively resulting in an $8\times 8$ basis for the conventional variational approach and a $16\times 16$ basis for the PEVA. \par

\begin{figure}
 \input{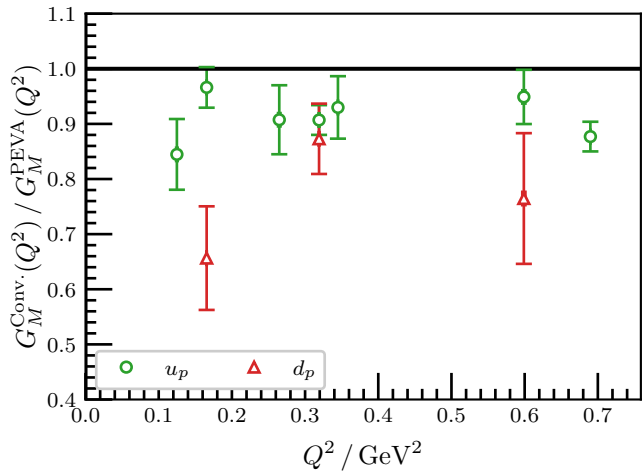}
 \caption{Ratios of plateau method results from the conventional approach and PEVA for the quark-connected contribution to $G_M(Q^2)$. Results are shown separately for up- and down-quark contributions and have been obtained on an ensemble with $M_\pi\approx156\mev$, $a=0.0933(13)\fm$ and $T\times L^3= 64a\times (32a)^3$. The figure has been originally published in ref.~\cite{Stokes:2018emx} and is reproduced under the Creative Commons Attribution 4.0 International license.}
 \label{fig:PEVA_G_M_ratio}
\end{figure}

\begin{figure}
 \input{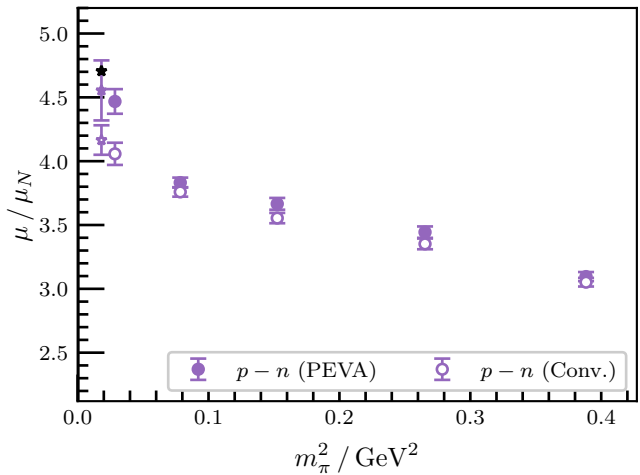}
 \caption{Comparison of results for the isovector magnetic moment from the conventional approach (open symbols) and the PEVA method (filled symbols) across several ensembles. The data have been finite-volume corrected using the corrections in ref.~\cite{Hall:2012pk}. The values at physical quark mass have been obtained from a chiral extrapolation. The figure has been originally published in ref.~\cite{Stokes:2018emx} and is reproduced under the Creative Commons Attribution 4.0 International license.}
 \label{fig:PEVA_G_M_vs_Mpisqr}
\end{figure}

In ref.~\cite{Stokes:2018emx} the method has been applied to the calculation of electromagnetic form factors while the most recent study using the PEVA approach in ref.~\cite{Stokes:2019zdd} extends this further to (elastic) form factors of the first two parity-odd excitations of the nucleon and its lowest-lying parity-even excitation. Here we shall focus on some of the results from ref.~\cite{Stokes:2018emx}. In this study the PEVA method has been first applied for $G_E(Q^2)$ for which no significant difference to results from the conventional approach has been observed. However, for the magnetic form factor PEVA was found to lead to additional reduction of the excited state contamination. This is shown in fig.~\ref{fig:PEVA_G_M_ratio} where the ratio $G_M^\mathrm{Conv.}(Q^2)/G_M^\mathrm{PEVA}(Q^2)$ of the result from the conventional over the PEVA approach is plotted for individual $u$- and $d$-quark contributions including only quark-connected diagrams. Still, a significant difference is only observed on the ensemble with the lightest quark mass in this study, which can be inferred from fig.~\ref{fig:PEVA_G_M_vs_Mpisqr} showing results for the isovector magnetic moment as a function of $M_\pi^2$. Note that the data in this figure have been corrected for finite volume effects using chiral perturbation theory \cite{Hall:2012pk} which is found to give a correction of $\sim 10\%$ for the isovector magnetic moment. \par

Although it is not obvious why one would expect contamination by opposite parity excited states for the magnetic form factor particularly at lighter pion masses, the authors of ref.~\cite{Stokes:2018emx} give some possible explanation for this feature from an investigation of the negative-parity spectrum, \ie they infer an increasing role of multiparticle states in the negative-parity spectrum at lighter pion mass. They further speculate that this could lead to a change in the coupling to the localized operators they use and thus affect the opposite-parity contamination in the ground state matrix element. However, if this explanation is correct, it is not {\it a priori} clear that the PEVA method would still yield a significant correction at light pion masses once multiparticle operators are actually included and it might in fact be more relevant for studies of parity-odd states as in the exploratory study in ref.~\cite{Stokes:2019zdd}. \par

\section{Summary and outlook}
\label{sec:summary}
In the last few years studies of nucleon structure at physical quark mass have become feasible, thus systematic effects due to the chiral extrapolation can be considered well under control. Furthermore, discretization effects are empirically found to be rather small and continuum extrapolations are now part of state-of-the-art calculations as well. The later also applies to some extent to finite size extrapolations, although in this case the picture may be less clear. On the other hand, excited states remain arguably as the most important source of systematic uncertainty in current lattice simulations. One reason for this is that they are strongly observable-dependent and can be large, which in practice makes it difficult to come up with a generally applicable approach to treat them. Moreover, excited state contamination is intimately related to the signal-to-noise problem, which is why there is usually a trade-off between statistical precision and accuracy related to this systematic. This is problematic as it can easily lead to an underestimation of the overall error. \par

In this review an overview of approaches to mitigate excited state contamination in contemporary lattice QCD calculations of nucleon structure has been presented. These methods can be roughly divided in three categories, \ie summed operator insertions, multi-state fits and applications of the variational method. While direct, quantitative comparisons of different methods taking into account the computational cost are rarely found in the literature and often difficult or even impossible to perform, each of these methods has its individual advantages and shortcomings that can be summarized as follows:

\begin{enumerate}
\item Methods based on the summation over the operator insertion aim at increased suppression of excited states. At least in the commonly used form this approach is simple to implement, rather robust and less affected by human bias than \eg multi-state fits, because there are not many tunable parameters involved. However, statistical errors are typically larger than for (naive) fits which might explain why the summation method has often been merely used as a crosscheck for other methods. Moreover, there is no a priori guarantee that the additional suppression of excited states is indeed sufficient for a given observable and target precision. \vspace{0.5\baselineskip}

\item Multi-state fits have become rather sophisticated over the last years and are by now predominantly used to obtain final results. They aim at explicitly correcting lattice data for the leading excited state contamination. Their main advantages are flexibility with respect to the choice of fit ansatz and -- in principle -- the possibility of tracking convergence of the resulting energy gaps. In practice, most implementations use information from the two-point functions to determine energy gaps and make heavy use of priors if more than two states are included, making them more prone to human bias. Moreover, fits naively relying on information on energies from two-point function should be considered unreliable as the fitted energy gaps notoriously fail to converge to the theoretically expected lowest-lying state and it has been shown that these fits may indeed miss the lowest-lying (multiparticle) states which can lead to large residual excited state contamination as observed for \eg axial form factors. \vspace{0.5\baselineskip}

\item The variational method is potentially the most powerful approach as it allows to systematically extract states and suppress excited state contamination in a given matrix element. However, in current implementations the efficacy of the method remains limited and strongly observable-dependent as only fairly small bases of operators have been used and multiparticle operators have not been included at all. The last point causes an issue similar to the one for multi-state fits using information on energy gaps from two-point functions, \ie a small basis particularly without multiparticle may simply miss certain low-lying excited states. Still, such states can yield significant residual contamination depending on the matrix element. \vspace{0.5\baselineskip}
\end{enumerate}

Since no approach is clearly favorable across multiple observables, it remains crucial to crosscheck whatever method is used to obtain final results and perform a careful assessment of the residual excited state contamination. In particular, it is insufficient to claim agreement between two methods if one of them has much larger errors while the result with the smaller statistical error is quoted as the final estimate. Ideally, one should study correlated differences and assign a systematic error due to residual excited state contamination. If multi-state fits are used on a sufficient number of ensembles, non-parametric criteria may be used to further test the plausibility of results. Additional care is required when using methods that are likely to be affected by human bias due to a large number of free parameters or that make model assumptions, such as using the spectrum from the two-point functions to determine energy gaps of the three-point function. \par

One way to systematically improve over existing analyses of excited states would be the inclusion of multiparticle operators in a variational analysis. This should yield a more reliable resolution of the spectrum and thus lead to a more comprehensive and observable independent removal of excited state contamination provided that a large enough basis is used. However, such an extension would still be demanding from a computational as well as technical point of view. Another possibility for future improvement would be a solution or mitigation of the exponential signal-to-noise problem itself. This might be achieved by using multilevel methods as introduced in refs.~\cite{Ce:2016idq,Ce:2016ajy} which have recently been further explored in the context of the hadronic vacuum polarization contribution to $g-2$ \cite{DallaBrida:2020cik}. \par

\section*{Acknowledgments}
I would like to thank the organizers of the 38th International Symposium on Lattice Field Theory for giving me the opportunity to write this review as part of the EPJ-A special issue ``Lattice Field Theory during the covid-19 pandemic'' in replacement of a plenary presentation. Discussions with Kyriakos Hadjiyiannakou, Harvey B. Meyer and Georg von Hippel are gratefully acknowledged. I am indebted to Hartmut Wittig for the careful reading of the manuscript. The author's work is funded by the Deutsche Forschungsgemeinschaft (DFG, German Research Foundation) -- Project No.~399400745 (HI~2048/1-2).
\bibliographystyle{epj.bst}
\bibliography{refs}

\end{document}